\documentclass[a4paper,12pt]{article}

\usepackage{hyperref}

\usepackage[utf8]{inputenc}
\usepackage{{amsmath,amsfonts,latexsym, amstext, amssymb, amsthm}}
\usepackage[english]{babel}
\usepackage{dsfont}
\usepackage{tabularx}
\usepackage{longtable}
\usepackage{comment}
\usepackage[small,bf]{caption}
\usepackage{pstricks}
\usepackage{pst-node}
\usepackage{pst-plot}

\usepackage{multirow}

\usepackage{cite}

\usepackage{tocloft}
\newcommand{\fpage}{\thepage}
\usepackage{fancyhdr}
\fancypagestyle{plain}{
\fancyhf{}
\fancyfoot[C]{\fpage}

}

\usepackage[a4paper,top=2.85cm,bottom=2.85cm,left=2.85cm,right=2.85cm,
marginparsep=0.25cm, 
marginparwidth=2.5cm 
]{geometry}

\usepackage{xspace} 
\usepackage{longtable}
\usepackage{multirow}
\usepackage{pdflscape}
\usepackage[textsize=footnotesize]{todonotes}

\usepackage{etex}
\usepackage{xcolor}

\newlength{\spinlength}
\newlength{\spinheight}
\newlength{\spindepth}
\newlength{\spinwidth}
\newcommand{\upspin}[2][t]{%
\ifthenelse{\equal{#1}{t}}{%
\rule[-\spindepth]{0pt}{\spinlength}
\rule[-\spindepth]{0.5\spinwidth}{0pt}
\psline[arrows=->,origin={0,-\sn\spindepth},xunit=\spinwidth,yunit=\spinlength,linewidth=0.25mm,linecolor=#2](0,0)(0,1)
\rule[\spindepth]{0.5\spinwidth}{0pt}
}%
{\begin{pspicture}(0,0)(0,0)\psline[arrows=->,xunit=\spinwidth,yunit=\spinlength,linewidth=0.25mm,linecolor=#2](0,-0.5)(0,0.5)
\end{pspicture}}}

\newcommand{\downspin}[2][t]{%
\ifthenelse{\equal{#1}{t}}{%
\rule[-\spindepth]{0pt}{\spinlength}
\rule[-\spindepth]{0.5\spinwidth}{0pt}

\psline[arrows=->,origin={0,-\sn\spindepth},xunit=\spinwidth,yunit=\spinlength,linewidth=0.25mm,linecolor=#2](0,1)(0,0)
\rule[-\spindepth]{0.5\spinwidth}{0pt}}%
{\begin{pspicture}(0,0)(0,0)
\psline[arrows=->,xunit=\spinwidth,yunit=\spinlength,linewidth=0.25mm,linecolor=#2](0,0.5)(0,-0.5)
\end{pspicture}}}

\newcommand{\fourspin}[3][t]{%
\ifthenelse{\equal{#1}{t}}{%
\rule[-\spindepth]{0pt}{\spinlength}
\rule[-\spindepth]{0.5\spinwidth}{0pt}
\psline[arrows=-,origin={0,-\sn\spindepth},xunit=\spinwidth,yunit=\spinlength,linewidth=0.25mm,linecolor=#2](0,0)(0,0.5)
\psline[arrows=-,origin={0,-\sn\spindepth},xunit=\spinwidth,yunit=\spinlength,linewidth=0.25mm,linecolor=#3](0,0.5)(0,1)
\rule[\spindepth]{0.5\spinwidth}{0pt}
\psdots[origin={0,-\sn\spindepth},xunit=\spinwidth,yunit=\spinlength,linewidth=0.25mm,linecolor=#2](0,0)(0,0.333)
\psdots[origin={0,-\sn\spindepth},xunit=\spinwidth,yunit=\spinlength,linewidth=0.25mm,linecolor=#3](0,0.666)(0,1)
\psdots[origin={0,-\sn\spindepth},xunit=\spinwidth,yunit=\spinlength,linewidth=0.125mm,linecolor=black](0,0)(0,0.333)
}%
{\begin{pspicture}(0,0)(0,0)
\psline[arrows=-,xunit=\spinwidth,yunit=\spinlength,linewidth=0.25mm,linecolor=#2](0,-0.5)(0,0)
\psline[arrows=-,xunit=\spinwidth,yunit=\spinlength,linewidth=0.25mm,linecolor=#3](0,0)(0,0.5)
\psdots[xunit=\spinwidth,yunit=\spinlength,linewidth=0.25mm,linecolor=#2](0,-0.5)(0,-0.1666)
\psdots[xunit=\spinwidth,yunit=\spinlength,linewidth=0.25mm,linecolor=#3](0,0.1666)(0,0.5)
\psdots[xunit=\spinwidth,yunit=\spinlength,linewidth=0.1mm,linecolor=black](0,-0.5)(0,-0.1666)
\end{pspicture}}}

\newcommand{\foursping}[3][t]{%
\ifthenelse{\equal{#1}{t}}{%
\rule[-\spindepth]{0pt}{\spinlength}
\rule[-\spindepth]{0.5\spinwidth}{0pt}
\psline[arrows=-,origin={0,-\sn\spindepth},xunit=\spinwidth,yunit=\spinlength,linewidth=0.25mm,linecolor=#2](0,0)(0,0.5)
\psline[arrows=-,origin={0,-\sn\spindepth},xunit=\spinwidth,yunit=\spinlength,linewidth=0.25mm,linecolor=#3](0,0.5)(0,1)
\rule[\spindepth]{0.5\spinwidth}{0pt}
\psdots[origin={0,-\sn\spindepth},xunit=\spinwidth,yunit=\spinlength,linewidth=0.25mm,linecolor=#2](0,0)(0,0.333)
\psdots[origin={0,-\sn\spindepth},xunit=\spinwidth,yunit=\spinlength,linewidth=0.25mm,linecolor=#3](0,0.666)(0,1)
\psdots[origin={0,-\sn\spindepth},xunit=\spinwidth,yunit=\spinlength,linewidth=0.125mm,linecolor=gray](0,0)(0,0.333)
}%
{\begin{pspicture}(0,0)(0,0)
\psline[arrows=-,xunit=\spinwidth,yunit=\spinlength,linewidth=0.25mm,linecolor=#2](0,-0.5)(0,0)
\psline[arrows=-,xunit=\spinwidth,yunit=\spinlength,linewidth=0.25mm,linecolor=#3](0,0)(0,0.5)
\psdots[xunit=\spinwidth,yunit=\spinlength,linewidth=0.25mm,linecolor=#2](0,-0.5)(0,-0.1666)
\psdots[xunit=\spinwidth,yunit=\spinlength,linewidth=0.25mm,linecolor=#3](0,0.1666)(0,0.5)
\psdots[xunit=\spinwidth,yunit=\spinlength,linewidth=0.1mm,linecolor=gray](0,-0.5)(0,-0.1666)
\end{pspicture}}}

\newcommand{\twospin}[3][t]{%
\ifthenelse{\equal{#1}{t}}{%
\rule[-\spindepth]{0pt}{\spinlength}
\rule[-\spindepth]{0.5\spinwidth}{0pt}
\psline[arrows=-,origin={0,-\sn\spindepth},xunit=\spinwidth,yunit=\spinlength,linewidth=0.25mm,linecolor=#2](0,0)(0,1)
\rule[\spindepth]{0.5\spinwidth}{0pt}
\psdots[origin={0,-\sn\spindepth},xunit=\spinwidth,yunit=\spinlength,linewidth=0.25mm,linecolor=#2](0,0)(0,1)
\psdots[origin={0,-\sn\spindepth},xunit=\spinwidth,yunit=\spinlength,linewidth=0.125mm,linecolor=#3](0,0)(0,1)
}%
{\begin{pspicture}(0,0)(0,0)\psline[arrows=-,xunit=\spinwidth,yunit=\spinlength,linewidth=0.25mm,linecolor=#2](0,-0.5)(0,0.5)
\psdots[xunit=\spinwidth,yunit=\spinlength,linewidth=0.25mm,linecolor=#2](0,-0.5)(0,0.5)
\psdots[xunit=\spinwidth,yunit=\spinlength,linewidth=0.1mm,linecolor=#3](0,-0.5)(0,0.5)
\end{pspicture}}}

\newgray{ogray}{0.85}
\newgray{hatchgray}{1}
\newgray{sgray}{0.8}
\newgray{hiddengray}{0.9}

\newcommand{\olcolor}{ogray}
\newcommand{\olfillstyle}{crosshatch*}
\newcommand{\olhatchcolor}{ogray}
\newcommand{\orcolor}{ogray}
\newcommand{\orfillstyle}{crosshatch*}
\newcommand{\orhatchcolor}{ogray}
\newcommand{\sfillstyle}{solid}

\newlength{\unit}
\newlength{\rad}
\newlength{\roff}
\newlength{\ri}
\psset{xunit=\unit,yunit=\unit,runit=\unit}
\newlength{\linew}
\newlength{\blacklinew}
\newlength{\dlinewidth}
\newlength{\doublesep}
\psset{doublesep=\doublesep}
\psset{linewidth=\linew}
\psset{dotscale=0.8}
\newlength{\auxlen}
\newlength{\linearc}
\newlength{\flinearc}
\newlength{\xa}
\newlength{\ya}
\newlength{\xb}
\newlength{\yb}
\newlength{\xc}
\newlength{\yc}
\newlength{\xd}
\newlength{\yd}
\newlength{\xe}
\newlength{\ye}

\newlength{\yf}

\newcommand{\uoneprop}[4][white]{%
\setlength{\xa}{0\unit}
\addtolength{\xa}{-1\unit}
\setlength{\xb}{-.6\unit}
\addtolength{\xb}{-0.5\dlinewidth}
\setlength{\xc}{0.6\unit}
\addtolength{\xc}{0.5\dlinewidth}
\setlength{\xd}{0\unit}
\addtolength{\xd}{1\unit}
\setlength{\ya}{0\unit}
\addtolength{\ya}{-1\unit}
\setlength{\yb}{0\unit}
\addtolength{\yb}{-0.5\dlinewidth}
\setlength{\yc}{0\unit}
\addtolength{\yc}{0.5\dlinewidth}
\setlength{\yd}{0\unit}
\addtolength{\yd}{1\unit}
\psset{doubleline=false}
\rput{0}(#2\unit,#3\unit){%
\pscustom[fillstyle=\sfillstyle,fillcolor=#1,linecolor=#1,linewidth=0pt]{%
\rotate{#4}
\psbezier[liftpen=1,linearc=\linearc](\xb,\yc)(0,\yc)(0,\yb)(\xb,\yb)
}
\pscustom[fillstyle=\sfillstyle,fillcolor=#1,linecolor=#1,linewidth=0pt]{%
\rotate{#4}
\psbezier[liftpen=2,linearc=\linearc](\xc,\yb)(0,\yb)(0,\yc)(\xc,\yc)
}
\pscustom{%
\rotate{#4}
\psbezier[linearc=\linearc](\xb,\yc)(-0.1,\yc)(-0.1,\yb)(\xb,\yb)
\psbezier[liftpen=2,linearc=\linearc](\xc,\yb)(0.1,\yb)(0.1,\yc)(\xc,\yc)
}
}
}

\newcommand{\ulinsert}[3][white]{%
\setlength{\xa}{#2\unit}
\addtolength{\xa}{0\unit}
\setlength{\xb}{#2\unit}
\addtolength{\xb}{1.5\unit}
\setlength{\xc}{#2\unit}
\addtolength{\xc}{3\unit}
\setlength{\ya}{#3\unit}
\addtolength{\ya}{0.5\dlinewidth}
\setlength{\yb}{#3\unit}
\addtolength{\yb}{-0.5\dlinewidth}
\setlength{\yc}{#3\unit}
\addtolength{\yc}{-1.5\unit}
\psset{doubleline=false}
\pscustom[fillstyle=\sfillstyle,fillcolor=#1,linecolor=#1,linewidth=0pt]{%
\psline[liftpen=1,linearc=\linearc](\xc,\yb)(\xb,\yb)(\xb,\yc)
\psline[liftpen=1](\xb,\ya)(\xc,\ya)}
\pscustom[fillstyle=\olfillstyle,fillcolor=\olcolor,hatchcolor=\olhatchcolor,
linecolor=\olcolor,linewidth=\linew]{%
\psline[linearc=\linearc](\xb,\yc)(\xb,\ya)
\psline[liftpen=1,linearc=2\linearc](\xb,\ya)(\xa,\ya)(\xa,\yc)}
\psline[linearc=\linearc,linewidth=\blacklinew](\xb,\yc)(\xb,\yb)(\xc,\yb)
\psline[linearc=2\linearc,linewidth=\blacklinew](\xa,\yc)(\xa,\ya)(\xb,\ya)
\psline[linewidth=\blacklinew](\xb,\ya)(\xc,\ya)
}
\newcommand{\dlinsert}[3][white]{%

\setlength{\xa}{#2\unit}
\addtolength{\xa}{0\unit}
\setlength{\xb}{#2\unit}
\addtolength{\xb}{1.5\unit}
\setlength{\xc}{#2\unit}
\addtolength{\xc}{3\unit}
\setlength{\ya}{#3\unit}
\addtolength{\ya}{-0.5\dlinewidth}
\setlength{\yb}{#3\unit}
\addtolength{\yb}{0.5\dlinewidth}
\setlength{\yc}{#3\unit}
\addtolength{\yc}{1.5\unit}
\psset{doubleline=false}
\pscustom[fillstyle=\sfillstyle,fillcolor=#1,linecolor=#1,linewidth=0pt]{%
\psline[linearc=\linearc](\xc,\yb)(\xb,\yb)(\xb,\yc)
\psline(\xb,\ya)(\xc,\ya)}
\pscustom[fillstyle=\olfillstyle,fillcolor=\olcolor,hatchcolor=\olhatchcolor,
linecolor=\olcolor,linewidth=\linew]{%
\psline[linearc=\linearc](\xb,\yc)(\xb,\ya)
\psline[liftpen=1,linearc=2\linearc](\xb,\ya)(\xa,\ya)(\xa,\yc)}
\psline[linearc=\linearc,linewidth=\blacklinew](\xb,\yc)(\xb,\yb)(\xc,\yb)
\psline[linearc=2\linearc,linewidth=\blacklinew](\xa,\yc)(\xa,\ya)(\xb,\ya)
\psline[linewidth=\blacklinew](\xb,\ya)(\xc,\ya)
}
\newcommand{\drinsert}[3][white]{%
\setlength{\xa}{#2\unit}
\addtolength{\xa}{0\unit}
\setlength{\xb}{#2\unit}
\addtolength{\xb}{-1.5\unit}
\setlength{\xc}{#2\unit}
\addtolength{\xc}{-3\unit}
\setlength{\ya}{#3\unit}
\addtolength{\ya}{-0.5\dlinewidth}
\setlength{\yb}{#3\unit}
\addtolength{\yb}{0.5\dlinewidth}
\setlength{\yc}{#3\unit}
\addtolength{\yc}{1.5\unit}
\psset{doubleline=false}
\pscustom[fillstyle=\sfillstyle,fillcolor=#1,linecolor=#1,linewidth=0pt]{%
\psline[linearc=\linearc](\xc,\yb)(\xb,\yb)(\xb,\yc)
\psline(\xb,\ya)(\xc,\ya)}
\pscustom[fillstyle=\orfillstyle,fillcolor=\orcolor,hatchcolor=\orhatchcolor,
linecolor=\orcolor,linewidth=\linew]{%
\psline[linearc=\linearc](\xb,\yc)(\xb,\ya)
\psline[liftpen=1,linearc=2\linearc](\xb,\ya)(\xa,\ya)(\xa,\yc)}
\psline[linearc=\linearc,linewidth=\blacklinew](\xb,\yc)(\xb,\yb)(\xc,\yb)
\psline[linearc=2\linearc,linewidth=\blacklinew](\xa,\yc)(\xa,\ya)(\xb,\ya)
\psline[linewidth=\blacklinew](\xb,\ya)(\xc,\ya)
}
\newcommand{\urinsert}[3][white]{%
\setlength{\xa}{#2\unit}
\addtolength{\xa}{0\unit}
\setlength{\xb}{#2\unit}
\addtolength{\xb}{-1.5\unit}
\setlength{\xc}{#2\unit}
\addtolength{\xc}{-3\unit}
\setlength{\ya}{#3\unit}
\addtolength{\ya}{0.5\dlinewidth}
\setlength{\yb}{#3\unit}
\addtolength{\yb}{-0.5\dlinewidth}
\setlength{\yc}{#3\unit}
\addtolength{\yc}{-1.5\unit}
\psset{doubleline=false}
\pscustom[fillstyle=\sfillstyle,fillcolor=#1,linecolor=#1,linewidth=0pt]{%
\psline[linearc=\linearc](\xc,\yb)(\xb,\yb)(\xb,\yc)
\psline[liftpen=1](\xb,\ya)(\xc,\ya)}
\pscustom[fillstyle=\orfillstyle,fillcolor=\orcolor,hatchcolor=\orhatchcolor,
linecolor=\orcolor,linewidth=\linew]{%
\psline[linearc=\linearc](\xb,\yc)(\xb,\ya)
\psline[liftpen=1,linearc=2\linearc](\xb,\ya)(\xa,\ya)(\xa,\yc)}
\psline[linearc=\linearc,linewidth=\blacklinew](\xb,\yc)(\xb,\yb)(\xc,\yb)
\psline[linearc=2\linearc,linewidth=\blacklinew](\xa,\yc)(\xa,\ya)(\xb,\ya)
\psline[linewidth=\blacklinew](\xb,\ya)(\xc,\ya)
}
\newcommand{\olvertex}[3][white]{%
\setlength{\xa}{#2\unit}
\addtolength{\xa}{0\unit}
\setlength{\xb}{#2\unit}
\addtolength{\xb}{1.5\unit}
\setlength{\xc}{#2\unit}
\addtolength{\xc}{3\unit}
\setlength{\ya}{#3\unit}
\addtolength{\ya}{1.5\unit}
\setlength{\yb}{#3\unit}
\addtolength{\yb}{0.5\dlinewidth}
\setlength{\yc}{#3\unit}
\addtolength{\yc}{-0.5\dlinewidth}
\setlength{\yd}{#3\unit}
\addtolength{\yd}{-1.5\unit}
\psset{doubleline=false}
\pscustom[fillstyle=\sfillstyle,fillcolor=#1,linecolor=#1,linewidth=0pt]{%
\psline[linearc=\linearc](\xc,\yb)(\xb,\yb)(\xb,\ya)
\psline[liftpen=1,linearc=\linearc](\xb,\yd)(\xb,\yc)(\xc,\yc)}
\pscustom[fillstyle=\olfillstyle,fillcolor=\olcolor,hatchcolor=\olhatchcolor,
linecolor=\olcolor,linewidth=0pt]{%
\psline[liftpen=0](\xa,\yd)(\xb,\yd)
\psline[liftpen=0](\xb,\ya)(\xa,\ya)
}
\psline[linecolor=\olcolor,linewidth=\linew](\xb,\ya)(\xb,\yd)
\psline[linearc=\linearc,linewidth=\blacklinew]{-C}(\xc,\yb)(\xb,\yb)(\xb,\ya)
\psline[liftpen=1,linearc=\linearc,linewidth=\blacklinew](\xb,\yd)(\xb,\yc)(\xc,\yc)

\psline[linewidth=\blacklinew]{C-}(\xa,\ya)(\xa,\yd)
}
\newcommand{\orvertex}[3][white]{%
\setlength{\xa}{#2\unit}
\addtolength{\xa}{0\unit}
\setlength{\xb}{#2\unit}
\addtolength{\xb}{-1.5\unit}
\setlength{\xc}{#2\unit}
\addtolength{\xc}{-3\unit}
\setlength{\ya}{#3\unit}
\addtolength{\ya}{1.5\unit}
\setlength{\yb}{#3\unit}
\addtolength{\yb}{0.5\dlinewidth}
\setlength{\yc}{#3\unit}
\addtolength{\yc}{-0.5\dlinewidth}
\setlength{\yd}{#3\unit}
\addtolength{\yd}{-1.5\unit}
\psset{doubleline=false}
\pscustom[fillstyle=\sfillstyle,fillcolor=#1,linecolor=#1,linewidth=0pt]{%
\psline[linearc=\linearc](\xc,\yb)(\xb,\yb)(\xb,\ya)
\psline[liftpen=1,linearc=\linearc](\xb,\yd)(\xb,\yc)(\xc,\yc)}
\pscustom[fillstyle=\orfillstyle,fillcolor=\orcolor,hatchcolor=\orhatchcolor,
linecolor=\orcolor,linewidth=0pt]{%
\psline[liftpen=0](\xb,\ya)(\xa,\ya)
\psline[liftpen=0](\xa,\yd)(\xb,\yd)
\psline[liftpen=0](\xb,\ya)(\xa,\ya)
}
\psline[linecolor=\orcolor,linewidth=\linew](\xb,\ya)(\xb,\yd)
\psline[linearc=\linearc,linewidth=\blacklinew]{-C}(\xc,\yb)(\xb,\yb)(\xb,\ya)
\psline[liftpen=1,linearc=\linearc,linewidth=\blacklinew](\xb,\yd)(\xb,\yc)(\xc,\yc)
\psline[linewidth=\blacklinew]{C-}(\xa,\ya)(\xa,\yd)
}

\newcommand{\uoutex}[3][white]{%
\setlength{\xb}{#2\unit}
\addtolength{\xb}{-1.5\unit}
\setlength{\xc}{#2\unit}
\setlength{\ya}{#3\unit}
\addtolength{\ya}{0.5\dlinewidth}
\setlength{\yb}{#3\unit}
\addtolength{\yb}{-0.5\dlinewidth}
\setlength{\yc}{#3\unit}
\addtolength{\yc}{-1.5\unit}
\psset{doubleline=false}
\pscustom[fillstyle=\sfillstyle,fillcolor=#1,linecolor=#1]{%
\psline[liftpen=1,linearc=\linearc](\xc,\yb)(\xb,\yb)(\xb,\yc)
\psline[liftpen=1](\xb,\ya)(\xc,\ya)}
\psline[linearc=\linearc,linewidth=\blacklinew](\xb,\yc)(\xb,\yb)(\xc,\yb)
\psline[linewidth=\blacklinew](\xb,\ya)(\xc,\ya)

}
\newcommand{\doutex}[3][white]{%
\setlength{\xb}{#2\unit}
\addtolength{\xb}{-1.5\unit}
\setlength{\xc}{#2\unit}
\setlength{\ya}{#3\unit}
\addtolength{\ya}{-0.5\dlinewidth}
\setlength{\yb}{#3\unit}
\addtolength{\yb}{0.5\dlinewidth}
\setlength{\yc}{#3\unit}
\addtolength{\yc}{1.5\unit}
\psset{doubleline=false}
\pscustom[fillstyle=\sfillstyle,fillcolor=#1,linecolor=#1]{%
\psline[linearc=\linearc](\xc,\yb)(\xb,\yb)(\xb,\yc)
\psline(\xb,\ya)(\xc,\ya)}
\psline[linearc=\linearc,linewidth=\blacklinew](\xb,\yc)(\xb,\yb)(\xc,\yb)
\psline[linewidth=\blacklinew](\xb,\ya)(\xc,\ya)
}
\newcommand{\dinex}[3][white]{%
\setlength{\xb}{#2\unit}
\addtolength{\xb}{1.5\unit}
\setlength{\xc}{#2\unit}
\setlength{\ya}{#3\unit}
\addtolength{\ya}{-0.5\dlinewidth}
\setlength{\yb}{#3\unit}
\addtolength{\yb}{0.5\dlinewidth}
\setlength{\yc}{#3\unit}
\addtolength{\yc}{1.5\unit}
\psset{doubleline=false}
\pscustom[fillstyle=\sfillstyle,fillcolor=#1,linecolor=#1]{%
\psline[linearc=\linearc](\xc,\yb)(\xb,\yb)(\xb,\yc)
\psline(\xb,\ya)(\xc,\ya)}
\psline[linearc=\linearc,linewidth=\blacklinew](\xb,\yc)(\xb,\yb)(\xc,\yb)
\psline[linewidth=\blacklinew](\xb,\ya)(\xc,\ya)
}
\newcommand{\uinex}[3][white]{%
\setlength{\xb}{#2\unit}
\addtolength{\xb}{1.5\unit}
\setlength{\xc}{#2\unit}
\setlength{\ya}{#3\unit}
\addtolength{\ya}{0.5\dlinewidth}
\setlength{\yb}{#3\unit}
\addtolength{\yb}{-0.5\dlinewidth}
\setlength{\yc}{#3\unit}
\addtolength{\yc}{-1.5\unit}
\psset{doubleline=false}
\pscustom[fillstyle=\sfillstyle,fillcolor=#1,linecolor=#1]{%
\psline[linearc=\linearc](\xc,\yb)(\xb,\yb)(\xb,\yc)
\psline[liftpen=1](\xb,\ya)(\xc,\ya)}
\psline[linearc=\linearc,linewidth=\blacklinew](\xb,\yc)(\xb,\yb)(\xc,\yb)
\psline[linewidth=\blacklinew](\xb,\ya)(\xc,\ya)
}
\newcommand{\ioutex}[3][white]{%
\setlength{\xb}{#2\unit}
\addtolength{\xb}{-1.5\unit}
\setlength{\xc}{#2\unit}
\setlength{\ya}{#3\unit}
\addtolength{\ya}{1.5\unit}
\setlength{\yb}{#3\unit}
\addtolength{\yb}{0.5\dlinewidth}
\setlength{\yc}{#3\unit}
\addtolength{\yc}{-0.5\dlinewidth}
\setlength{\yd}{#3\unit}
\addtolength{\yd}{-1.5\unit}
\psset{doubleline=false}
\pscustom[fillstyle=\sfillstyle,fillcolor=#1,linecolor=#1]{%
\psline[linearc=\linearc](\xc,\yb)(\xb,\yb)(\xb,\ya)
\psline[liftpen=1,linearc=\linearc](\xb,\yd)(\xb,\yc)(\xc,\yc)}
\psline[linecolor=#1,linewidth=\blacklinew](\xb,\ya)(\xb,\yd)
\psline[linearc=\linearc,linewidth=\blacklinew]{-C}(\xc,\yb)(\xb,\yb)(\xb,\ya)
\psline[liftpen=1,linearc=\linearc,linewidth=\blacklinew](\xb,\yd)(\xb,\yc)(\xc,\yc)
}
\newcommand{\iinex}[3][white]{%
\setlength{\xb}{#2\unit}
\addtolength{\xb}{1.5\unit}
\setlength{\xc}{#2\unit}
\setlength{\ya}{#3\unit}
\addtolength{\ya}{1.5\unit}
\setlength{\yb}{#3\unit}
\addtolength{\yb}{0.5\dlinewidth}
\setlength{\yc}{#3\unit}
\addtolength{\yc}{-0.5\dlinewidth}
\setlength{\yd}{#3\unit}
\addtolength{\yd}{-1.5\unit}
\psset{doubleline=false}
\pscustom[fillstyle=\sfillstyle,fillcolor=#1,linecolor=#1,linewidth=0pt]{%
\psline[linearc=\linearc](\xc,\yb)(\xb,\yb)(\xb,\ya)
\psline[liftpen=1,linearc=\linearc](\xb,\yd)(\xb,\yc)(\xc,\yc)}
\psline[linecolor=#1,linewidth=\blacklinew](\xb,\ya)(\xb,\yd)
\psline[linearc=\linearc,linewidth=\blacklinew]{-C}(\xc,\yb)(\xb,\yb)(\xb,\ya)
\psline[liftpen=1,linearc=\linearc,linewidth=\blacklinew](\xb,\yd)(\xb,\yc)(\xc,\yc)
}

\usepackage{rotating}
\newlength{\arlength}
\newlength{\arheight}

\makeatletter
\DeclareRobustCommand*{\bfseries}{%
  \not@math@alphabet\bfseries\mathbf
  \fontseries\bfdefault\selectfont
  \boldmath
}
\makeatother

\DeclareMathOperator{\diladensity}{\mathfrak{D}}
\newcommand{\joinsymbol}{\star}

\newcommand{\sDO}{d}
\newcommand{\Sthree}{\mathbb{S}_3}
\newcommand{\tablefrac}[2]{\frac{#1}{#2}}

\newcommand{\specpar}[1]{s_{#1}}

\newcommand{\ZZ}{\ensuremath{\mathbb{Z}}}
\newcommand{\NN}{\ensuremath{\mathbb{N}}}

\newcommand{\UN}{\ensuremath{U(N)}\xspace}
\newcommand{\SUN}{\ensuremath{SU(N)}\xspace}
\newcommand{\U}[1]{\ensuremath{U(#1)}\xspace}
\newcommand{\SU}[1]{\ensuremath{SU(#1)}\xspace}
\newcommand{\SO}[1]{\ensuremath{SO(#1)}\xspace}

\newcommand{\su}[1]{\ensuremath{\mathfrak{su}(#1)}\xspace}

\newcommand{\complexi}{i}

\newcommand{\comm}[3][]{{}[{}#2{}\,{\overset{#1}{,}}\,{}#3{}]{}}
\newcommand{\acomm}[3][]{{}\{{}#2{}\,{\overset{#1}{,}}\,{}#3{}\}{}}

\newcommand{\YM}{{\mathrm{\scriptscriptstyle YM}}}
\newcommand{\maxset}[1]{\max{(#1)}}
\newcommand{\minset}[1]{\min{(#1)}}
\DeclareMathOperator{\tr}{tr}

\DeclareMathOperator{\phaneq}{\phantom{{}=}}

\newcommand{\colors}{s}

\newcommand{\e}{\operatorname{e}}
\newcommand{\cstar}{\ensuremath{\ast}}

\newcommand{\ket}[1]{\mathopen{\mid}{}#1{}\mathclose{\rangle}}

\newcommand{\bra}[1]{\mathopen{\langle}{}#1{}\mathclose{\mid}}
\newcommand{\starcomm}[2]{\comm[\cstar]{#1}{#2}}
\newcommand{\staracomm}[2]{\acomm[\cstar]{#1}{#2}}

\newcommand{\vac}{\operatorname{\mid} 0 \, \operatorname{\rangle}}

\newcommand{\eqndot}{\, .}
\newcommand{\eqncom}{\, ,}

\newcommand{\de}{\operatorname{d}\!}

\newcommand{\cA}{\mathcal{A}}

\newcommand{\cD}{\mathcal{D}}
\newcommand{\cE}{\mathcal{E}}
\newcommand{\cF}{\mathcal{F}}

\newcommand{\cN}{\mathcal{N}}
\newcommand{\cO}{\mathcal{O}}
\newcommand{\cP}{\mathcal{P}}
\newcommand{\cQ}{\mathcal{Q}}

\newcommand{\ba}{\mathbf{a}}
\newcommand{\bb}{\mathbf{b}}
\newcommand{\bc}{\mathbf{c}}

\newcommand{\alphadot}{\dot{\alpha}}
\newcommand{\betadot}{\dot{\beta}}

\newcommand{\Nfour}{$\mathcal{N}=4$\xspace}

\newcommand{\NfSYMt}{\Nfour SYM theory\xspace}

\newcommand{\ferm}{\psi}

\newcommand{\aosc}{\ba}
\newcommand{\aoscdag}{\aosc^\dagger}
\newcommand{\bosc}{\bb}
\newcommand{\boscdag}{\bosc^\dagger}
\newcommand{\cosc}{\bc}
\newcommand{\coscdag}{\cosc^\dagger}

\newcommand{\akind}[1][ ]{a^{#1}}
\newcommand{\bkind}[1][ ]{b^{#1}}
\newcommand{\ckind}[1][ ]{c^{#1}}

\newcommand{\akindsite}[2][ ]{a^{#1}_{(#2)}}
\newcommand{\bkindsite}[2][ ]{b^{#1}_{(#2)}}
\newcommand{\ckindsite}[2][ ]{c^{#1}_{(#2)}}

\DeclareMathOperator{\T}{T}

\newlength{\eqoff}

\DeclareMathOperator{\D}{D}

\usepackage{feynmp}
\newcommand{\feynmpdefinitionscaps}{

\fmfcmd{%
linejoin:=rounded;
}

\fmfcmd{%
style_def plain_n expr p =
 linecap:=butt;
 cdraw p;
 linecap:=rounded;
enddef;
style_def plain_h expr p =
 linecap:=butt;
 cdraw subpath (0, 1length p ) of p;
 linecap:=rounded;
 cdraw subpath (0.95length p ,length p ) of p;
enddef;
style_def plain_t expr p =
 linecap:=rounded;
 cdraw subpath (0, 0.05length p ) of p;
 linecap:=butt;
 cdraw subpath (0,length p ) of p;
 linecap:=rounded;
enddef;
style_def plain_ht expr p =
 linecap:=rounded;
 cdraw p;
enddef;
style_def iplain_n expr p =
 linecap:=squared;
 cdraw p;
 linecap:=rounded;
enddef;
style_def iplain_h expr p =
 linecap:=squared;
 cdraw subpath (0, 0.5length p ) of p;
 linecap:=rounded;
 cdraw subpath (0.5length p ,length p ) of p;
enddef;
style_def iplain_t expr p =
 linecap:=rounded;
 cdraw subpath (0, 0.5length p ) of p;
 linecap:=squared;
 cdraw subpath (0.5length p,length p ) of p;
 linecap:=rounded;
enddef;
style_def iplain_ht expr p =
 linecap:=rounded;
 cdraw p;
enddef;
style_def interrupted_plain_n expr p =
 draw_plain_h subpath (0, 0.25length p ) of p;
 draw_plain_t subpath (0.75length p ,length p ) of p;
enddef;
style_def interrupted_plain_h expr p =
 draw_plain_h subpath (0, 0.25length p ) of p;
 draw_plain_ht subpath (0.75length p ,length p ) of p;
enddef;
style_def interrupted_plain_t expr p =
 draw_plain_ht subpath (0, 0.25length p ) of p;
 draw_plain_t subpath (0.75length p ,length p ) of p;
enddef;
style_def interrupted_plain_ht expr p =
 draw_plain_ht subpath (0, 0.25length p ) of p;
 draw_plain_ht subpath (0.75length p ,length p ) of p;
enddef;
style_def iinterrupted_plain_n expr p =
 draw_iplain_h subpath (0, 0.25length p ) of p;
 draw_iplain_t subpath (0.75length p ,length p ) of p;
enddef;
style_def iinterrupted_plain_h expr p =
 draw_iplain_h subpath (0, 0.25length p ) of p;
 draw_iplain_ht subpath (0.75length p ,length p ) of p;
enddef;
style_def iinterrupted_plain_t expr p =
 draw_iplain_ht subpath (0, 0.25length p ) of p;
 draw_iplain_t subpath (0.75length p ,length p ) of p;
enddef;
style_def iinterrupted_plain_ht expr p =
 draw_iplain_ht subpath (0, 0.25length p ) of p;
 draw_iplain_ht subpath (0.75length p ,length p ) of p;
enddef;
vardef shift_p expr p =  
p  shifted 4.5 (unitvector direction (length (p)/2) of p rotated +90)
enddef;
style_def leftrightarrows expr p =
  shrink 0.5;
  cfill arrow (shift_p (p));
  cfill arrow (shift_p (reverse p));
  endshrink;
enddef;%
style_def leftrightarrows_interrupted expr p =
  shrink 0.5;
  cfill arrow (shift_p (subpath (0, 0.4length p ) of p));
  cfill arrow (shift_p (reverse (subpath (0, 0.4length p ) of p)));
  cfill arrow (shift_p (subpath (0.6length p ,length p ) of p));
  cfill arrow (shift_p (reverse (subpath (0.6length p ,length p ) of p)));
  endshrink;
enddef;%
style_def leftrightarrows_interruptedold expr p =
  shrink 0.5;
  cfill arrow (shift_p (subpath (0, 0.25length p ) of p));
  cfill arrow (shift_p (reverse (subpath (0, 0.25length p ) of p)));
  cfill arrow (shift_p (subpath (0.75length p ,length p ) of p));
  cfill arrow (shift_p (reverse (subpath (0.75length p ,length p ) of p)));
  endshrink;
enddef;%
style_def leftrightarrows_start expr p =
  shrink 0.5;
  cfill arrow (shift_p (subpath (0, 0.1length p ) of p));
  cfill arrow (shift_p (reverse (subpath (0, 0.1length p ) of p)));
  endshrink;
enddef;}
}
\newcommand{\feynmpdefinitionsdiags}{

\fmfcmd{%
thin := 1pt; 
thick := 2thin;
arrow_len := 4mm;
arrow_ang := 15;
curly_len := 3mm;
dash_len := 1.8mm; 
dot_len := 0.75mm; 
wiggly_len := 2mm; 
wiggly_slope := 60;
zigzag_len := 2mm;
zigzag_width := 2thick;
decor_size := 5mm;
dot_size := 2thick;
}

\fmfcmd{%
style_def plain_sarrow expr p =
  cdraw p;
  shrink (0.55); 
  cfill (arrow p);
  endshrink;
enddef;
style_def dashes_sarrow expr p =
  draw_dashes p;
  shrink (0.55);
  cfill (arrow p);
  endshrink;
enddef;
style_def dots_sarrow expr p =
  draw_dots p;
  shrink (0.55);
  cfill (arrow p);
  endshrink;
enddef;
style_def plain_srarrow expr p =
  cdraw p;
  shrink (0.55);
  cfill (arrow (reverse p));
  endshrink;
enddef;
style_def dashes_srarrow expr p =
  draw_dashes p;
  shrink (0.55);
  cfill (arrow (reverse p));
  endshrink;
enddef;
style_def dots_srarrow expr p =
  draw_dots p;
  shrink (0.55);
  cfill (arrow (reverse p));
  endshrink;
enddef;
marksize=2mm;
def draw_mark(expr p,a) =
  begingroup
    save t,tip,dma,dmb; pair tip,dma,dmb;
    t=arctime a of p;
    tip =marksize*unitvector direction t of p;
    dma =marksize*unitvector direction t of p rotated -45;
    dmb =marksize*unitvector direction t of p rotated 45;
    linejoin:=beveled;
    draw (-.5dma.. .5tip-- -.5dmb) shifted point t of p;
  endgroup
enddef;
style_def derplain expr p =
    save amid;
    amid=.5*arclength p;
    draw_mark(p, amid);
    draw p;
enddef;
}
}

\usepackage{graphicx}

\usepackage{xkeyval}
\makeatletter

\define@boolkey{FDiagram}{schannel}[true]{}
\define@boolkey{FDiagram}{tchannel}[true]{}
\define@boolkey{FDiagram}{xchannel}[true]{}
\define@boolkey{FDiagram}{leftSE}[true]{}
\define@boolkey{FDiagram}{rightSE}[true]{}
\define@boolkey{FDiagram}{long}[true]{}
\define@boolkey{FDiagram}{longup}[true]{}
\define@boolkey{FDiagram}{oldlabels}[true]{}
\define@key{FDiagram}{styleleftbottom}%
 {\def\FDiagramstyleleftbottom{#1}}
\define@key{FDiagram}{stylerightbottom}%
 {\def\FDiagramstylerightbottom{#1}}
\define@key{FDiagram}{stylelefttop}%
 {\def\FDiagramstylelefttop{#1}}
\define@key{FDiagram}{stylerighttop}%
 {\def\FDiagramstylerighttop{#1}}
\define@key{FDiagram}{stylemid}%
 {\def\FDiagramstylemid{#1}}
\define@key{FDiagram}{labelleftbottom}%
 {\def\FDiagramlabelleftbottom{#1}}
\define@key{FDiagram}{labelrightbottom}%
 {\def\FDiagramlabelrightbottom{#1}}
\define@key{FDiagram}{labellefttop}%
 {\def\FDiagramlabellefttop{#1}}
\define@key{FDiagram}{labelrighttop}%
 {\def\FDiagramlabelrighttop{#1}}
\define@key{FDiagram}{labelmid}%
 {\def\FDiagramlabelmid{#1}}
\presetkeys{FDiagram}{%
schannel=false,%
tchannel=false,%
xchannel=false,%
leftSE=false,%
rightSE=false,%
long=false,%
longup=false,%
oldlabels=false,%
labelleftbottom=$\phantom{0}$,%
labelrightbottom=$\phantom{0}$,%
labellefttop=$\phantom{0}$,%
labelrighttop=$\phantom{0}$,%
labelmid=$\phantom{0}$,%
styleleftbottom=plain,%
styleleftbottom=plain,%
stylerightbottom=plain,%
stylelefttop=plain,%
stylerighttop=plain,%
stylemid=plain}{}
\newcommand*\FDiagram[6][]{%
 \setkeys{FDiagram}{#1}%
\settoheight{\eqoff}{$\times$}%
\setlength{\eqoff}{0.5\eqoff}%
\addtolength{\eqoff}{-12.0\unitlength}%
\raisebox{\eqoff}{%
\fmfframe(2,2)(2,2){%
\begin{fmfchar*}(12,20)
\fmfbottom{vb1,vbb1,v1,vb2,vbb2,vb3,vb4,v2,vbb5,vb5}
\fmftop{vt1,vtt1,v3,vt2,vtt2,vt3,vt4,v4,vtt5,vt5}
\ifKV@FDiagram@schannel
\fmf{#2,left=0.3,tension=1}{v1,vc1}
\fmf{#3,right=0.3,tension=1}{v2,vc1}
\fmf{#4,tension=2}{vc1,vc2}
\fmf{#5,left=0.3,foreground=(0.65,,0.65,,0.65)}{vc2,v3}
\fmf{#6,right=0.3,foreground=(0.65,,0.65,,0.65)}{vc2,v4}
\else\fi
\ifKV@FDiagram@tchannel
\fmf{#2,left=0,tension=1}{v1,vc1}
\fmf{#3,right=0,tension=1}{v2,vc2}
\fmf{#4,tension=0}{vc1,vc2}
\fmf{#5,foreground=(0.65,,0.65,,0.65)}{vc1,v3}
\fmf{#6,foreground=(0.65,,0.65,,0.65)}{vc2,v4}
\else\fi
\ifKV@FDiagram@xchannel
\fmf{#2,left=0.25,tension=1}{v1,vc1}
\fmf{#3,right=0.25,tension=1}{v2,vc1}
\fmf{#5,left=0.25,foreground=(0.65,,0.65,,0.65)}{vc1,v3}
\fmf{#6,right=0.25,foreground=(0.65,,0.65,,0.65)}{vc1,v4}
\else\fi
\ifKV@FDiagram@leftSE
\fmf{#2,left=0,tension=1}{v1,vc1}
\fmf{#2,right=0,tension=1,foreground=(0.65,,0.65,,0.65)}{vc1,v3}
\fmf{#3,foreground=(0.65,,0.65,,0.65)}{v2,v4}
\fmfv{decor.shape=circle,decor.filled=full,decor.size=15}{vc1}
\else\fi
\ifKV@FDiagram@rightSE
\fmf{#3,left=0,tension=1}{v2,vc1}
\fmf{#3,right=0,tension=1,foreground=(0.65,,0.65,,0.65)}{vc1,v4}
\fmf{#2,foreground=(0.65,,0.65,,0.65)}{v1,v3}
\fmfv{decor.shape=circle,decor.filled=full,decor.size=15}{vc1}
\else\fi
\fmffreeze
\fmfposition
\fmfipath{p[]}
\fmfipair{vm[]}
\fmfcmd{pair verta, vertb, vertc, vertd, vertca, vertcb; verta = vloc(__v1); vertb = vloc(__v2); vertc = vloc(__v3); vertd = vloc(__v4); vertca = vloc(__vc1); vertcb = vloc(__vc2);}
\ifKV@FDiagram@oldlabels
\fmfiv{label=\FDiagramlabelleftbottom,l.a=+120,l.dist=0.07w}{verta}
\fmfiv{label=\FDiagramlabelrightbottom,l.a=+60,l.dist=0.07w}{vertb}
\ifKV@FDiagram@tchannel
\fmfiv{label=\FDiagramlabelmid,l.a=30,l.dist=0.18w}{vertca}
\else
\fmfiv{label=\FDiagramlabelmid,l.a=60,l.dist=0.20w}{vertca}
\fi
\fmfiv{label=\FDiagramlabellefttop,l.a=-120,l.dist=0.07w}{vertc}
\fmfiv{label=\FDiagramlabelrighttop,l.a=-60,l.dist=0.07w}{vertd}
\else
\fmfiv{label=\FDiagramlabelleftbottom,l.a=+120,l.dist=0.07w}{verta}
\fmfiv{label=\FDiagramlabelrightbottom,l.a=+60,l.dist=0.07w}{vertb}
\ifKV@FDiagram@tchannel
\fmfiv{label=\FDiagramlabellefttop,l.a=+120,l.dist=0.07w}{vertca}
\fmfiv{label=\FDiagramlabelrighttop,l.a=+60,l.dist=0.07w}{vertcb}
\else
\fi
\ifKV@FDiagram@schannel
\fmfiv{label=\FDiagramlabellefttop,l.a=+165,l.dist=0.22w}{vertcb}
\fmfiv{label=\FDiagramlabelrighttop,l.a=+15,l.dist=0.22w}{vertcb}
\else
\fi
\ifKV@FDiagram@xchannel
\fmfiv{label=\FDiagramlabellefttop,l.a=+158,l.dist=0.20w}{vertca+(0,0.025h)}
\fmfiv{label=\FDiagramlabelrighttop,l.a=+22,l.dist=0.20w}{vertca+(0,0.025h)}
\else
\fi
\ifKV@FDiagram@leftSE
\fmfiv{label=\FDiagramlabellefttop,l.a=+120,l.dist=0.07w}{verta+(0,0.6125h)}
\else
\fi
\ifKV@FDiagram@rightSE
\fmfiv{label=\FDiagramlabelrighttop,l.a=+60,l.dist=0.07w}{vertb+(0,0.6125h)}
\else
\fi
\fi
\fmfdraw
\ifKV@FDiagram@long
\fmf{plain,width=1mm}{vb1,vb5}
\else 
\fmf{plain,width=1mm}{v1,v2}
\fi
\ifKV@FDiagram@longup
\fmf{plain,width=1mm,foreground=(0.65,,0.65,,0.65)}{vt1,vt5}
\fi
\end{fmfchar*}%
}}%
}
\title{The Complete One-Loop Dilatation Operator of the planar $\beta$-deformed $\mathcal{N}=4$ SYM theory}
\author{Jan Fokken, Christoph Sieg and Matthias Wilhelm}

\begin{document}

\begin{fmffile}{diagrams}
\feynmpdefinitionscaps
\feynmpdefinitionsdiags


\begingroup\parindent0pt
\begin{flushright}\footnotesize
\texttt{HU-MATH-2013-22}\\
\texttt{HU-EP-13/72}
\end{flushright}
\vspace*{2em}
\centering
\begingroup\LARGE
\bf
The complete one-loop dilatation operator of planar real $\beta$-deformed \\
$\mathcal{N}=4$ SYM theory%
\par\endgroup
\vspace{1.5em}
\begingroup\large
{\bf Jan Fokken,
Christoph Sieg,
Matthias Wilhelm}
\par\endgroup
\vspace{1em}
\begingroup\itshape
Institut f\"ur Mathematik und Institut f\"ur Physik\\
Humboldt-Universit\"at zu Berlin\\
IRIS Geb\"aude \\
Zum Grossen Windkanal 6 \\
12489 Berlin
\par\endgroup
\vspace{1em}
\begingroup\ttfamily
fokken, csieg, mwilhelm@physik.hu-berlin.de \\
\par\endgroup
\vspace{1.5em}
\endgroup
\thispagestyle{empty}

\paragraph{Abstract.}
We determine the missing finite-size corrections to the asymptotic one-loop dilatation operator of the real $\beta$-deformed $\mathcal{N}=4$ SYM theory for the gauge groups \UN and \SUN in the 't Hooft limit. In the \SUN case, the absence of the \U1 field components leads to a new kind of finite-size effect, which we call prewrapping. We classify which states are potentially affected by prewrapping at generic loop orders and comment on the necessity to include it into the integrability-based description. As a further result, we identify classes of $n$-point correlation functions which at all loop orders in the planar theory are given by the values of their undeformed counterparts. Finally, we determine the superconformal multiplet structure and one-loop anomalous dimensions of all single-trace states with classical scaling dimension $\Delta_0 \leq 4.5$.


\newpage

\setcounter{tocdepth}{2}
\par\noindent\hrulefill\par\vskip 0em
\tableofcontents
\par\noindent\hrulefill\par\vskip-4.3em


\section{Introduction}

\subsection{General setup and conformality}

The $\text{AdS}/\text{CFT}$ correspondence \cite{Maldacena:1997re,Gubser:1998bc,Witten:1998qj} predicts dualities between certain string theories in anti-de Sitter (AdS) space and conformal field theories (CFTs). Its most prominent example relates type IIB string theory in $\text{AdS}_5\times\text{S}^5$ with $N$ units of five-form flux to the four-dimensional maximally ($\mathcal{N}=4$) supersymmetric Yang-Mills (SYM) theory with gauge group \SUN. It is most accessible in the 't Hooft (planar) limit \cite{'tHooft:1973jz}, where $N\to\infty$ and the Yang-Mills coupling constant $g_\YM\to0$ such that the 't Hooft coupling $\lambda=g_\YM^2 N$ is kept fixed: the string theory becomes free, and in the gauge theory non-planar vacuum diagrams are suppressed.\footnote{Non-planar non-vacuum diagrams may, however, become planar when connected to external states, and thus may contribute in the 't Hooft limit \cite{Sieg:2005kd}. They give rise to so-called finite-size corrections, which are the main object of this work.}
Starting from this maximally (super)symmetric setup, further examples can be found e.g.\ by discrete orbifold projections \cite{Kachru:1998ys,Lawrence:1998ja} or by applying deformations which depend on continuous parameters \cite{Lunin:2005jy,Frolov:2005ty,Frolov:2005dj,Frolov:2005iq}.

The prime example of such continuous deformations is the correspondence between the $\cN=1$ supersymmetric real $\beta$-deformed \NfSYMt ($\beta$-de\-for\-ma\-tion) and the type II B string theory in the Lunin-Maldacena background \cite{Lunin:2005jy}. In $\cN=1$ superspace, the gauge-theory action reads
\begin{equation}
\label{eq: superspace action}
\begin{aligned}
 S &= \frac{1}{ 2 g^2_\YM}\int \! \de^4 x  \de^2 \theta \tr\left( W^\alpha W_\alpha\right) + \int \! \de^4 x \de^4 \theta \tr \big( \e^{- g_\YM V} \bar{\Phi}_i \e^{g_\YM V}\Phi^i\big) \\
& \phaneq {}+{} i g_\YM \int \! \de^4 x  \de^2 \theta \tr \big(\Phi^1\Phi^2\Phi^3 \e^{-i  \frac{\beta}{2}} 
{}-{} \Phi^1\Phi^3\Phi^2 \e^{i \frac{\beta}{2}} \big) + \text{h.c.} \eqncom
\end{aligned}
\end{equation}
which reduces to the one of \NfSYMt if one sets the real parameter $\beta$ to zero. The above deformation is a special case of the more general exactly marginal Leigh-Strassler deformations \cite{LS95}. 
In \cite{Lunin:2005jy}, Lunin and Maldacena discussed the corresponding string theory background, which can be constructed by applying a TsT transformation, i.e.\ a T-duality, a shift (s) of an angular variable and another T-duality, to the $\text{S}^5$-factor in $\text{AdS}_5\times\text{S}^5$. This breaks the isometry group $\SO{6}$ of the $\text{S}^5$ to its $\U{1}_{Q^1}\times\U{1}_{Q^2}\times\U{1}_R$ Cartan subgroup. In \cite{Lunin:2005jy}, it was also found that the $\beta$-deformation can be formulated by replacing all products in the undeformed superspace action by noncommutative \cstar-products that introduce phase factors depending on the $\U{1}_{Q^1}\times\U{1}_{Q^2}$ Cartan charges of the respective fields.

All fields of the \NfSYMt transform in the adjoint representation of the gauge group. In the action, their representation matrices are contracted forming a single trace in colour space. Moreover, in the interactions each of these matrix products can be rephrased in terms of a commutator. Hence, if one considers $\UN$ as gauge group, the $\U1$ components of all fields decouple, and the $\UN$ theory is essentially the same as the $\SUN$ theory.

The $\beta$-deformation, however, does distinguish between the gauge groups \UN and \SUN. While the \U1 component of the vector superfield $V$ still decouples, this is no longer the case for the matter superfields $\Phi^i$,$\bar\Phi^i$ of flavours $i=1,2,3$. Moreover, the \UN theory is not even conformally invariant. This can be seen e.g.\ in a component expansion of the action \eqref{eq: superspace action}, where quantum corrections induce the running of a quartic double-trace coupling. While its tree-level value vanishes in the $\UN$ theory, it is at its non-vanishing IR fix point value in the $\SUN$ theory \cite{Hollowood:2004ek}. This value is found by integrating out the F-term auxiliary fields from the component expansion of \eqref{eq: superspace action}, cf.\ \cite{Fokken:2013aea}. Although coming with a prefactor of $\frac{1}{N}$, these double-trace terms do contribute in the 't Hooft limit. As we have argued in detail in \cite{Fokken:2013aea}, they were, however, neglected in the proof of conformal invariance in \cite{Ananth:2006ac}, which only considers planar single-trace couplings. 

\enlargethispage{0.5\baselineskip}
In a gauge theory without conformal symmetry, the (anomalous) scaling dimensions of gauge invariant composite operators depend on the $\beta$-functions of the couplings and thus are renormalisation-scheme dependent starting from the second loop order.\footnote{In \cite{Fokken:2013aea}, we have shown that the nonsupersymmetric three-parameter $\gamma_i$-deformation, which was proposed as candidate gauge theory of a generalisation of the Lunin-Maldacena $\text{AdS}/\text{CFT}$ setup in \cite{Frolov:2005dj}, is not conformally invariant -- not even in the 't Hooft limit. Double-trace couplings are induced whose $\beta$-functions have no fixed points as functions of $\lambda$, neither for gauge group \UN nor \SUN. As already remarked in \cite{Fokken:2013aea}, these couplings affect the planar spectrum of the theory, which is hence sensitive to the breakdown of conformal invariance. In \cite{Fokken:2014soa}, we will give an explicit example of a planar anomalous dimension that depends on one of these couplings and that is hence renormalisation-scheme dependent. In the later work \cite{Jin:2013baa}, the running of these couplings was confirmed. However, the author of \cite{Jin:2013baa} nevertheless claims that the $\gamma_i$-deformation is `conformally invariant in the planar limit'.} In a CFT, however, the $\beta$-functions vanish and with them the re\-nor\-ma\-li\-sa\-tion-scheme dependence. The anomalous dimensions are then observables, and the $\text{AdS}/\text{CFT}$ correspondence conjectures that they match the energies of respective string states in the gravitational theory.

\subsection{The dilatation operator in \texorpdfstring{\NfSYMt}{N=4 SYM theory}}

The scaling dimensions can be measured as eigenvalues of the generator of dilatations, known as the dilatation operator. In the 't Hooft limit, it admits a perturbative expansion in the effective planar coupling $g$: 
\begin{equation}
 D=D_0+g^2 D_2+\cO(g^3)\eqncom \qquad g=\frac{\sqrt\lambda}{4\pi} \eqncom
\end{equation}
where only the classical piece $D_0$ and the one-loop correction $D_2$ are shown. These operators, as well as higher-order corrections, can (in principle) be computed via Feynman diagrams, see \cite{ReviewofAdSCFTint1.1,Sieg:2010jt} for reviews. Moreover, it is sufficient to consider the action of $D$ on the subset of gauge-invariant composite operators with a single colour-trace. On these single-trace operators, which can be represented as cyclic spin chains, $D_2$ acts as an integrable Hamiltonian. This integrability appears to persist also beyond the first loop order. The conjecture of all-loop integrability has led to enormous progress in checking and understanding the $\text{AdS}/\text{CFT}$ correspondence, see the review collection \cite{ReviewofAdSCFTint} for a comprehensive list of references. 

In the basis of cyclic spin-chain states of length $L$, the dilatation operator $D$ can be written in terms of a (site-independent) density $\diladensity$ as
\begin{equation}\label{eq: dilatation operator on spin chain}
 D_{2K}=\sum_{i=1}^L D_{i,i+1,\dots,i+K} \eqncom \qquad D_{i,i+1.\dots,i+K}=\underbrace{\mathds{1}\otimes\dots\otimes\mathds{1}}_{i-1}\otimes\diladensity_{2K}\otimes\underbrace{\mathds{1}\otimes\dots\otimes\mathds{1}}_{L-i-K}\eqncom
\end{equation}
where $i+L$ is identified with $i$. Here, $D_{2K}$ denotes the contribution at order $g^{2K}$ and, for the sake of simplicity, we have neglected length-changing contributions appearing at loop orders $K>1$. Note that the density $\diladensity_{2K}$ has an interaction range $R=K+1$, i.e.\ a maximum of $R$ neighbouring sites interact with each other. In particular, the one-loop density $\diladensity_{2}$ contains at most nearest-neighbour interactions. For the \NfSYMt, $\diladensity_{2}^{\cN=4}$ was found by Beisert in \cite{Beisert03}.

Since $R$ increases by one with each additional loop order, an obvious problem occurs when $R$ exceeds the length $L$ of the state, which happens at loop-orders $K\geq L$. This necessitates finite-size corrections, i.e.\ length-dependent corrections to the asymptotic (length-independent) density.
In the field-theory picture of \NfSYMt, such corrections occur in the form of wrapping diagrams and were analysed in detail in \cite{Sieg:2005kd}. They originate from the fact that in the 't Hooft limit diagrams with external legs may contribute even if their $N$-power is naively subleading. The decision whether a diagram contributes can a priori only be made for diagrams in which all colour lines are closed, i.e.\ external lines have to be connected to external states (composite operators). In the notation of \cite{Sieg:2005kd}, a connected diagram with external legs and without external states is called planar if it contributes at leading order in the $\frac{1}{N}$-expansion when its external legs are planarly contracted with a single-trace vertex. Besides these diagrams, in the 't Hooft limit there may be contributions from non-planar connected diagrams, which effectively are multi-trace interactions. Such an interaction enhances its $N$-power if one of its colour-trace factors is fully and planarly contracted with a colour-trace of the same length in an external state.
 
In the integrability-based description of \NfSYMt, wrapping effects are incorporated in terms of L\"uscher corrections, Y-system and the thermodynamic Bethe ansatz (TBA). They all correct the result from the asymptotic Bethe ansatz and match the available field-theory data -- see \cite{Janik:2010kd,Gromov:2010kf,Bajnok:2010ke,Ahn:2010ka,Sieg:2010jt} for reviews. More recently, these formalisms were further developed to the so-called finite non-linear integral equations (FiNLIE) \cite{Gromov:2012eu} and the quantum spectral curve (QSC) \cite{Gromov:2013pga}.

\subsection{The dilatation operator in the \texorpdfstring{$\beta$}{beta}-deformation}

The $\beta$-deformation is closely connected to the \NfSYMt, such that many of the methods and also some of the results can be adopted. In particular, it is claimed to be as integrable as its undeformed parent theory, see \cite{ReviewofAdSCFTint4.2} for a review.

At the asymptotic level, the formulation of the $\beta$-deformation in terms of noncommutative Moyal-like \cstar-products allows the adaptation of a particular theorem derived for spacetime noncommutative field theories in \cite{Filk96}: the contribution of a planar diagram in the deformed theory is given by its undeformed counterpart times a phase factor that is determined by the order and charges of the external fields alone.
Beisert and Roiban used this theorem in \cite{BR05} to propose a planar one-loop dilatation-operator density in the deformed theory. In addition, they deformed the asymptotic Bethe ansatz by introducing twisted boundary conditions.

Subsequently, also wrapping corrections were investigated. They can conveniently be studied for so-called single-impurity operators, which are single-trace operators containing $L-1$ chiral scalar fields of one flavour and a single chiral scalar field of a different flavour. For $L\geq3$, the respective field-theory results of \cite{Fiamberti:2008sn} were reproduced in \cite{Gunnesson:2009nn} for $\beta=\frac{1}{2}$ and in \cite{Gromov:2010dy} and \cite{Arutyunov:2010gu} for generic $\beta$.\footnote{See \cite{Arutyunov:2010gu,Bajnok:2010ud} for higher-order results.} They are insensitive to the choice of $\UN$ or $\SUN$ as gauge group.

For the $L=2$ single-impurity operator, however, a sensitivity to that choice was observed already some years earlier in \cite{FG05}: for gauge group \SUN its one-loop anomalous dimension vanishes identically whereas for \UN it is non-vanishing.
In \cite{Frolov:2005iq}, it was noted that the latter result is reproduced by the one-loop dilatation operator as well as the asymptotic Bethe ansatz, both proposed in \cite{BR05}.
But up to now the results of \cite{BR05} could not be modified to incorporate the conformally invariant $\SUN$ theory, which is the CFT candidate for the $\text{AdS}/\text{CFT}$ correspondence.\footnote{See \cite{Frolov:2005iq} for some comments concerning the differences between $\UN$ and $\SUN$ gauge group in the deformed $\text{AdS}/\text{CFT}$ correspondence.}

\enlargethispage{-1\baselineskip}
The situation becomes worse at the level of wrapping corrections. According to field-theory calculations, the two-loop anomalous dimension of the $L=2$ single-impurity state vanishes for gauge group $\SUN$ \cite{Penati:2005hp}. The corresponding integrability-based result of \cite{Arutyunov:2010gu}, however, is logarithmically divergent,\footnote{Such a divergence was encountered earlier in the expressions for the ground-state energy of the TBA \cite{Frolov:2009in}. In \cite{deLeeuw:2012hp}, it was found that the divergent ground-state energy vanishes in the undeformed theory when a regulating twist is introduced in the $\text{AdS}_5$ directions. This regularisation extends to the ground state of the supersymmetric deformations \cite{FrolovPC}.} which cannot even be correct for gauge group $\UN$.

In this paper, we address the above problems from a field-theory perspective.

\subsection{Organisation of our paper}

This work is organised as follows.

In Section \ref{sec: asymptotic dilatation operator}, we give a short review of the $\beta$-deformation and Filk's theorem in \cite{Filk96}, which was used to derive the one-loop dilatation-operator density in \cite{BR05}. We analyse in which cases this theorem can be adapted to the $\beta$-deformation and find that this adaptability is generically limited by the occurrence of finite-size effects. This implies that the proposal of \cite{BR05} is a priori valid only asymptotically. For a certain class of operators, however, the theorem can be generalised to include these effects. We find that their $n$-point correlation functions are given by the values of their undeformed counterparts at all loop orders in the planar theory. 

In Section \ref{sec: A new type of finite-size effect}, we identify a new type of finite-size effect, which captures the differences between the planar correlation functions of the $\UN$ and $\SUN$ $\beta$-deformation. In particular, this effect accounts for the aforementioned sensitivity of certain anomalous dimensions to the choice of the gauge group.
As it starts to affect single-trace operators one loop order earlier than wrapping, we call it \emph{prewrapping}. It is caused by the double-trace structure in the $\SUN$ propagator and has no net effect in the undeformed theory. We classify which types of operators may be affected by prewrapping.

In Section \ref{sec: complete one-loop dilatation operator}, we determine the  missing finite-size corrections to the asymptotic dilatation operator of \cite{BR05}. For gauge group \SUN these are the aforementioned prewrapping corrections at $L=2$, whereas for gauge group \UN they are the ordinary wrapping corrections at $L=1$. We thus obtain the complete planar one-loop dilatation operator of the $\beta$-deformation.

In Section \ref{sec: spectrum}, we identify all $\cN=1$ superconformal primary states of the $\beta$-deformation with classical scaling dimension $\Delta_0\leq 4.5$. We then apply the one-loop dilatation operator of Section \ref{sec: complete one-loop dilatation operator} to these states and compute their anomalous dimensions for both gauge groups, stressing in particular the differences between the two cases.

Our summary and outlook can be found in Section \ref{sec: Conclusion and outlook}. There, we also comment on the implications our findings have on integrability. If the $\beta$-deformation is indeed as integrable as \NfSYMt, a consistent incorporation of prewrapping into the integrability-based descriptions must exist. In particular, this should cure the divergence encountered in \cite{Arutyunov:2010gu}.

Several appendices contain details of the calculations as well as a table of primary states and their anomalous dimensions supplementing the analysis of Section \ref{sec: spectrum}.

\section{Filk's theorem in the \texorpdfstring{$\beta$}{beta}-deformation}
\label{sec: asymptotic dilatation operator}

In this section, we review the \cstar-product formulation of the $\beta$-deformation and a theorem that was first derived for spacetime noncommutative field theories by Filk in \cite{Filk96}. 
We analyse the subtleties which arise when it is adapted to the $\beta$-deformation and determine the limits of its applicability.
In the asymptotic regime, we review how it can be used to derive the dilatation operator of \cite{BR05}. Beyond that, we show that certain classes of $n$-point correlation functions are at all loop orders in the planar theory given by the values of their undeformed counterparts. 

The $\beta$-deformation can be realised via a noncommutative Moyal-like \cstar-product \cite{Lunin:2005jy}, which for two fields $A$ and $B$ is defined as 
\begin{equation}\label{eq: def star product}
 A\cstar B= A B \e^{{}\frac{\complexi}{2}\left(\mathbf{q}_{A}{}\wedge{}\mathbf{q}_{B}\right){}}\eqndot
\end{equation}
Here $\mathbf{q}_{A}=(q_A^1,q_A^2,q_A^3)$ and $\mathbf{q}_{B}=(q_B^1,q_B^2,q_B^3)$ are the charge vectors of the fields associated with the Cartan  subgroup of the $\SU{4}_R$ symmetry group of the undeformed theory, see Table~\ref{tab: su(4) charges}. Their antisymmetric product is defined as 
\begin{equation}\label{eq: antisymmetric product}
 \mathbf{q}_{A}\wedge \mathbf{q}_{B}= -\beta \sum_{a,b,c=1}^3 \epsilon_{abc} q_A^a q_B^b \eqncom
\end{equation}
where $\epsilon$ is the three-dimensional antisymmetric tensor normalised to $\epsilon_{123}=1$. 
In fact, the antisymmetric product \eqref{eq: antisymmetric product} depends only on $Q^1=q^1-q^2$ and $Q^2=q^2-q^3$, in terms of which it reads $\mathbf{q}_{A}\wedge \mathbf{q}_{B}=-\beta(Q^1_AQ^2_B-Q^1_BQ^2_A)
$. It is insensitive to the $\U1_R$ charge $r=\frac{2}{3}(q^1+q^2+q^3)$. Hence, the \cstar-product \eqref{eq: def star product} can be used also in the superfield formulation, leading to the action \eqref{eq: superspace action}. The basis $(Q^1,Q^2,r)$ of the \su4 Cartan charges was originally used in \cite{Lunin:2005jy} and also appears in our representation-theoretical considerations in Section \ref{sec: spectrum} and Appendix~\ref{app: representation content}. 
\begin{table}[htbp]
\centering
$\begin{array}{c|c|ccc|cccc|ccc}
B&A_{\mu}&\phi^1&\phi^2&\phi^3 & \psi^1_{\alpha}&\psi^2_{\alpha}&\psi^3_{\alpha}&\psi^4_{\alpha} &F^1 & F^2 & F^3\\
 & & & & & & & &&&&\\[-0.4cm]
\hline
 & & & & & & & &&&&\\[-0.4cm]
q^1_B & 0 & 1 & 0 & 0 &  +\frac12 & -\frac12 & -\frac12 & +\frac12 &0 & -1 & -1 \\
 & & & & & & & &&&&\\[-0.4cm]
q^2_B &  0 & 0 & 1 & 0 &-\frac12 & +\frac12 & -\frac12 & +\frac12 & -1 & 0 & -1 \\
 & & & & & & & &&&&\\[-0.4cm]
q^3_B &  0 & 0 & 0 & 1 &-\frac12 & -\frac12 & +\frac12 & +\frac12 & -1 & -1 & 0 \\
 & & & & & & & &&&&\\[-0.4cm] \hline
 & & & & & & & &&&&\\[-0.4cm]
Q^1_B &  0 & 1 & -1 & 0 &1 & -1 & 0 & 0 & 1 & -1 & 0\\
 & & & & & & & &&&&\\[-0.4cm]
Q^2_B &  0 & 0 & 1 & -1 & 0 & 1 & -1 & 0 &0 & 1 & -1\\
 & & & & & & & &&&&\\[-0.4cm]
r_B &  0 & \frac{2}{3} & \frac{2}{3} & \frac{2}{3} &-\frac{1}{3} & -\frac{1}{3} & -\frac{1}{3} & 1 & -\frac{4}{3} & -\frac{4}{3} & -\frac{4}{3}\\
\end{array}$
\caption{Cartan charges of the fields, including the F-term auxiliary fields $F^i$, in two different bases. The respective anti-fields carry the opposite charges.}
\label{tab: su(4) charges}
\end{table}

The action of the $\beta$-deformation is obtained from the one of \NfSYMt by replacing all products by \cstar-products and thus all commutators by \cstar-commutators:
\begin{equation}\label{eq: def star commutator}
 \starcomm{A}{B}=A\cstar B-B\cstar A \eqndot
\end{equation}
For gauge group \SUN, however, this has to be done in the superspace action \eqref{eq: superspace action}, or in the component expansion before the auxiliary fields are integrated out. In this case, a double-trace coupling is generated which is at its fix-point value, cf.\ \cite{Fokken:2013aea}. Although formally suppressed by $\frac{1}{N}$, it is necessary for the conformal invariance also of the planar theory.

Introducing \cstar-products in the component field action without auxiliary fields misses this coupling at tree-level and induces its running at loop-level.
In the conventions of \cite{Fokken:2013aea}, the Euclidean component action\footnote{The double-trace term in the action was written explicitly in \cite{Jin:2012np} but follows also directly from the procedure mentioned much earlier in \cite{FG05}.} reads
\begin{equation}\label{eq: component action}
\begin{aligned}
S
 &=\int\de^4x\,\Bigl[\tr\Big(
 -\frac{1}{4} F^{\mu\nu}F_{\mu\nu}- (\D^{\mu}\bar\phi_j)\D_{\mu}\phi^j
 +i \bar\psi^{\dot\alpha}_A \D_{\dot\alpha}{}^\alpha\psi_\alpha^A\\
 &\hphantom{{}={}\int\de^4x\,\Bigl[\tr\Big({}}{}
 +g_\YM\Bigl(
 \frac{i}{2}\epsilon_{ijk}\phi^i \staracomm{\psi^{\alpha j}}{\psi^k_\alpha}
 +\phi^j \acomm{\bar\psi^{\dot\alpha}_4}{
 \bar\psi_{\dot\alpha j}}+\text{h.c.}
 \Bigr)\\
 &\phantom{{}={}\int\de^4x\,\Bigl[\tr\Big({}}{}
 -\frac{g^2_\YM}{4}
 \comm{\bar\phi_j}{\phi^j}\comm{\bar\phi_k}{\phi^k}
 +\frac{g^2_\YM}{2}
 \starcomm{\bar\phi_j}{\bar\phi_k}\starcomm{\phi^j}{\phi^k}
 \Big)
 \\
 &{}\phantom{{}={}\int\de^4x\,\Bigl[{}}{}
 -\frac{\colors}{N}\frac{g^2_\YM}{2}\tr\bigl(\starcomm{\bar\phi_j}{\bar\phi_k}\bigr)
 \tr\bigl(\starcomm{\phi^j}{\phi^k}\bigr)
\Bigr] \eqncom
\end{aligned}
\end{equation}
with spacetime indices $\mu,\nu=0,1,2,3$, spinor indices $\alpha=1,2$, $\dot\alpha=\dot1,\dot2$, flavour indices $i,j,k=1,2,3$ and $A=1,2,3,4$, as well as the gauge-group parameter
\begin{equation}\label{eq: color s}
 \colors=\begin{cases}
    0\quad\text{ for }\UN\eqncom \\
    1\quad\text{ for }\SUN\eqndot
   \end{cases}
\end{equation}
We have kept only those \cstar-products that do introduce net deformations. 
In particular, all interactions in which the gauge field $A_\mu$ and the gluino $\psi^4$ occur are undeformed since these fields are uncharged under $(Q^1,Q^2)$.

The \cstar-product of the $\beta$-deformation \eqref{eq: def star product} is similar to the Moyal \cstar-product used to formulate a specific type of spacetime noncommutative field theories.\footnote{See \cite{Szabo:2001kg} for a review.} This similarity allows to adapt a particular theorem for this class of noncommutative field theories derived by Filk in \cite{Filk96}: the deformed version of a planar Feynman diagram is equal to its undeformed counterpart times a phase factor which is determined by the order and charges of the external fields alone.
A completely explicit formulation of the relation between the deformed and undeformed diagrams of colour-ordered amplitudes is given in \cite{Khoze05}. If the fields entering such an amplitude are $A_1,A_2,\dots, A_n$ in cyclic order, its phase factor is the one of $A_1\cstar A_2\cstar\cdots\cstar A_n$. This relation can be depicted as
\begin{equation}\label{eq: relation for colour ordered amplitudes}
 \begin{aligned}
\settoheight{\eqoff}{$+$}%
\setlength{\eqoff}{0.5\eqoff}%
\addtolength{\eqoff}{-9\unit}%
\raisebox{\eqoff}{%
\begin{pspicture}(-2,1)(11,-17)
\rput(4.5,-8){%
\rotatebox{90}{%
\begin{pspicture}(-1,-2)(17,11)
\uinex{2}{9}
\iinex{2}{6}
\dinex{2}{0}
\setlength{\ya}{9\unit}
\addtolength{\ya}{0.5\dlinewidth}
\setlength{\yb}{0\unit}
\addtolength{\yb}{-0.5\dlinewidth}
\setlength{\xc}{7.5\unit}
\setlength{\yc}{4.5\unit}
\addtolength{\yc}{-0.5\dlinewidth}
\setlength{\xd}{8.5\unit}
\setlength{\yd}{4.5\unit}
\addtolength{\yd}{0.5\dlinewidth}
\psline[linewidth=\blacklinew](3.5,\ya)(12.5,\ya)
\psline[linestyle=dotted,linewidth=\blacklinew](3.5,4)(3.5,2)
\psline[linewidth=\blacklinew](12.5,\yb)(3.5,\yb)
\doutex{14}{0}
\ioutex{14}{6}
\uoutex{14}{9}
\psline[linestyle=dotted,linewidth=\blacklinew](12.5,4)(12.5,2)
\setlength{\xa}{3.5\unit}
\addtolength{\xa}{\dlinewidth}
\setlength{\xb}{12.5\unit}
\addtolength{\xb}{-\dlinewidth}
\setlength{\ya}{9\unit}
\addtolength{\ya}{-0.5\dlinewidth}
\setlength{\yb}{0\unit}
\addtolength{\yb}{0.5\dlinewidth}
\pscustom[linecolor=gray,fillstyle=solid,fillcolor=gray,linearc=\linearc]{%
\psline(\xa,4.5)(\xa,\ya)(\xb,\ya)(\xb,4.5)
\psline[liftpen=2](\xb,4.5)(\xb,\yb)(\xa,\yb)(\xa,4.5)
}
\end{pspicture}}
\rput(-2,1.25){%
\rput[b](-9,14.5){$A_{1}$}
\rput[b](-6,14.5){$A_{2}$}
\rput[b](0,14.5){$A_{i}$}
\rput[t](0,1){$A_{i+1}$}
\rput[t](-6,1){$A_{n-1}$}
\rput[t](-9,1){$A_{n}$}}
}
\rput(4.5,-6.75){planar}
\rput(4.5,-9.25){$\beta$}
\end{pspicture}}
\,
&=\,
\settoheight{\eqoff}{$+$}%
\setlength{\eqoff}{0.5\eqoff}%
\addtolength{\eqoff}{-9\unit}%
\raisebox{\eqoff}{%
\begin{pspicture}(-2,1)(11,-17)
\rput(4.5,-8){%
\rotatebox{90}{%
\begin{pspicture}(-1,-2)(17,11)
\uinex{2}{9}
\iinex{2}{6}
\dinex{2}{0}
\setlength{\ya}{9\unit}
\addtolength{\ya}{0.5\dlinewidth}
\setlength{\yb}{0\unit}
\addtolength{\yb}{-0.5\dlinewidth}
\setlength{\xc}{7.5\unit}
\setlength{\yc}{4.5\unit}
\addtolength{\yc}{-0.5\dlinewidth}
\setlength{\xd}{8.5\unit}
\setlength{\yd}{4.5\unit}
\addtolength{\yd}{0.5\dlinewidth}
\psline[linewidth=\blacklinew](3.5,\ya)(12.5,\ya)
\psline[linestyle=dotted,linewidth=\blacklinew](3.5,4)(3.5,2)
\psline[linewidth=\blacklinew](12.5,\yb)(3.5,\yb)
\doutex{14}{0}
\ioutex{14}{6}
\uoutex{14}{9}
\psline[linestyle=dotted,linewidth=\blacklinew](12.5,4)(12.5,2)
\setlength{\xa}{3.5\unit}
\addtolength{\xa}{\dlinewidth}
\setlength{\xb}{12.5\unit}
\addtolength{\xb}{-\dlinewidth}
\setlength{\ya}{9\unit}
\addtolength{\ya}{-0.5\dlinewidth}
\setlength{\yb}{0\unit}
\addtolength{\yb}{0.5\dlinewidth}
\pscustom[linecolor=gray,fillstyle=solid,fillcolor=gray,linearc=\linearc]{%
\psline(\xa,4.5)(\xa,\ya)(\xb,\ya)(\xb,4.5)
\psline[liftpen=2](\xb,4.5)(\xb,\yb)(\xa,\yb)(\xa,4.5)
}
\end{pspicture}}
\rput(-2,1.25){%
\rput[b](-9,14.5){$A_{1}$}
\rput[b](-6,14.5){$A_{2}$}
\rput[b](0,14.5){$A_{i}$}
\rput[t](0,1){$A_{i+1}$}
\rput[t](-6,1){$A_{n-1}$}
\rput[t](-9,1){$A_{n}$}}
}
\rput(4.5,-6.75){planar}
\rput(4.5,-9.25){$\mathcal{N}=4$}
\end{pspicture}}
\times
\,\,\,
\Phi\left(%
A_{1}{}\ast{}A_{2}
\ast{}\dots{}\ast{}A_{n}
\right)\eqncom
\end{aligned}
\end{equation}
where $\Phi$ denotes the phase factor of the \cstar-product.

A direct application of relation \eqref{eq: relation for colour ordered amplitudes} to the planar diagrams contributing to $\diladensity_2$ led to the following proposal for the one-loop dilatation-operator density of the deformed theory \cite{BR05}: 
\begin{equation}\label{eq: deformation of D_2}
\begin{aligned}
(\diladensity_2^\beta)_{A_iA_j}^{A_kA_l}&= \Phi(A_k\cstar A_l\cstar A_j\cstar A_i)(\diladensity_2^{ \cN=4 })_{A_iA_j}^{A_kA_l}\\&=\e^{\frac{\complexi}{2} (\mathbf{q}_{A_k} \wedge \mathbf{q}_{A_l}- \mathbf{q}_{A_i} \wedge \mathbf{q}_{A_j})}(\diladensity_2^{\cN=4})_{A_iA_j}^{A_kA_l} \eqncom
\end{aligned}
\end{equation} 
where $A_i$, $A_j$ denote the incoming and $A_k$, $A_l$ the outgoing fields. 
They are taken from the alphabet 
\begin{equation}\label{eq: alphabet}
 \cA =\{ \D^k\phi^i, \D^k\bar\phi^i, \D^k\psi^A_{\alpha}, \D^k\bar\psi^A_{\dot\alpha},\D^k\cF_{\alpha\beta},\D^k\bar\cF_{\dot\alpha\dot\beta} \}
\eqncom
\end{equation}
where the field strength $F_{\mu\nu}$ has been split into its self-dual and anti-self-dual part $\cF$ and $\bar\cF$, respectively, and all Minkowski indices are translated to spinor indices using the relations in Appendix~\ref{subsec: Mathematica implementation of the harmonic action}. The abbreviation $\D^k\bar\psi^2_{\dot\alpha}$ stands for expressions with arbitrarily many covariant derivatives acting on $\bar\psi^2_{\dot\alpha}$ and totally symmetrised in both kinds of spinor indices.

Although the dilatation-operator density \eqref{eq: deformation of D_2} seems to follow immediately from relation \eqref{eq: relation for colour ordered amplitudes}, one has to be very careful when adapting Filk's theorem to the $\beta$-deformation.
The $\beta$-deformation is not a spacetime noncommutative field theory, and also its Moyal-like \cstar-product is distinctly different from the one in \cite{Filk96}. In spacetime noncommutativity, the phase as well as the ordering principle of fields entering an interaction depend on the spacetime coordinates or, after Fourier transformation, the momenta. In the \cstar-product of the $\beta$-deformation, the phase depends on the $(Q^1,Q^2)$ charges of the fields, while the ordering principle is colour arrangement. Colour and $(Q^1,Q^2)$ charge are, however, independent of each other. Traces, i.e.\ colour-neutral object, may be $(Q^1,Q^2)$-charged, as e.g.\ the trace factors in the double-trace term in \eqref{eq: component action}. 
In this case, the prescription of replacing all products by \cstar-products becomes generically ill-defined since it is in conflict with the cyclic invariance of the trace.\footnote{The \cstar-deformation of the double-trace term in the component action \eqref{eq: component action} is, however, unambiguous because it arises from the \cstar-deformation of charge-neutral single-trace terms in the $\cN=1$ superspace action \eqref{eq: superspace action} or in its off-shell component expansion (before the auxiliary fields $F^i$ are integrated out). 
Accordingly, working in superspace or off-shell component space, all vertices are of flavour-neutral single-trace type.
} 
 This can exemplarily be seen for $\tr [\phi^i \phi^j]=\tr[ \phi^j \phi^i]$, whereas in the \cstar-deformed case $\tr[\phi^i \cstar \phi^j]\neq\tr[ \phi^j \cstar \phi^i]$ for $i\neq j$.
The fact that colour and $(Q^1,Q^2)$ charge are not connected also manifests itself in a problem concerning external states and the notion of planarity. The commonly repeated statement that ``only planar diagrams contribute in the planar / 't Hooft limit'' is true only for diagrams in which all external legs have been connected to external states (composite operators). Before the connection to external states, some of the contributing diagrams are non-planar. These are the diagrams giving rise to the finite-size corrections \cite{Sieg:2005kd}.
As external states are not explicitly covered in \cite{Filk96}, a priori all diagrams have to be considered without external states, leaving subdiagrams of elementary interactions. In particular, the resulting subdiagrams of finite-size corrections are non-planar and hence are not captured by \eqref{eq: relation for colour ordered amplitudes}.

We can, however, extend the applicability of \eqref{eq: relation for colour ordered amplitudes} beyond the subdiagrams of elementary interactions. In theories with spacetime noncommutativity, external states can immediately be incorporated. They can be added to the action as local interactions at which momentum is conserved. Hence, as far as the deformation is concerned, they are on equal footing with the elementary vertices. In particular, momentum conservation implies that the deformation via the Moyal \cstar-product is well defined since the resulting phase factor is invariant under a cyclic relabelling of the external momenta. In spacetime-noncommutative theories, there is hence no need for treating any external state separately from the elementary vertices when applying Filk's theorem. The above considerations can be adapted in parts to the $\beta$-deformation. The similarity between the $\beta$-deformation and spacetime noncommutativity holds only if colour and $(Q^1,Q^2)$ neutrality coincide, i.e.\ the fields within a single colour trace must have vanishing $(Q^1,Q^2)$ net charge. This concerns external states consisting of one or several colour traces each of which is uncharged under the $U(1)_{Q^1}\times U(1)_{Q^2}$ flavour symmetry. Relation \eqref{eq: relation for colour ordered amplitudes} then applies to the diagrams including such states, where on the left hand side the elementary vertices as well as the external states are \cstar-deformed.

In the $\beta$-deformation, the extension of relation \eqref{eq: relation for colour ordered amplitudes} to
diagrams containing external states with $(Q^1,Q^2)$-charge-neutral trace factors has severe consequences. 
Since it is not necessary to remove such external states, diagrams which are planar only due to the presence of these states also obey \eqref{eq: relation for colour ordered amplitudes}, even if their subdiagrams of elementary interactions are non-planar. In particular, any gauge-invariant correlation function of such external states is independent of the deformation. This follows immediately from evaluating \eqref{eq: relation for colour ordered amplitudes} with no ($n=0$) external legs. It implies that e.g.\ the anomalous scaling dimensions and the structure constants for external states (composite operators) of this class are at all loops independent of the deformation parameter $\beta$ and are directly given by their $\mathcal{N}=4$
SYM counterparts. 

One example in the subclass of $(Q^1,Q^2)$-charge-neutral single-trace operators is the Konishi operator $\sum^3_{i=1}\tr[\phi^i\bar\phi^i]$, whose anomalous dimension in the undeformed theory has recently been determined up to eight loops in the framework of integrability \cite{Gromov:2013pga}.\footnote{The anomalous dimension was obtained via one of the $\cN=4$ descendants of the Konishi operator which is not an $\cN=1$ descendant. While in the $\beta$-deformation the anomalous dimension of this descendant is altered, the \NfSYMt result remains valid for the Konishi primary.}  By the above argument, this result is valid also in the $\beta$-deformation.
Another example is the family of chiral primary operators with length $L=3k$ and $(q^1,q^2,q^3)=(k,k,k)$:
\begin{equation}\label{eq: O_k}
 \cO_k= \tr[(\phi^1)^k(\phi^2)^k(\phi^3)^k]+\text{all permutations}\eqndot
\end{equation}
In contrast to the Konishi operator, the operators $\cO_k$ themselves are altered by the \cstar-deformation.\footnote{Cf. the alternative deformation prescription in \cite{FG05}, which differs only by an overall factor.}
In \cite{David:2013oha}, the three-point functions $\langle \cO_k \cO_{k^\prime} \cO_{k^{\prime\prime}} \rangle$ of the deformed operators were investigated at one-loop order in the planar gauge theory and at strong coupling in the Lunin-Maldacena background. In these regimes, they were found to be independent of $\beta$. From the above argument it is clear that these three-point functions as well as all their (higher) $n$-point functions are independent of $\beta$ at all loop orders.%
\footnote{%
See \cite{Frolov:2005iq} for a generalisation of earlier arguments given in \cite{Berenstein:2000ux,Berenstein:2000hy} for rational $\beta$ that the anomalous dimensions of the operators \eqref{eq: O_k} vanish as for their undeformed counterparts in $\mathcal{N}=4$ SYM theory.}

If one removes the external single-trace operators in a planar two-point diagram, one either obtains a planar single-trace diagram as in \eqref{eq: relation for colour ordered amplitudes} or a non-planar double-trace diagram. 
The latter type of diagram generates the finite-size effects. The applicability of \cite{Filk96} to external single-trace operators with vanishing $(Q^1,Q^2)$ charge leads to the following relation for double-trace interactions that contribute in the planar limit:\footnote{%
This relation was also obtained in \cite{Jin:2012np} for the dominant contribution to certain multi-trace amplitudes at large but finite $N$.
}
\begin{equation}\label{eq: relation for double-trace diagrams}
 \begin{aligned}
\settoheight{\eqoff}{$+$}%
\setlength{\eqoff}{0.5\eqoff}%
\addtolength{\eqoff}{-9\unit}%
\raisebox{\eqoff}{%
\begin{pspicture}(-2,1)(11,-17)
\rput(4.5,-8){%
\rotatebox{90}{%
\begin{pspicture}(-1,-2)(17,11)
\uinex{2}{9}
\iinex{2}{6}
\dinex{2}{0}
\setlength{\ya}{9\unit}
\addtolength{\ya}{0.5\dlinewidth}
\setlength{\yb}{0\unit}
\addtolength{\yb}{-0.5\dlinewidth}
\setlength{\xc}{7.0\unit}
\setlength{\yc}{4.5\unit}
\addtolength{\yc}{-0.5\dlinewidth}
\setlength{\xd}{9.0\unit}
\setlength{\yd}{4.5\unit}
\addtolength{\yd}{0.5\dlinewidth}
\psline[linearc=\linearc](3.5,\ya)(\xc,\ya)(\xc,\yb)(3.5,\yb)
\psline[linestyle=dotted](3.5,4.5)(3.5,1.5)
\psline[linearc=\linearc](12.5,\ya)(\xd,\ya)(\xd,\yb)(12.5,\yb)
\doutex{14}{0}
\ioutex{14}{6}
\uoutex{14}{9}
\psline[linestyle=dotted](12.5,4.5)(12.5,1.5)
\setlength{\xa}{3.5\unit}
\addtolength{\xa}{\dlinewidth}
\setlength{\xb}{7.0\unit}
\addtolength{\xb}{-\dlinewidth}
\setlength{\ya}{9\unit}
\addtolength{\ya}{-0.5\dlinewidth}
\setlength{\yb}{0\unit}
\addtolength{\yb}{0.5\dlinewidth}
\pscustom[linecolor=gray,fillstyle=solid,fillcolor=gray,linearc=\linearc]{%
\psline(\xa,4.5)(\xa,\ya)(\xb,\ya)(\xb,4.5)
\psline[liftpen=2](\xb,4.5)(\xb,\yb)(\xa,\yb)(\xa,4.5)
}
\setlength{\xa}{12.5\unit}
\addtolength{\xa}{-\dlinewidth}
\setlength{\xb}{9.0\unit}
\addtolength{\xb}{\dlinewidth}
\setlength{\ya}{9\unit}
\addtolength{\ya}{-0.5\dlinewidth}
\setlength{\yb}{0\unit}
\addtolength{\yb}{0.5\dlinewidth}
\pscustom[linecolor=gray,fillstyle=solid,fillcolor=gray,linearc=\linearc]{%
\psline(\xa,4.5)(\xa,\ya)(\xb,\ya)(\xb,4.5)
\psline[liftpen=2](\xb,4.5)(\xb,\yb)(\xa,\yb)(\xa,4.5)
}
\pnode(7.25,4.5){left}
\pnode(8.75,4.5){right}
\ncline[arrows=->,linecolor=black]{left}{right}
\end{pspicture}}
\rput(-2,1.25){%
\rput[b](-9,14.5){$A_{1}$}
\rput[b](-6,14.5){$A_{2}$}
\rput[b](0,14.5){$A_{i}$}
\rput[t](0,1){$A_{i+1}$}
\rput[t](-6,1){$A_{n-1}$}
\rput[t](-9,1){$A_{n}$}}
}
\rput(4.5,-5.25){$\beta$}
\rput(4.5,-10.75){$\beta$}
 \rput[l](5.0,-8.125){$\scriptstyle \delta \mathbf{Q} =0$}
\end{pspicture}}
\,
&=\,
\settoheight{\eqoff}{$+$}%
\setlength{\eqoff}{0.5\eqoff}%
\addtolength{\eqoff}{-9\unit}%
\raisebox{\eqoff}{%
\begin{pspicture}(-2,1)(11,-17)
\rput(4.5,-8){%
\rotatebox{90}{%
\begin{pspicture}(-1,-2)(17,11)
\uinex{2}{9}
\iinex{2}{6}
\dinex{2}{0}
\setlength{\ya}{9\unit}
\addtolength{\ya}{0.5\dlinewidth}
\setlength{\yb}{0\unit}
\addtolength{\yb}{-0.5\dlinewidth}
\setlength{\xc}{7.0\unit}
\setlength{\yc}{4.5\unit}
\addtolength{\yc}{-0.5\dlinewidth}
\setlength{\xd}{9.0\unit}
\setlength{\yd}{4.5\unit}
\addtolength{\yd}{0.5\dlinewidth}
\psline[linearc=\linearc](3.5,\ya)(\xc,\ya)(\xc,\yb)(3.5,\yb)
\psline[linestyle=dotted](3.5,4.5)(3.5,1.5)
\psline[linearc=\linearc](12.5,\ya)(\xd,\ya)(\xd,\yb)(12.5,\yb)
\doutex{14}{0}
\ioutex{14}{6}
\uoutex{14}{9}
\psline[linestyle=dotted](12.5,4.5)(12.5,1.5)
\setlength{\xa}{3.5\unit}
\addtolength{\xa}{\dlinewidth}
\setlength{\xb}{7.0\unit}
\addtolength{\xb}{-\dlinewidth}
\setlength{\ya}{9\unit}
\addtolength{\ya}{-0.5\dlinewidth}
\setlength{\yb}{0\unit}
\addtolength{\yb}{0.5\dlinewidth}
\pscustom[linecolor=gray,fillstyle=solid,fillcolor=gray,linearc=\linearc]{%
\psline(\xa,4.5)(\xa,\ya)(\xb,\ya)(\xb,4.5)
\psline[liftpen=2](\xb,4.5)(\xb,\yb)(\xa,\yb)(\xa,4.5)
}
\setlength{\xa}{12.5\unit}
\addtolength{\xa}{-\dlinewidth}
\setlength{\xb}{9.0\unit}
\addtolength{\xb}{\dlinewidth}
\setlength{\ya}{9\unit}
\addtolength{\ya}{-0.5\dlinewidth}
\setlength{\yb}{0\unit}
\addtolength{\yb}{0.5\dlinewidth}
\pscustom[linecolor=gray,fillstyle=solid,fillcolor=gray,linearc=\linearc]{%
\psline(\xa,4.5)(\xa,\ya)(\xb,\ya)(\xb,4.5)
\psline[liftpen=2](\xb,4.5)(\xb,\yb)(\xa,\yb)(\xa,4.5)
}
\pnode(7.25,4.5){left}
\pnode(8.75,4.5){right}
\ncline[arrows=->,linecolor=black]{left}{right}
\end{pspicture}}
\rput(-2,1.25){%
\rput[b](-9,14.5){$A_{1}$}
\rput[b](-6,14.5){$A_{2}$}
\rput[b](0,14.5){$A_{i}$}
\rput[t](0,1){$A_{i+1}$}
\rput[t](-6,1){$A_{n-1}$}
\rput[t](-9,1){$A_{n}$}}
}
\rput(4.5,-5.25){$\cN=4$}
\rput(4.5,-10.75){$\cN=4$}
 \rput[l](5.0,-8.125){$\scriptstyle \delta \mathbf{Q} =0$}
\end{pspicture}}
\times
\,\,\,
\underbrace{\Phi\left(
A_{1}{}
\ast{}\dots{}\ast{}A_{i}
\right)
\Phi\left(
A_{i+1}{}
\ast{}\dots{}\ast{}A_{n}
\right)\rule[-0.25cm]{0pt}{0.5cm}}_{\rule[0pt]{0pt}{0.5cm}\displaystyle  \Phi\left(
A_{1}{}
\ast{}\dots{}\ast{}A_{i}{}\ast{}
A_{i+1}{}
\ast{}\dots{}\ast{}A_{n}
\right)}\eqncom
\end{aligned}
\end{equation}
where $\delta \mathbf{Q}=(\delta Q^1,\delta Q^2)$ denotes the charge flow between the separate traces. As in \eqref{eq: relation for colour ordered amplitudes}, the grey-shaded regions represent arbitrary planar interactions.

Regardless of the above generalisations concerning $(Q^1,Q^2)$-neutral single-trace operators, finite-size corrections  still set the limit of applicability of relation \eqref{eq: relation for colour ordered amplitudes} for $(Q^1,Q^2)$-charged ones.
These corrections can, however, not contribute if the maximal interaction range $R$ of the connected subdiagrams at a given loop-order $K$ is strictly smaller than the length $L$ of the single-trace operator.
In particular, \eqref{eq: relation for colour ordered amplitudes} can be applied to the $K$-loop asymptotic dilatation-operator density in \eqref{eq: dilatation operator on spin chain}, where asymptotic now means $L>R=K+1$.\footnote{Note that the $K$-loop asymptotic dilatation-operator density $\diladensity_{2K}$ incorporates interaction subdiagrams of ranges $R^\prime\leq K+1$. In the cases where $R^\prime< K+1$, it acts as identity on the remaining $K+1-R^\prime$ spin-chain sites, which corresponds to bare propagators in the two-point function. Relation \eqref{eq: relation for double-trace diagrams} applied for $i=2$, together with the fact that $\Phi(A\cstar\bar A)=1$, guarantees that the phase factor obtained from the interaction subdiagram is exactly the one obtained from applying \eqref{eq: relation for colour ordered amplitudes} to $\diladensity_{2K}$. Disconnected interaction subdiagrams are also captured by  relation  \eqref{eq: relation for double-trace diagrams}. They do, however, not contribute to the dilatation operator, since they do not have overall UV divergences.}

In the special case of the one-loop dilatation operator, our previous discussion guarantees that \eqref{eq: deformation of D_2} is correct asymptotically, i.e. for $L\geq3$, but it does not include finite-size effects, which may -- in principle -- contribute at $L\leq2$. In Section \ref{sec: complete one-loop dilatation operator}, we calculate the corresponding finite-size corrections. For gauge group \UN, these are the conventional wrapping corrections at $L=1$. For gauge group $\SUN$, a new type of finite-size effect occurs at $L=2$, which we discuss in the next section.

\section{A new type of finite-size effect}
\label{sec: A new type of finite-size effect}

In this section, we describe a new type of finite-size effect which appears in particular in the $\beta$-deformation with gauge group $\SUN$, while the undeformed $\mathcal{N}=4$ SYM theory is insensitive to it. Moreover, we classify which operators are potentially affected.

As explained before, finite-size corrections arise in the 't Hooft limit from certain multi-trace colour structures, whose apparently suppressed $N$-powers are enhanced to the leading order when attached to external states. In a theory with $U(N)$ gauge group and only single-trace vertices, such as the $\mathcal{N}=4$ SYM theory, these multi-trace structures can only be generated by the wrapping effect, i.e.\ by interactions that wrap at least once around an external state. This is only possible if the loop-order $K$ of the interaction equals or exceeds the length $L$ of the composite single-trace operator, i.e.\ $K\geq L$. In the following, we will argue that for $\SUN$ gauge group an additional source for multi-trace structures and hence finite-size corrections exists: the propagator.

We denote by $(\T^a)^i{}_j$ the generators of the gauge group with fundamental indices $i,j=1,\dots,N$ and adjoint index $a=s,\dots,N^2-1$. Recall that $s=0$ for \UN and $s=1$ for \SUN, as defined in \eqref{eq: color s}. The generators are normalised such that 
\begin{align}
 \tr(\T^a\T^b)=\delta^{ab} \eqndot
\end{align}
They obey the following important identity:
\begin{equation}\label{eq: Ts summed over a}
\sum_{a=\colors}^{N^2-1}(\T^a)^i{}_j(\T^a)^k{}_l
=\delta^i_l\delta^k_j-\frac{\colors}{N}\delta^i_j\delta^k_l \eqncom
\end{equation}
which occurs as the colour part of the propagator of adjoint fields. The second term is only present in the \SUN case, where it removes a contribution from the \U1 generator contained in the first term. In double-line notation, \eqref{eq: Ts summed over a} can be depicted as 
\setlength{\fboxrule}{0pt} 
\begin{align}
\delta^i_l\delta^k_j - \frac{\colors}{N} \delta^i_j\delta^k_l {} &= {} \, %
\settoheight{\eqoff}{$\times$}%
\setlength{\eqoff}{0.5\eqoff}%
\addtolength{\eqoff}{-4.5\unitlength}%
\raisebox{\eqoff}{\fbox{%
\fmfframe(1,2)(1,2){%
 \begin{fmfgraph*}(20,5)
  \fmfstraight
  \fmfpen{10}
  \fmfleft{i}
  \fmfright{o}
  \fmf{plain_n}{i,o}
  \fmffreeze
  \fmfdraw
  \fmfpen{8}
  \fmf{iplain_n,foreground=white}{i,o}
  \fmfdraw
  \fmf{leftrightarrows,foreground=black}{i,o}
  \fmfdraw
  \fmfiv{l=$\scriptstyle i$,l.a=90,l.d=7}{vloc(__i)}
  \fmfiv{l=$\scriptstyle j$,l.a=-90,l.d=7}{vloc(__i)}
  \fmfiv{l=$\scriptstyle l$,l.a=90,l.d=7}{vloc(__o)}
  \fmfiv{l=$\scriptstyle k$,l.a=-90,l.d=7}{vloc(__o)}
  \fmfdraw
 \end{fmfgraph*}%
}}}\,{}%
-{}\frac{\colors}{N}{}\,%
\settoheight{\eqoff}{$\times$}%
\setlength{\eqoff}{0.5\eqoff}%
\addtolength{\eqoff}{-4.5\unitlength}%
\raisebox{\eqoff}{\fbox{%
\fmfframe(1,2)(1,2){%
 \begin{fmfgraph*}(20,5)
  \fmfstraight
  \fmfpen{10}
  \fmfleft{i}
  \fmfright{o}
  \fmf{interrupted_plain_n}{i,o}
  \fmffreeze
  \fmfdraw
  \fmfpen{8}
  \fmf{iinterrupted_plain_n,foreground=white}{i,o}
  \fmfdraw
  \fmf{leftrightarrows_interruptedold,foreground=black}{i,o}
  \fmfdraw
  \fmfiv{l=$\scriptstyle i$,l.a=90,l.d=7}{vloc(__i)}
  \fmfiv{l=$\scriptstyle j$,l.a=-90,l.d=7}{vloc(__i)}
  \fmfiv{l=$\scriptstyle l$,l.a=90,l.d=7}{vloc(__o)}
  \fmfiv{l=$\scriptstyle k$,l.a=-90,l.d=7}{vloc(__o)}
  \fmfdraw
 \end{fmfgraph*}%
}}}{}
\eqndot \label{eq: double-line SUN propagator}
\end{align}
Let us investigate in which cases the $\frac{1}{N}$ prefactor of the second term in \eqref{eq: double-line SUN propagator} is enhanced in $N$ such that it contributes at the same leading order as the first term. In a generic diagram contributing to the $n$-point function of gauge-invariant composite operators, the colour lines in \eqref{eq: double-line SUN propagator} are closed by other propagators, vertices and operators. This allows two distinct connections of the indices $i,j,k,l$ in \eqref{eq: double-line SUN propagator}: either $i$ is connected with $l$ and $k$ with $j$ or $i$ with $j$ and $k$ with $l$.
In the first case, we obtain
\begin{equation}
\underbrace{\fbox{
\settoheight{\eqoff}{$\times$}%
\setlength{\eqoff}{0.5\eqoff}%
\addtolength{\eqoff}{-10.5\unitlength}%
\raisebox{\eqoff}{%
\fmfframe(1.5,8)(0,8){
 \begin{fmfgraph*}(20,5)
  \fmfstraight
  \fmfpen{10}
  \fmfleft{i}
  \fmfright{o}
  \fmf{plain_n}{i,o}
  \fmfdraw
  \fmfpen{8}
  \fmf{iplain_n,foreground=white}{i,o}
  \fmfdraw
  \fmf{leftrightarrows,foreground=black}{i,o}
  \fmfdraw
 \fmfpen{1}
 \fmfcmd{z0=(0.5 vloc(__i)+0.5 vloc(__o)+(0,-30)); draw (vloc(__i)+(0,-4.5)){dir 180}...z0...{dir 180}(vloc(__o)+(0,-4.5)) withcolor (0.65,0.65,0.65);}
 \fmfcmd{z1=(0.5 vloc(__i)+0.5 vloc(__o)+(0,+30)); draw (vloc(__i)+(0,+4.5)){dir 180}...z1...{dir 180}(vloc(__o)+(0,+4.5)) withcolor (0.65,0.65,0.65);}
  \fmfiv{l=$\scriptstyle i$,l.a=90,l.d=7}{vloc(__i)}
  \fmfiv{l=$\scriptstyle j$,l.a=-90,l.d=7}{vloc(__i)}
  \fmfiv{l=$\scriptstyle l$,l.a=90,l.d=7}{vloc(__o)}
  \fmfiv{l=$\scriptstyle k$,l.a=-90,l.d=7}{vloc(__o)}
  \fmfdraw
 \end{fmfgraph*}
}
}}
}_{N^2}
\,{}-{}\frac{\colors}{N}{}\,
\underbrace{\fbox{
\settoheight{\eqoff}{$\times$}%
\setlength{\eqoff}{0.5\eqoff}%
\addtolength{\eqoff}{-10.5\unitlength}%
\raisebox{\eqoff}{%
\fmfframe(1.5,8)(0,8){
 \begin{fmfgraph*}(20,5)
  \fmfstraight
  \fmfpen{10}
  \fmfleft{i}
  \fmfright{o}
  \fmf{interrupted_plain_n}{i,o}
  \fmffreeze
  \fmfdraw
  \fmfpen{8}
  \fmf{iinterrupted_plain_n,foreground=white}{i,o}
  \fmfdraw
  \fmf{leftrightarrows_interruptedold,foreground=black}{i,o}
  \fmfdraw 
\fmfpen{1}
 \fmfcmd{z0=(0.5 vloc(__i)+0.5 vloc(__o)+(0,-30)); draw (vloc(__i)+(0,-4.5)){dir 180}...z0...{dir 180}(vloc(__o)+(0,-4.5)) withcolor (0.65,0.65,0.65);}
 \fmfcmd{z1=(0.5 vloc(__i)+0.5 vloc(__o)+(0,+30)); draw (vloc(__i)+(0,+4.5)){dir 180}...z1...{dir 180}(vloc(__o)+(0,+4.5)) withcolor (0.65,0.65,0.65);}
  \fmfiv{l=$\scriptstyle i$,l.a=90,l.d=7}{vloc(__i)}
  \fmfiv{l=$\scriptstyle j$,l.a=-90,l.d=7}{vloc(__i)}
  \fmfiv{l=$\scriptstyle l$,l.a=90,l.d=7}{vloc(__o)}
  \fmfiv{l=$\scriptstyle k$,l.a=-90,l.d=7}{vloc(__o)}
  \fmfdraw
 \end{fmfgraph*}
}
}}
}_{N^1}
\,{}\propto{}\left(1-\frac{\colors}{N^2}\right)N^2 \eqndot
\end{equation}
The second term is suppressed by $\frac{1}{N^2}$ since besides the $\frac{1}{N}$ prefactor it has one index loop less than the first term.
In the second case, however, we have 
\begin{equation}
\underbrace{%
\settoheight{\eqoff}{$\times$}%
\setlength{\eqoff}{0.5\eqoff}%
\addtolength{\eqoff}{-4.5\unitlength}%
\raisebox{\eqoff}{\fbox{%
\fmfframe(3,2)(3,2){%
 \begin{fmfgraph*}(20,5)
  \fmfstraight
  \fmfpen{10}
  \fmfleft{i}
  \fmfright{o}
  \fmf{plain_n}{i,o}
  \fmfdraw
  \fmfpen{8}
  \fmf{iplain_n,foreground=white}{i,o}
  \fmfdraw
  \fmf{leftrightarrows,foreground=black}{i,o}
  \fmfdraw
 \fmfpen{1}
 \fmfcmd{z3=(vloc(__i)+(-10,0)); draw (vloc(__i)+(0,-4.5)){dir 180}...z3...(vloc(__i)+(0,4.5)){dir 0} withcolor (0.65,0.65,0.65);}
 \fmfcmd{z4=(vloc(__o)+(10,0)); draw (vloc(__o)+(0,-4.5)){dir 0}...z4...(vloc(__o)+(0,4.5)){dir 180} withcolor (0.65,0.65,0.65);}
  \fmfiv{l=$\scriptstyle i$,l.a=90,l.d=7}{vloc(__i)}
  \fmfiv{l=$\scriptstyle j$,l.a=-90,l.d=7}{vloc(__i)}
  \fmfiv{l=$\scriptstyle l$,l.a=90,l.d=7}{vloc(__o)}
  \fmfiv{l=$\scriptstyle k$,l.a=-90,l.d=7}{vloc(__o)}
  \fmfdraw
 \end{fmfgraph*}%
}%
}}%
}_{N^1}\,%
{}-{}\frac{\colors}{N}\,%
\underbrace{%
\settoheight{\eqoff}{$\times$}%
\setlength{\eqoff}{0.5\eqoff}%
\addtolength{\eqoff}{-4.5\unitlength}%
\raisebox{\eqoff}{\fbox{%
\fmfframe(3,2)(3,2){%
 \begin{fmfgraph*}(20,5)
  \fmfstraight
  \fmfpen{10}
  \fmfleft{i}
  \fmfright{o}
  \fmf{interrupted_plain_n}{i,o}
  \fmffreeze
  \fmfdraw
  \fmfpen{8}
  \fmf{iinterrupted_plain_n,foreground=white}{i,o}
  \fmfdraw
  \fmf{leftrightarrows_interruptedold,foreground=black}{i,o}
  \fmfdraw 
\fmfpen{1}
 \fmfcmd{z3=(vloc(__i)+(-10,0)); draw (vloc(__i)+(0,-4.5)){dir 180}...z3...(vloc(__i)+(0,4.5)){dir 0} withcolor (0.65,0.65,0.65);}
 \fmfcmd{z4=(vloc(__o)+(10,0)); draw (vloc(__o)+(0,-4.5)){dir 0}...z4...(vloc(__o)+(0,4.5)){dir 180} withcolor (0.65,0.65,0.65);}
  \fmfiv{l=$\scriptstyle i$,l.a=90,l.d=7}{vloc(__i)}
  \fmfiv{l=$\scriptstyle j$,l.a=-90,l.d=7}{vloc(__i)}
  \fmfiv{l=$\scriptstyle l$,l.a=90,l.d=7}{vloc(__o)}
  \fmfiv{l=$\scriptstyle k$,l.a=-90,l.d=7}{vloc(__o)}
  \fmfdraw
 \end{fmfgraph*}%
}%
}}%
}_{N^2}\,{}\propto{}\left(1-\colors\right)N\eqndot
\end{equation}
Here, the second term is of the same order in $N$ as the first term since its prefactor $\frac{1}{N}$ is compensated by a factor $N$ from an additional index loop compared to the first term.

These results can be easily interpreted: the second term in \eqref{eq: double-line SUN propagator} contributes in leading order precisely if only the $\U1$ component propagates in the first term. This can only happen if by cutting a single propagator the diagram of the gauge-invariant $n$-point function decomposes into two parts, i.e.\ if it is one-particle reducible. For gauge group $\SUN$, i.e.\ for $\colors=1$, such diagrams do not contribute, as the second (double-trace) contribution in \eqref{eq: double-line SUN propagator} cancels the first (single-trace) contribution. For the particular case of two-point functions, the affected diagrams are of s-channel\footnote{We trust that the reader will not confuse the Mandelstam variable s with the gauge group parameter $\colors$ previously defined in \eqref{eq: color s}.} type and have the generic form 
\begin{equation}
\label{eq: generic prewrapping diagram}
\settoheight{\eqoff}{$+$}%
\setlength{\eqoff}{1.5\eqoff}%
\addtolength{\eqoff}{-9\unit}%
\raisebox{\eqoff}{\fbox{%
\rotatebox{90}{%
\begin{pspicture}(0,-1.5)(16,10.5)
\ulinsert{0}{9}
\olvertex{0}{6}
\psline[linestyle=dotted,linewidth=\blacklinew](0,4)(0,2)
\psline[linestyle=dotted,linecolor=ogray,linewidth=\blacklinew](0.75,4)(0.75,2)
\psline[linestyle=dotted,linewidth=\blacklinew](1.5,4)(1.5,2)
\dlinsert{0}{0}
\urinsert{16}{9}
\orvertex{16}{6}
\psline[linestyle=dotted,linewidth=\blacklinew](14.5,4)(14.5,2)
\psline[linestyle=dotted,linecolor=ogray,linewidth=\blacklinew](15.25,4)(15.25,2)
\psline[linestyle=dotted,linewidth=\blacklinew](16,4)(16,2)
\drinsert{16}{0}
\uinex{2}{9}
\iinex{2}{6}
\dinex{2}{0}
\setlength{\ya}{9\unit}
\addtolength{\ya}{0.5\dlinewidth}
\setlength{\yb}{0\unit}
\addtolength{\yb}{-0.5\dlinewidth}
\setlength{\xc}{7.0\unit}
\setlength{\yc}{4.5\unit}
\addtolength{\yc}{-0.5\dlinewidth}
\setlength{\xd}{9.0\unit}
\setlength{\yd}{4.5\unit}
\addtolength{\yd}{0.5\dlinewidth}
\psbezier[linewidth=\blacklinew](3.5,\ya)(5.5,\ya)(5.5,\yd)(\xc,\yd)
\psbezier[linewidth=\blacklinew](3.5,\yb)(5.5,\yb)(5.5,\yc)(\xc,\yc)
\psline[linestyle=dotted,linewidth=\blacklinew](3.5,4)(3.5,2)
\psbezier[linewidth=\blacklinew](12.5,\ya)(10.5,\ya)(10.5,\yd)(\xd,\yd)
\psbezier[linewidth=\blacklinew](12.5,\yb)(10.5,\yb)(10.5,\yc)(\xd,\yc)
\doutex{14}{0}
\ioutex{14}{6}
\uoutex{14}{9}
\psline[linestyle=dotted,linewidth=\blacklinew](12.5,4)(12.5,2)
\setlength{\xa}{3.5\unit}
\addtolength{\xa}{\dlinewidth}
\setlength{\xb}{6.5\unit}
\addtolength{\xb}{-\dlinewidth}
\setlength{\ya}{8\unit}
\addtolength{\ya}{-0.5\dlinewidth}
\setlength{\yb}{1\unit}
\addtolength{\yb}{0.5\dlinewidth}
\psline[linecolor=gray,fillstyle=solid,fillcolor=gray,linearc=\linearc](\xa,\ya)(\xb,4.5)(\xa,\yb)
\setlength{\xa}{12.5\unit}
\addtolength{\xa}{-\dlinewidth}
\setlength{\xb}{9.5\unit}
\addtolength{\xb}{\dlinewidth}
\setlength{\ya}{8\unit}
\addtolength{\ya}{-0.5\dlinewidth}
\setlength{\yb}{1\unit}
\addtolength{\yb}{0.5\dlinewidth}
\psline[linecolor=gray,fillstyle=solid,fillcolor=gray,linearc=\linearc](\xa,\ya)(\xb,4.5)(\xa,\yb)
\psset{linecolor=black,linewidth=\blacklinew}
\psline[doubleline=true,doublesep=4.\blacklinew](7.0,4.5)(9.0,4.5)
\end{pspicture}}}}%
\,{}-{}\displaystyle\frac{\colors}{N}\, %
\settoheight{\eqoff}{$+$}%
\setlength{\eqoff}{1.5\eqoff}%
\addtolength{\eqoff}{-9\unit}%
\raisebox{\eqoff}{\fbox{%
\rotatebox{90}{%
\begin{pspicture}(0,-1.5)(16,10.5)
\ulinsert{0}{9}
\olvertex{0}{6}
\psline[linestyle=dotted,linewidth=\blacklinew](0,4)(0,2)
\psline[linestyle=dotted,linecolor=ogray,linewidth=\blacklinew](0.75,4)(0.75,2)
\psline[linestyle=dotted,linewidth=\blacklinew](1.5,4)(1.5,2)
\dlinsert{0}{0}
\urinsert{16}{9}
\orvertex{16}{6}
\psline[linestyle=dotted,linewidth=\blacklinew](14.5,4)(14.5,2)
\psline[linestyle=dotted,linecolor=ogray,linewidth=\blacklinew](15.25,4)(15.25,2)
\psline[linestyle=dotted,linewidth=\blacklinew](16,4)(16,2)
\drinsert{16}{0}
\uinex{2}{9}
\iinex{2}{6}
\dinex{2}{0}
\setlength{\ya}{9\unit}
\addtolength{\ya}{0.5\dlinewidth}
\setlength{\yb}{0\unit}
\addtolength{\yb}{-0.5\dlinewidth}
\setlength{\xc}{7.0\unit}
\setlength{\yc}{4.5\unit}
\addtolength{\yc}{-0.5\dlinewidth}
\setlength{\xd}{9.0\unit}
\setlength{\yd}{4.5\unit}
\addtolength{\yd}{0.5\dlinewidth}
\psbezier[linewidth=\blacklinew](3.5,\ya)(5.5,\ya)(5.5,\yd)(\xc,\yd)
\psbezier[linewidth=\blacklinew](3.5,\yb)(5.5,\yb)(5.5,\yc)(\xc,\yc)
\psline[linestyle=dotted,linewidth=\blacklinew](3.5,4)(3.5,2)
\psbezier[linewidth=\blacklinew](12.5,\ya)(10.5,\ya)(10.5,\yd)(\xd,\yd)
\psbezier[linewidth=\blacklinew](12.5,\yb)(10.5,\yb)(10.5,\yc)(\xd,\yc)
\doutex{14}{0}
\ioutex{14}{6}
\uoutex{14}{9}
\psline[linestyle=dotted,linewidth=\blacklinew](12.5,4)(12.5,2)
\setlength{\xa}{3.5\unit}
\addtolength{\xa}{\dlinewidth}
\setlength{\xb}{6.5\unit}
\addtolength{\xb}{-\dlinewidth}
\setlength{\ya}{8\unit}
\addtolength{\ya}{-0.5\dlinewidth}
\setlength{\yb}{1\unit}
\addtolength{\yb}{0.5\dlinewidth}
\psline[linecolor=gray,fillstyle=solid,fillcolor=gray,linearc=\linearc](\xa,\ya)(\xb,4.5)(\xa,\yb)
\setlength{\xa}{12.5\unit}
\addtolength{\xa}{-\dlinewidth}
\setlength{\xb}{9.5\unit}
\addtolength{\xb}{\dlinewidth}
\setlength{\ya}{8\unit}
\addtolength{\ya}{-0.5\dlinewidth}
\setlength{\yb}{1\unit}
\addtolength{\yb}{0.5\dlinewidth}
\psline[linecolor=gray,fillstyle=solid,fillcolor=gray,linearc=\linearc](\xa,\ya)(\xb,4.5)(\xa,\yb)
\psset{linecolor=black,linewidth=\blacklinew}
\uoneprop{8}{4.5}{0}
\end{pspicture}}}}
\,{}\propto{}
\,\,\,(1-\colors)N^{2L-1}\eqncom
\end{equation}
where the two external states of length $L$ are depicted in light grey and the area in dark grey stands for possible additional planar interactions. The interactions have to reduce the $L$ fields of each of the two operators into a single field. Since the reduction of two fields into a single field comes with one factor of the effective planar coupling $g$,\footnote{This consideration holds for all $v$-valent vertices in the actions  \eqref{eq: superspace action} and \eqref{eq: component action} since they are of order $\cO(g^{v-2})$. } the diagrams of type \eqref{eq: generic prewrapping diagram} are at least of order $\cO(g^{2L-2})$. This is one loop order lower than the leading wrapping order and hence we call this new finite-size effect \emph{prewrapping}.\footnote{For the sake of simplicity, we are neglecting length-changing interactions in the main text. These are, however, also affected and easy to incorporate. For a two-point function connecting an operator of length $L$ with one of length $L^\prime$, the critical order simply becomes $g^{L+L^\prime-2}$.} The consequence of prewrapping is the vanishing of all s-channel diagrams in the $\SUN$ case ($\colors=1$) in contrast to the $\UN$ case ($\colors=0$), cf.\ \eqref{eq: generic prewrapping diagram}.

Note that the quartic scalar interactions from the action \eqref{eq: component action} also fit into the above analysis. In fact, the superspace action contains only cubic interactions between chiral superfields. This is still the case in the component expansion including auxiliary fields. The quartic scalar vertices only appear when the auxiliary fields are integrated out. 
In particular, the elimination of the $F^i$ auxiliary fields generates the double-trace term in the action \eqref{eq: component action} via \eqref{eq: Ts summed over a}.
This term is just another description of the prewrapping effect caused by the propagator of the $F^i$ auxiliary fields.
It is thus perfectly acceptable and, for a homogeneous description of prewrapping, also advisable to conduct the analysis in one of the two former formulations.

Since the undeformed theory is insensitive to the difference between the gauge groups $\UN$ and $\SUN$, prewrapping must not have a net effect there. This means the contributions to both colour structures in \eqref{eq: generic prewrapping diagram} must vanish separately. 
The occurring cancellation among the contributing diagrams with different flavour structure can be most easily seen at the example of $L=2$ states at one loop. In this case, the cyclicity of the trace symmetrises the operators with respect to their flavour degrees of freedom, while the commutator interaction reducing these two fields to the s-channel propagator is antisymmetric.\footnote{For two fermions, the state is of course antisymmetric and the interaction is symmetric.}
The same argument goes through for longer states if we consider the last remaining pair of fields in the aforementioned reduction to a single field. 
In the deformed theory, whose \cstar-commutator interactions are not antisymmetric, this cancellation ceases to happen. 

Based on the above consideration, we can classify for which operators prewrapping can occur. 
In prewrapping diagrams, the whole composite operator is reduced to a single field, which hence must carry the complete $\su4$ Cartan charge of that operator. In particular, a field (or auxiliary field) with such charges must exist, cf.\ Table~\ref{tab: su(4) charges}. In addition, its $(Q^1,Q^2)$ charge has to be non-vanishing; fields with vanishing $(Q^1,Q^2)$ charge have undeformed interactions, which leads to an automatic cancellation as in \NfSYMt.

Let us apply these criteria to the closed subsectors of the theory, which for the undeformed theory were classified in \cite{Beisert03}. It becomes an easy combinatorial exercise to give all operators that are potentially affected.
For the compact closed subsectors, the results are displayed in Table~\ref{tab: prewrapping in compact subsectors}.
It shows that all respective candidates for prewrapping are obtained by acting 
on the $L=2$ single-impurity state $\tr[\phi^2\phi^3]$ with charge conjugation and/or a $\ZZ_3$ symmetry. The latter symmetry cyclically relabels the scalars $\phi^i$ and fermions $\psi^i$, as well as their conjugates, leaving the action invariant.\footnote{This can most easily be seen in the superspace formulation \eqref{eq: superspace action}.} Recall that the sensitivity of this state to the choice of the gauge group was already observed in \cite{FG05}, and we can now identify this phenomenon as a manifestation of prewrapping.

\begin{table}[htbp]
\centering
\begin{tabular}{l|l|l}
 Subsector & Fields & Prewrapping candidates \\ \hline
 $\SU2$ & $\phi^1,\phi^2$ & $\tr[\phi^1\phi^2]$ \\
 $\SU2$ & $\phi^1,\bar\phi^2$ & none \\ \hline
 $\U{1|1}$ & $\phi^1,\psi^4_1$ & none \\
 $\U{1|1}$ & $\phi^1,\psi^1_1$ & none \\
 $\U{1|1}$ & $\bar\phi^1,\psi^2_1$ & none \\
 $\U{1|1}$ & $\bar\phi^1,\psi^3_1$ & none \\ \hline
 $\U{1|2}$ & $\phi^1,\phi^2,\psi^4_1$ & $\tr[\phi^1\phi^2]$ \\
 $\U{1|2}$ & $\bar\phi^2,\bar\phi^3,\psi^1_1$ & $\tr[\bar\phi^2\bar\phi^3]$ \\
 $\U{1|2}$ & $\phi^1,\bar\phi^2,\psi^1_1$ & none\\
 $\U{1|2}$ & $\phi^1,\bar\phi^3,\psi^1_1$ & none \\ \hline
 $\U{1|3}$ & $\phi^1,\phi^2,\phi^3,\psi^4_1$ & $\tr[\phi^1\phi^2]+\ZZ_3$ \\
 $\U{1|3}$ & $\phi^1,\bar\phi^2,\bar\phi^3,\psi^1_1$ & $\tr[\bar\phi^2\bar\phi^3]$ \\ \hline
 $\SU{2|3}$ & $\phi^1,\phi^2,\phi^3,\psi^4_1,\psi^4_2$ & $\tr[\phi^1\phi^2]+\ZZ_3$ \\
 $\SU{2|3}$ & $\phi^1,\bar\phi^2,\bar\phi^3,\psi^1_1,\psi^1_2$ & $\tr[\bar\phi^2\bar\phi^3]$ 
\end{tabular}
\caption{Candidates for prewrapping in the compact closed subsectors. We have omitted subsectors that are related to the above via the $\ZZ_3$ symmetry and/or charge conjugation.}
\label{tab: prewrapping in compact subsectors}
\end{table}

The analogous analysis for the noncompact closed subsectors is equally straightforward and we only want to stress one noteworthy feature. 
In closed subsectors that restrict the flavour content, no combination of $\su4$ charged fields $\{\phi^i,\bar\phi^i,\psi^A_{\alpha},\bar\psi^A_{\dot\alpha}\}$ exists whose total $\su4$ charge vanishes.\footnote{%
Note that our criteria are insensitive to the difference between the closed subsectors with unrestricted flavour content, namely $\U{1,1|4}$ and $\U{1,2|4}$, and the full theory. The latter is treated below.
}
As a consequence, the aforementioned criteria can only be fulfilled by operators containing a finite number of these fields. They may, however, contain an arbitrary number of field strengths and covariant derivatives if those are admitted in the sector.

In the full theory, large families of operators exist that are candidates for prewrapping, e.g.\ 
\begin{equation}\label{eq: candidate}
 \tr\left[\phi^2\phi^3(\phi^1\bar\phi^1)^i(\phi^2\bar\phi^2)^j(\phi^3\bar\phi^3)^k(\psi^1\bar\psi^1)^l(\psi^2\bar\psi^2)^m(\psi^3\bar\psi^3)^n(\psi^4\bar\psi^4)^o\cF^p\bar\cF^q\right]\eqncom
\end{equation}
where $i,j,k,l,m,n,o,p,q\in \NN_0$ and for simplicity we have suppressed spinor indices and covariant derivatives, which can additionally act on all of the fields.

\section{The complete one-loop dilatation operator} 
\label{sec: complete one-loop dilatation operator}

In this section, we determine the missing finite-size corrections and obtain the complete one-loop dilatation operator for the planar $\beta$-deformation -- first for gauge group \SUN and then for gauge group \UN.

\subsection{Gauge group \texorpdfstring{\SUN}{SU(N)}}

The dilatation operator is most directly extracted from the UV divergences of correlation functions involving either one or two composite operators. 
According to the previous discussion, it is hence affected by prewrapping if the gauge group is $SU(N)$. In particular, this means that the asymptotic one-loop result \eqref{eq: deformation of D_2} can be extended to the full one-loop result in the $\beta$-deformation with $SU(N)$ gauge group by removing all non-vanishing s-channel contributions for states with length $L=2$. In the undeformed \NfSYMt, this is not necessary since cancellations between different s-channel diagrams result in a vanishing net contribution, as already discussed in the previous section. 

In the following, we consider the operator $\cO = \tr [\psi^1_\alpha \phi^2]$  as an example and evaluate the result of the asymptotic dilatation operator \eqref{eq: deformation of D_2} acting on it. 
We explicitly show how the corresponding s-channel diagrams cancel in the undeformed theory, while in the $\beta$-deformation a net contribution persists. This is the correct s-channel contribution in the $U(N)$ theory, but it has to be removed in the $SU(N)$ theory. After the example, we show how -- with a simple prescription -- this can efficiently be done for all one-loop states.

The operator $\cO = \tr [\psi^1_\alpha \phi^2]$ maps to the cyclic spin-chain state $\ket{\cO}$, and we act on it with the one-loop dilatation operator $D_2$. On this state, $D_2$ is the sum of two insertions $D_{2}=D_{12}+D_{21}$, where $D_{ij}$ denotes the dilatation-operator density $\diladensity_2$ acting on the two legs of $\ket{\cO}$ in the specific order $ij$, cf.\ \eqref{eq: dilatation operator on spin chain}. Hence,
\begin{equation}
 \label{eq: matrix element}
 D_2\ket{\cO} = D_{12}\ket{\cO}+D_{21}\ket{\cO} 
\end{equation}
is a linear combination of certain $L=2$ states.

In the undeformed theory, \eqref{eq: matrix element} can be calculated using the so-called harmonic action \cite{Beisert03} for $\diladensity_2^{\cN=4}$, see Appendix~\ref{subsec: Mathematica implementation of the harmonic action} for details.\footnote{Note the factor $2$ difference in our conventions in comparison to \cite{Beisert03}.} We focus on the matrix element $\bra{\cO}D_2\ket{\cO}$. It is the sum of the following four contributions:
\begin{equation}\label{eq: contributions}
 \begin{aligned}
 (\diladensity_2^{\cN=4})_{\psi^1\phi^2}^{\psi^1\phi^2}= +3\eqncom \quad
 (\diladensity_2^{\cN=4})_{\psi^1\phi^2}^{\phi^2\psi^1}= -1\eqncom \quad
 (\diladensity_2^{\cN=4})_{\phi^2\psi^1}^{\phi^2\psi^1}= +3\eqncom \quad
 (\diladensity_2^{\cN=4})_{\phi^2\psi^1}^{\psi^1\phi^2}= -1 \eqncom
\end{aligned}
\end{equation}
They can also be understood in terms of Feynman diagrams. The first two contributions are given by
\setlength{\fboxrule}{0pt} 
\begin{equation}
\begin{aligned}
 (\diladensity_2^{\cN=4})_{\psi^1\phi^2}^{\psi^1\phi^2}&=
 \underbrace{\frac{1}{2}\fbox{\FDiagram[labelleftbottom=$\scriptstyle \psi^1$,
 	  labelrightbottom=$\scriptstyle \phi^2$,
 	  labellefttop=$\scriptstyle \psi^1$,
 	  labelrighttop=$\scriptstyle \phi^2$,
	  leftSE,long,longup]{dashes_sarrow}{plain_sarrow}{}{}{}}}_{+2}
\, + \,
  \underbrace{\frac{1}{2}\FDiagram[
 	  labelleftbottom=$\scriptstyle \psi^1$,
 	  labelrightbottom=$\scriptstyle \phi^2$,
 	  labellefttop=$\scriptstyle \psi^1$,
 	  labelrighttop=$\scriptstyle \phi^2$,
	  rightSE,long,longup]{dashes_sarrow}{plain_sarrow}{}{}{}}_{+1}
\, + \,
  \underbrace{\FDiagram[styleleftbottom=dashes,
 	  stylerightbottom=plain,
 	  stylemid=dashes,
 	  stylelefttop=dashes,
 	  stylerighttop=plain,
 	  labelleftbottom=$\scriptstyle \psi^1$,
 	  labelrightbottom=$\scriptstyle \phi^2$,
 	  labellefttop=$\scriptstyle \psi^1$,
 	  labelrighttop=$\scriptstyle \phi^2$,
	  tchannel,long,longup]{dashes_sarrow}{plain_sarrow}{wiggly}{dashes_sarrow}{plain_sarrow}}_{-1}
\, + \,
\underbrace{\FDiagram[styleleftbottom=dashes,
	  stylerightbottom=plain,
	  stylemid=dashes,
	  stylelefttop=dashes,
	  stylerighttop=plain,
	  labelleftbottom=$\scriptstyle \psi^1$,
	  labelrightbottom=$\scriptstyle \phi^2$,
	  labellefttop=$\scriptstyle \psi^1$,
	  labelrighttop=$\scriptstyle \phi^2$,
	  labelmid=$\scriptstyle \psi^3$,
	  schannel,long,longup]{dashes_sarrow}{plain_sarrow}{dashes_srarrow}{dashes_sarrow}{plain_sarrow}}_{+1} \eqncom \\
(\diladensity_2^{\cN=4})^{\phi^2\psi_1^1}_{\psi^1\phi^2}&=
\underbrace{\FDiagram[styleleftbottom=dashes,
	  stylerightbottom=plain,
	  stylemid=dashes,
	  stylelefttop=dashes,
	  stylerighttop=plain,
	  labelleftbottom=$\scriptstyle \psi^1$,
	  labelrightbottom=$\scriptstyle \phi^2$,
	  labellefttop=$\scriptstyle \phi^2$,
	  labelrighttop=$\scriptstyle \psi^1$,
	  labelmid=$\scriptstyle \psi^3$,
	  schannel,long,longup]{dashes_sarrow}{plain_sarrow}{dashes_srarrow}{plain_sarrow}{dashes_sarrow}}_{-1}
\eqncom
\end{aligned}
\end{equation}
where scalars are depicted by solid lines, fermions by dashed lines, gauge fields by wiggly lines, the `blob' represents one-loop self-energy insertions and the composite operators are depicted by bold horizontal lines. 
Underneath the diagrams, we have displayed the respective individual contributions to the harmonic action, which were calculated from the black parts of the diagrams\footnote{Note that the black parts as well as the labelling directly correspond to the diagrams of operator renormalisation, which is the most direct way to obtain the action of $\diladensity_2$ as operator on the spin chain, cf.\ \cite{Sieg:2010jt}. The respective diagrams can be obtained from those of the two-point function, which are depicted by the grey completion, by amputating the outgoing operators and propagators.\label{footnote: two-point function}} via the Feynman rules given in detail in \cite{Fokken:2013aea}. 
The extension of the bold horizontal lines beyond the points where the elementary field lines originate indicates that we have only kept those terms that would also contribute when connected to an arbitrarily long operator. These are the single-trace terms with the correct colour order.
The third and forth contribution in \eqref{eq: contributions} are given by the reflections of the above diagrams with respect to the vertical axis. The reflected diagrams give the same contributions as the unreflected ones.
The four contributions from the s-channel diagram do indeed cancel each other such that $\diladensity_2^{\cN=4}$ yields the correct result for
the \NfSYMt.

In the deformed theory, the corresponding cancellation between the different contributions from the deformed \emph{asymptotic} dilatation operator density does not occur. 
The four contributions in \eqref{eq: contributions} acquire the phases $1$, $\e^{i \beta }$, $1$ and $\e^{-i \beta }$, respectively, as follows from both, \eqref{eq: deformation of D_2} and the explicit Feynman diagram calculation. The net contribution from the s-channel diagrams is non-vanishing and given by
\begin{equation}
 1-\e^{i \beta }+1-\e^{-i \beta }=4\sin^2\tfrac{\beta}{2} 
\eqndot
\end{equation}
In the $SU(N)$ theory, this contribution has to vanish because of prewrapping.

A priori, this discrepancy requires the computation and subtraction of all deformed one-loop s-channel Feynman diagrams involving $L=2$ operators. Fortunately, this is not necessary. In the remainder of this subsection, we argue that a surprising short-cut is available.\footnote{Note that this short-cut works in the supersymmetric $\beta$-deformation but fails to work in the nonsupersymmetric $\gamma_i$-deformation.} It relies on the relatively small number and simple structure of Feynman diagrams at one-loop level and is proven in three steps. 

First, we show that for certain pairs of fields in the operators the automatic cancellations between different s-channel contributions take place as in the undeformed theory. Second, we identify pairs of fields which receive spurious s-channel contributions. These contributions can be removed by setting the deformation parameter $\beta$ to zero, which restores the cancellations from the undeformed theory. Third, we show that this procedure does not alter the contributions of any non-s-channel interactions. It can hence be applied to the sum of all contributions, i.e. at the level of $\diladensity_2$.

Recall that in the $\beta$-deformation only interactions between matter-type superfields $\{\Phi^i, \bar\Phi^i\}$ -- or their respective  components $\{ \phi^i, \psi^i_\alpha, F^i, \bar\phi^i, \bar\psi^i_{\dot\alpha},\bar F^i \}$ -- are deformed. Interactions involving at least one vector superfield $V$ -- or its gauge-type on-shell components in Wess-Zumino gauge  $\{ A_\mu, \psi^4_\alpha, \bar\psi^4_{\dot\alpha} \}$ -- are undeformed. The contributions from s-channel diagrams in which both vertices are undeformed cancel as in the undeformed theory. Moreover, s-channel diagrams involving one deformed and one undeformed vertex automatically have a vanishing contribution also in the deformed theory, as the combination of symmetric state (operator) and commutator-type vertex in either initial or final state suffices for a cancellation. Hence, a non-vanishing net contribution can only come from s-channel diagrams in which both vertices are deformed, implying that all fields are of matter type. As an immediate consequence, the dilatation operator obtained from \eqref{eq: deformation of D_2} gives the correct result if at least one of the external fields (i.e.\ fields in the initial or final state) is of gauge type.

We have depicted all s-channel diagrams with only matter-type fields in the first row of Table~\ref{tab: deformed possibilities}.
They describe interactions of two incoming matter fields $\{ \phi^i,\psi^i_\alpha \}$ or anti-matter fields $\{ \bar\phi^i,\bar\psi^i_{\dot\alpha} \}$ which become two outgoing matter or anti-matter fields, respectively.%
\footnote{Note that in the picture of the two-point function these diagrams are connecting two matter fields $\{ \phi^i,\psi^i_\alpha \}$ of an $L=2$ operator $\cO$ with two \emph{anti-matter} fields $\{ \bar\phi^i,\bar\psi^i_{\dot\alpha} \}$ of a second $L=2$ operator $\bar\cO^\prime$, or, respectively, anti-matter fields in the former to matter fields in the latter. } 
We can remove their contributions to the dilatation operator by setting the deformation parameter $\beta$ to zero whenever these combinations of external fields occur. This restores the cancellations of the undeformed theory. 

\begin{table}[tbp]
\centering
\begin{tabular}{|c|@{\quad}c@{\quad}|@{\quad}c@{\quad}|}
\hline
 & in components & $\mathcal{N}=1$ \\\hline
\begin{minipage}[c]{2.4cm}
\vspace*{0.4\baselineskip}
\centering {\bf s-channel} \\
\vspace*{0.2\baselineskip}
\begin{tabular}{@{}c@{}@{}l}
 $+${ } & vertical \& \\
 & horizontal \\ 
 & reflections \\
  $+${ } & twists
\end{tabular}%
\vspace*{0.2\baselineskip}
\end{minipage}%
 & %
$\FDiagram[labelleftbottom=$\scriptstyle \psi^i$,
 	  labelrightbottom=$\scriptstyle \psi^j$,
 	  labellefttop=$\scriptstyle \psi^i$,
 	  labelrighttop=$\scriptstyle \psi^j$,
	  labelmid=$\scriptstyle \phi^k$,
	  schannel,long,longup]{dashes_sarrow}{dashes_sarrow}{plain_srarrow}{dashes_sarrow}{dashes_sarrow}$ \,\,\,\,\,\, %
$\FDiagram[labelleftbottom=$\scriptstyle \phi^i$,
 	  labelrightbottom=$\scriptstyle \psi^j$,
 	  labellefttop=$\scriptstyle \phi^i$,
 	  labelrighttop=$\scriptstyle \psi^j$,
	  labelmid=$\scriptstyle \psi^k$,
	  schannel,long,longup]{plain_sarrow}{dashes_sarrow}{dashes_srarrow}{plain_sarrow}{dashes_sarrow}$ \,\,\,\,\,\, %
$\FDiagram[labelleftbottom=$\scriptstyle \phi^i$,
 	  labelrightbottom=$\scriptstyle \phi^j$,
 	  labellefttop=$\scriptstyle \phi^i$,
 	  labelrighttop=$\scriptstyle \phi^j$,
	  xchannel,long,longup]{plain_sarrow}{plain_sarrow}{}{plain_sarrow}{plain_sarrow}%
\!\!=\!\!\FDiagram[
 	  labelleftbottom=$\scriptstyle \phi^i$,
 	  labelrightbottom=$\scriptstyle \phi^j$,
 	  labellefttop=$\scriptstyle \phi^i$,
 	  labelrighttop=$\scriptstyle \phi^j$,
	  labelmid=$\scriptstyle F^k$,
	  schannel,long,longup]{plain_sarrow}{plain_sarrow}{dots_srarrow}{plain_sarrow}{plain_sarrow}$ %
& $\FDiagram[labelleftbottom=$\scriptstyle \varPhi^i$,
 	  labelrightbottom=$\scriptstyle \varPhi^j$,
 	  labellefttop=$\scriptstyle \varPhi^i$,
 	  labelrighttop=$\scriptstyle \varPhi^j$,
	  labelmid=$\scriptstyle \varPhi^k$,
	  schannel,long,longup]{plain_sarrow}{plain_sarrow}{plain_srarrow}{plain_sarrow}{plain_sarrow}$ \\ \hline
\begin{minipage}[c]{2.4cm}
\vspace*{0.4\baselineskip}
\centering {\bf t-channel} \\
\vspace*{0.2\baselineskip}
\begin{tabular}{@{}c@{}@{}l}
 $+${ } & vertical \& \\
 & horizontal \\ 
 & reflections \\
 &
\end{tabular}%
\vspace*{0.2\baselineskip}
\end{minipage}%
 & %
$\FDiagram[labelleftbottom=$\scriptstyle \psi^i$,
 	  labelrightbottom=$\scriptstyle \bar\psi^j$,
 	  labellefttop=$\scriptstyle \bar\psi^j$,
 	  labelrighttop=$\scriptstyle \psi^i$,
	  labelmid=$\scriptstyle \phi^k$,
tchannel,long,longup]{dashes_sarrow}{dashes_srarrow}{plain_srarrow}{dashes_srarrow}{dashes_sarrow}$ \,\,\,\,\,\, %
$\FDiagram[labelleftbottom=$\scriptstyle \phi^i$,
 	  labelrightbottom=$\scriptstyle \bar\psi^j$,
 	  labellefttop=$\scriptstyle \bar\psi^j$,
 	  labelrighttop=$\scriptstyle \phi^i$,
	  labelmid=$\scriptstyle \bar\psi^k$,
	  tchannel,long,longup]{plain_sarrow}{dashes_srarrow}{dashes_srarrow}{dashes_srarrow}{plain_sarrow}$ \,\,\,\,\,\, %
$\FDiagram[labelleftbottom=$\scriptstyle \phi^i$,
 	  labelrightbottom=$\scriptstyle \bar\phi^j$,
 	  labellefttop=$\scriptstyle \bar\phi^j$,
 	  labelrighttop=$\scriptstyle \phi^i$,
	  xchannel,long,longup]{plain_sarrow}{plain_srarrow}{}{plain_srarrow}{plain_sarrow}%
\!\!=\!\!\FDiagram[
 	  labelleftbottom=$\scriptstyle \phi^i$,
 	  labelrightbottom=$\scriptstyle \bar\phi^j$,
 	  labellefttop=$\scriptstyle \bar\phi^j$,
 	  labelrighttop=$\scriptstyle \phi^i$,
	  labelmid=$\scriptstyle F^k$,
	  tchannel,long,longup]{plain_sarrow}{plain_srarrow}{dots_srarrow}{plain_srarrow}{plain_sarrow}$ & %
$\FDiagram[labelleftbottom=$\scriptstyle \varPhi^i$,
  	  labelrightbottom=$\scriptstyle \bar\varPhi^j$,
  	  labellefttop=$\scriptstyle \bar\varPhi^j$,
  	  labelrighttop=$\scriptstyle \varPhi^i$,
 	  labelmid=$\scriptstyle \bar\varPhi^k$,
 	  tchannel,long,longup]{plain_sarrow}{plain_srarrow}{plain_srarrow}{plain_srarrow}{plain_sarrow}$ \\\hline
\end{tabular}
\caption{Asymptotic range $R=2$ diagrams with two deformed vertices. Scalars and chiral superfields are depicted by solid lines, fermions by dashed lines, $F$ auxiliary fields by dotted lines, gauge fields by wiggly lines and the composite operators are depicted by bold horizontal lines. Twist signifies the vertical  reflection of only the upper half of a diagram. Scalars are treated on the same footing as the matter fermions, as the quartic vertices can be rewritten as cubic vertices with `propagating' auxiliary fields. Covariant derivatives are suppressed in the notation.}
\label{tab: deformed possibilities}
\end{table}

We want to be able to apply this procedure to the dilatation-operator density $\diladensity_2$ instead of only to individual diagrams. Hence, we have to justify that non-s-channel diagrams with the above configurations of four external matter-type fields either do not exist or are not affected. The latter is the case if the diagrams are independent of $\beta$, i.e.\ undeformed. Therefore, we have to analyse only deformed diagrams with matter-type external fields. They necessarily contain also internal fields of only matter type and are a priori of s-channel, t-channel or self-energy type. We have depicted the respective t-channel diagrams in the second row of Table~\ref{tab: deformed possibilities}. They do not exist for those combinations of external fields for which s-channel diagrams occur and thus are not altered by our procedure. It remains to be shown that the contributions from the self-energy-type diagrams are not affected either. As their subdiagrams of elementary interactions have range $R=1$ and are connected to an operator of length $L=2$, we can apply \eqref{eq: relation for colour ordered amplitudes} with $n=2$, which immediately\footnote{Recall that the \cstar-product of a fields with its conjugate reduces to the ordinary product.} shows that their contributions are independent of $\beta$. 

As the vertices only depend on the flavours of the fields involved, the same analysis is true for covariant derivatives acting on these fields. In particular, it holds for those combinations forming the alphabet $\cA$ given in \eqref{eq: alphabet}. For the translation of the above considerations, we define the following two subsets of $\cA$:
\begin{equation}
\begin{aligned}
 \cA_\text{matter}&=\{ \D^k\phi^1, \D^k\phi^2, \D^k\phi^3, \D^k\psi^1_\alpha, \D^k\psi^2_\alpha, \D^k\psi^3_\alpha \} \eqncom \\
 \bar\cA_\text{matter}&=\{ \D^k\bar\phi^1, \D^k\bar\phi^2, \D^k\bar\phi^3, \D^k\bar\psi^1_{\dot\alpha}, \D^k\bar\psi^2_{\dot\alpha},\D^k\bar\psi^3_{\dot\alpha} \}
\eqndot
\end{aligned} 
\end{equation}

According to the above discussion, the complete one-loop dilatation operator of the planar $\beta$-deformation with gauge group \SUN is given by the following density:
\begin{equation}
 (\diladensity_2^\beta)_{A_iA_j}^{A_kA_l}= \e^{\frac{\complexi}{2} (\mathbf{q}_{A_k} \wedge \mathbf{q}_{A_l}- \mathbf{q}_{A_i} \wedge \mathbf{q}_{A_j})} \rule[-0.96cm]{0.1mm}{1.415cm}\!{\phantom{|}}_{\substack{\\[0.2cm]
\beta=0\text{ if }L=2\text{ and \phantom{...............}}\\ 
(A_i,A_j,A_k,A_l\in\cA_\text{matter}\text{ or \phantom{)}}\\  
\phantom{{}({}} A_i,A_j,A_k,A_l\in\bar\cA_\text{matter}) \phantom{\text{ or }}}}  (\diladensity_2^{\cN=4})_{A_iA_j}^{A_kA_l}
\eqncom
\end{equation}
where the rule for the implementation of prewrapping introduces an explicit dependence on the operator length $L$, as expected for finite-size effects.

\subsection{Gauge group \texorpdfstring{\UN}{U(N)}}

For gauge group \UN, the asymptotic result \eqref{eq: deformation of D_2} is valid for $L\geq2$. However, the $L=1$ states, which correspond to the \U1 modes of the fields, acquire anomalous dimensions. For the matter-type fields, the respective eigenvalues $E$ of the one-loop dilatation operator, which are the anomalous dimensions divided by $g^2$, read 
\begin{equation}
 E_{\tr\phi^i}=E_{\tr\bar\phi^i}=E_{\tr\psi^i_\alpha}=E_{\tr\bar\psi^i_{\dot\alpha}}=4 \sin^2 \tfrac{\beta}{2} \eqndot 
\end{equation}
The result for the scalar fields can be directly obtained from the self-energy diagrams given in the appendix of \cite{Fokken:2013aea}. Supersymmetry demands that the respective result for the fermions is the same, and we have confirmed this by an explicit calculation using the Feynman rules of \cite{Fokken:2013aea}. The $\U1$ components of the gluino and gauge field still decouple in the $\beta$-deformation and hence the following anomalous dimensions vanish:
\begin{equation}
 E_{\tr\cF_{\alpha\beta}}=E_{\tr\cF_{\dot\alpha\dot\beta}}=E_{\tr\psi^4_\alpha}=E_{\tr\bar\psi^4_{\dot\alpha}}=0 \eqndot
\end{equation}

\section{The spectrum}
\label{sec: spectrum}

In this section, we employ the one-loop dilatation operator to compute the anomalous dimensions of all single-trace operators with classical scaling dimension $\Delta_0\leq 4.5$.
The results are structured according to primary states and for gauge group $\SUN$ are shown in Table~\ref{tab: anomalous dimensions}, as well as Table~\ref{tab: anomalous dimensions 922} in Appendix~\ref{app: spectrum tables}. An analogous table for \NfSYMt can be found in \cite{Beisert03}. Table~\ref{tab: anomalous dimensions UN} enlists the primary states of the \UN theory that do either not exist in the \SUN theory or that differ in their anomalous dimensions. In the following, we introduce the notation and derive the results presented in the tables. We conclude with a discussion for which states the one-loop spectra of the \UN and \SUN theory differ, and derive an all-loop result for one such state.

We have determined the $\cN=1$ multiplet content of the $\beta$-deformation in analogy to the \NfSYMt case \cite{Bianchi:2003wx, Bianchi:2006ti} by applying the Eratosthenes super-sieve al\-go\-rithm \cite{Bianchi:2003wx} to the refined partition function. The required $\cN=1$ characters can e.g.\ be found in \cite{Dolan:2008qi,Dobrev:2012me}. A more detailed description of this procedure is presented in Appendix~\ref{app: representation content}. We have computed the anomalous dimensions by acting with the dilatation operator on each basis of states with specified quantum numbers, subsequently diagonalising the resulting (block-diagonal) mixing matrix; see Appendix~\ref{subsec: Mathematica implementation of the harmonic action} for an explicit expression of the harmonic action. Finally, we have assigned the anomalous dimensions to the multiplets.

In the following, we structure the above results according to the symmetry of the spectrum. The $\beta$-deformation breaks the $\SU{4}_R$ R-symmetry group of \NfSYMt to $\U{1}_{Q^1}\times \U{1}_{Q^2}\times \U{1}_r$. The corresponding conserved charges span the Cartan subalgebra of $\su{4}_R$, see Section \ref{sec: asymptotic dilatation operator}. Furthermore, the action is invariant under a $\ZZ_3$ symmetry that cyclically rotates the three matter superfields into each other and leaves the vectorfield invariant. This symmetry is supplemented by the  exchange of two matter superfields and the simultaneous replacement of $\beta$ by $2\pi-\beta$. As the one-loop spectrum is invariant under this transformation of $\beta$, it is invariant under the resulting larger $\Sthree$ symmetry. Finally, the one-loop spectrum is invariant under charge conjugation, which exchanges the $\su2$ and $\overline{\mathfrak{su}}(2)$ spins $j$ and $\bar\jmath$ and sends the $\su{4}_R$ Cartan charges to their negatives.

In the Tables~\ref{tab: anomalous dimensions}, \ref{tab: anomalous dimensions UN} and \ref{tab: anomalous dimensions 922}, we label primary states by the classical scaling dimension $\Delta_0$, the spins $[j,\bar\jmath]$, the $\su{4}_R$ Cartan charges\footnote{See Table~\ref{tab: su(4) charges} for the translation between $q^1,q^2,q^3$ and $Q^1,Q^2,r$.} $(q^1,q^2,q^3)$ and the length $L$, which is only preserved at one-loop order.\footnote{For a comparison with the spectrum of \NfSYMt in \cite{Beisert03}, one should keep in mind that our spins $j,\bar \jmath$ are half integers whereas the spins $s_1,s_2$ of \cite{Beisert03} are integers, our $E$ is twice the one of \cite{Beisert03} and the charges $q^1,q^2,q^3$ are related to the $\su4$ Dynkin labels $q_1,p,q_2$ as 
\begin{equation}
 q^1=\frac{1}{2}(q_1-q_2) \eqncom \quad q^2=-\frac{1}{2}(q_1+q_2) \eqncom \quad q^3=-\frac{1}{2}(q_1+2p+q_2) \eqndot
\end{equation}%
}
If several primary states of the $\cN =1 $ superconformal group are related via the $\Sthree$ symmetry, we only give the one with the highest $q^3$ charge. Moreover, if two primary states are related by charge conjugation, we only give the one with the higher $\su2$ spin $j$ and subordinately the highest $q^3$ charge. The anomalous dimensions divided by $g^2$ are given as the solutions of polynomial equations of the form
\begin{equation}\label{eq: notation}
 E^n=\sum_{k=0}^{n-1} a_k E^k=a_{n-1}E^{n-1}+\dots+a_1 E + a_0 \eqncom
\end{equation}
which are abbreviated as $\{a_{n-1},\dots , a_1,a_0\}$. In particular, $\{a_0\}$ means $E=a_0$. Moreover, we capture the dependence on the deformation parameter $\beta$ in terms of $s_n=\sin^2\frac{n\beta}{2}$. For example, $\{12,-32s_2\}$ stands for the two solutions of the quadratic equation $E^2=  12E-32\sin^2\beta$, namely $E=6\pm 2 \sqrt{9-8\sin^2\beta}$.

\begin{table}[htbp]
\begin{tabular}{c|@{\rule[-1.25ex]{0pt}{3.25ex} }l|l| p{10.2cm} } 
$\Delta_0$ &  $[j,\bar\jmath]_{(q^1,q^2,q^3)}$ &$L$ & $E$ \\ \hline
$2$ & $[0,0]_{(-1,0,1)}  $ & $2$ & $\{8\specpar{1}\}$  \\ 
$2$ & $[0,0]_{(0,0,0)}  $ & $2$ & $\{0\}$, $\{0\}$, $\{12\}$  \\ 
$2$ & $[0,0]_{(0,0,2)}  $ & $2$ & $\{0\}$  \\ 
$2$ & $[0,0]_{(0,1,1)}  $ & $2$ & $\{0\}$  \\ 
$\tablefrac{5}{2}$ &  $[\tablefrac{1}{2},0]_{(-\frac{1}{2},\frac{1}{2},\frac{1}{2})}  $ &$2$ & $\{12,-32\specpar{1}\}$ \\
$\tablefrac{5}{2}$ & $[\tablefrac{1}{2},0]_{(\frac{1}{2},\frac{1}{2},\frac{3}{2})}  $ & $2$ & $\{0\}$  \\ 
$3$ & $[0,0]_{(-1,0,2)}  $ & $3$ & $\{8\specpar{1}\}$  \\ 
$3$ & $[0,0]_{(-1,1,1)}  $ & $3$ & $\{12,-32\specpar{2}\}$  \\ 
$3$ & $[0,0]_{(0,0,1)}  $ & $3$ & $\{12,-32\specpar{1}\}^\joinsymbol$, $\{20,-96,128\specpar{1}\}$  \\ 
$3$ & $[0,0]_{(0,0,3)}  $ & $3$ & $\{0\}$  \\ 
$3$ & $[0,0]_{(0,1,2)}  $ & $3$ & $\{8\specpar{1}\}^\joinsymbol$  \\ 
$3$ & $[0,0]_{(1,1,1)}  $ & $2$ & $\{0\}$  \\ 
$3$ & $[0,0]_{(1,1,1)}  $ & $3$ & $\{0\}$, $\{12\}^\joinsymbol$  \\ 
$3$ & $[\tablefrac{1}{2},\tablefrac{1}{2}]_{(-1,0,1)}  $ & $2$ & $\{4(3-\specpar{1})\}$  \\ 
$3$ & $[\tablefrac{1}{2},\tablefrac{1}{2}]_{(0,0,0)}  $ & $2$ & $\{0\}$, $\{12\}$, $\{12\}$, $\{12\}$  \\ 
$3$ & $[1,0]_{(0,0,1)}  $ & $2$ & $\{12\}$  \\ 
$\tablefrac{7}{2}$ & $[\tablefrac{1}{2},0]_{(-\frac{3}{2},-\frac{1}{2},\frac{3}{2})}  $ & $3$ & $\{4(3-2\specpar{1})\}$  \\ 
$\tablefrac{7}{2}$ & $[\tablefrac{1}{2},0]_{(-\frac{3}{2},\frac{1}{2},\frac{1}{2})}  $ & $3$ & $\{20,-96,128\specpar{2})\}$  \\ 
$\tablefrac{7}{2}$ & $[\tablefrac{1}{2},0]_{(-\frac{1}{2},-\frac{1}{2},\frac{1}{2})}  $ & $3$ & $\{8\}$, $\{8\}$, $\{12\}$, $\{12\}$, $\{12\}^\joinsymbol$, $\{12,-32\specpar{1}\}$  \\ 
$\tablefrac{7}{2}$ & $[\tablefrac{1}{2},0]_{(-\frac{1}{2},\frac{1}{2},\frac{3}{2})}  $ & $3$ & $\{4(3-\specpar{1})\}^\joinsymbol$, $\{20,-32(3+\specpar{1}),32(8\specpar{1}+\specpar{2})\}$  \\ 
$\tablefrac{7}{2}$ & $[\tablefrac{1}{2},0]_{(\frac{1}{2},\frac{1}{2},\frac{1}{2})}  $ & $3$ & $\{0\}$, $\{0\}$, $\{8\}$, $\{8\}$, $\{8\}$, $\{12\}$, $\{12\}$, $\{12\}^\joinsymbol$, $\{12\}^\joinsymbol$, $\{12\}^\joinsymbol$  \\ 
$\tablefrac{7}{2}$ & $[\tablefrac{1}{2},0]_{(\frac{1}{2},\frac{1}{2},\frac{5}{2})}  $ & $3$ & $\{0\}$  \\ 
$\tablefrac{7}{2}$ & $[\tablefrac{1}{2},0]_{(\frac{1}{2},\frac{3}{2},\frac{3}{2})}  $ & $3$ & $\{12,-32\specpar{1}\}^\joinsymbol$  \\ 
$\tablefrac{7}{2}$ & $[1,\tablefrac{1}{2}]_{(-\frac{1}{2},-\frac{1}{2},\frac{3}{2})}  $ & $2$ & $\{12\}$  \\ 
$\tablefrac{7}{2}$ & $[1,\tablefrac{1}{2}]_{(-\frac{1}{2},\frac{1}{2},\frac{1}{2})}  $ & $2$ & $\{12\}$, $\{12\}$  \\ 
$\tablefrac{7}{2}$ & $[\tablefrac{3}{2},0]_{(\frac{1}{2},\frac{1}{2},\frac{1}{2})}  $ & $2$ & $\{12\}$ 
\end{tabular}
\caption{Anomalous dimensions of all primary states with classical scaling dimension $\Delta_0< 4$ for gauge group $\SUN$ in the notation introduced after \eqref{eq: notation}. The dependence on the deformation parameter is encoded in $\specpar{n}=\sin^2\frac{n\beta}{2}$. Highest-weight states of the free theory that join lower-lying multiplets in the interacting theory are marked with a $\joinsymbol$. Each state has to be supplemented by its images under the $\Sthree$ symmetry and charge conjugation.}
\label{tab: anomalous dimensions}
\end{table}

\begin{table}[htbp]
\begin{tabular}{c|@{\rule[-1.25ex]{0pt}{3.25ex} }l|l| l | l }
$\Delta_0$ &  $[j,\bar\jmath]_{(q^1,q^2,q^3)}$ &$L$ & $E_{\UN}$ & $E_{\SUN}$ \\ \hline
$1$  & $[0,0]_{(0,0,1)}  $ & $1$& $\{4\specpar{1}\}$ & $-$\\ 
$\tablefrac{3}{2}$  & $[\tablefrac12,0]_{(\frac{1}{2},\frac{1}{2},\frac{1}{2})}  $ & $1$& $\{0\}$ & $-$\\ 
$2$ &  $[0,0]_{(0,1,1)}  $ & $2$ &$\{8\specpar{1}\}$ & $\{0\}$
\end{tabular}
\caption{Comparison of the anomalous dimensions of all primary states with classical scaling dimension $\Delta_0 \leq 4.5$ that differ for gauge groups $\UN$ and $\SUN$. The notation is introduced after \eqref{eq: notation}. Each state has to be supplemented by its images under the $\Sthree$ symmetry and charge conjugation.}
\label{tab: anomalous dimensions UN}
\end{table}

Certain representations are only irreducible in the interacting theory, where they have an anomalous dimension. In the free theory, where their highest weight states are at the unitarity threshold, these representations are reducible and split into several irreducible ones, see Appendix~\ref{subsec: review of N=1} for details. In Tables~\ref{tab: anomalous dimensions}, \ref{tab: anomalous dimensions UN} and \ref{tab: anomalous dimensions 922}, we have marked the corresponding highest-weight states of the free theory that have to be dropped in the interacting theory with a $\joinsymbol$. This happens e.g.\ for the Konishi multiplet, whose highest-weight state is contained in Table~\ref{tab: anomalous dimensions} as the third state in the second line with $E=12$. In the interacting theory, the second state in the $13^{\text{th}}$ line of Table~\ref{tab: anomalous dimensions} is a descendant of this state. Note that representations with vanishing one-loop anomalous dimensions -- as e.g.\ the first and second in state in the second line of Table~\ref{tab: anomalous dimensions} -- may in general still move away from the unitary threshold at higher loop orders. For the latter two states this is, however, not the case. In the undeformed theory, they are descendants of the $L=2$ protected state and thus have vanishing all-loop anomalous dimensions. As they have vanishing $\U1_{Q^1}\times\U1_{Q^2}$ charge, the argument from Section \ref{sec: asymptotic dilatation operator} can be applied and their anomalous dimensions stay zero at all loop orders in the $\beta$-deformation. Analogous considerations hold for all tabled multiplets at the unitary threshold with vanishing one-loop anomalous dimensions. 

For $\Delta_0\leq4.5$, and presumably also for higher $\Delta_0$, only one multiplet is affected by prewrapping at one-loop level, namely $\tr[\phi^2\phi^3]$ (and its images under $\Sthree$ and charge conjugation). Naively, one might expect a whole tower of affected multiplets, built from $n\geq0$ covariant derivatives distributed on $\tr[\phi^2\phi^3]$ in a similar fashion as the $SL(2)$ sector is built on $\tr[\phi^3\phi^3]$. For the conformal primary states\footnote{Similar to the situation in \NfSYMt, only the primary state with $n=0$  is a highest-weight state under the full symmetry group. For the $n\geq1$ states, the true highest-weight state is a fermionic state with classical scaling dimension $n+\frac{3}{2}$.} corresponding to $n\geq1$, however, contributions from s-channel diagrams vanish due to cancellations in the spacetime part that are independent of the deformation, see Appendix~\ref{app: argument} for details.

At two-loop order, the two-point function of the state $\tr[\phi^2\phi^3]$ was investigated in \cite{Penati:2005hp}. While the two-point function receives finite $\frac{1}{N}$ corrections, its anomalous dimension remains zero for gauge group \SUN.\footnote{Note that this is incorrectly summarised in \cite{Frolov:2005iq}.} We now show that this state is even protected at all orders in planar perturbation theory. In the $\cN=1$ superspace formulation, the superfield $\tr[\Phi^2\Phi^3]$ contains $\tr[\phi^2\phi^3]$ as its lowest component in the $\theta$-expansion. Its anomalous dimension can be extracted from the overall UV divergence of the correlation function $\langle \Phi^2(x)\Phi^3(y)\tr[\Phi^2\Phi^3](0)\rangle$. According to the finiteness-conditions of \cite{Sieg:2010tz}, an $\cN=1$ superspace Feynman diagram of range $R=2$ that contributes to this correlation function can only have an overall UV divergence if at least one of its vertices is not part of a loop. This condition is only fulfilled by the diagrams of s-channel type, shown at one-loop in the upper right corner of Table~\ref{tab: deformed possibilities}. At higher loops, arbitrary interactions supplement the lower (black) half of this diagram. For gauge group \SUN, these diagrams vanish by the prewrapping effect as discussed before. Moreover, the remaining $R=1$ diagrams are self-energy corrections of the elementary superfields, which are finite. The above correlation function is therefore finite and the anomalous dimension vanishes.

\section{Conclusion and outlook}
\label{sec: Conclusion and outlook}

In this paper, we have analysed finite-size corrections in the real $\beta$-deformed $\mathcal{N}=4$ SYM theory in the 't Hooft limit. We have constructed the complete one-loop dilatation operator by incorporating these corrections into the proposal of \cite{BR05}. The latter (asymptotic) result was obtained by applying Filk's theorem from spacetime noncommutative field theory to the undeformed dilatation-operator density, i.e.\ to the combination of respective planar single-trace Feynman diagrams.

We have analysed in detail the limits and implications of Filk's theorem when applied in the $\beta$-deformation. 
We have found that generic external states have to be removed from a diagram before the theorem is applicable.
If the diagram is associated with a finite-size effect, it becomes non-planar after this truncation. Hence, finite-size corrections in general
invalidate the results relying on Filk's theorem.
External multi-trace states in which all traces are neutral under the $(Q^1,Q^2)$ global charge, however, need not be removed from the planar $\mathcal{N}=4$ SYM diagrams, and Filk's theorem can be applied to the entire diagrams undergoing the deformation. 
This implies that all $n$-point correlation functions of the resulting deformed states are identical to their counterparts in the undeformed $\mathcal{N}=4$ SYM theory at any loop order.
In particular, the anomalous dimension of the Konishi primary operator $\sum^3_{i=1}\tr[\phi^i\bar\phi^i]$ is undeformed. 

From the findings summarised in the above paragraph, it follows that the one-loop dilatation operator of \cite{BR05} is valid only up to finite-size corrections.  
In the non-conformal $U(N)$ theory, it has to be supplemented by the anomalous dimensions of the $L=1$ states, which are affected by the well-known finite-size effect of wrapping. In the conformal $SU(N)$ theory, we have identified a new type of finite-size effect, which has to be taken into account for certain $L=2$ states.
It is caused by the $SU(N)$ propagators of the adjoint fields and starts to affect states of length $L$ at loop order $K=L-1$. Since this is one loop order lower than the critical wrapping order $K=L$, we call it \emph{prewrapping}.
We have identified criteria for states which may be affected by it. 
In all compact closed subsectors, prewrapping at all loop orders affects only the state $\tr[\phi^2\phi^3]$ (and its five images under the $\mathds{Z}_3$ symmetry and charge conjugation).
In the full theory, prewrapping candidates with generic lengths $L\geq3$ exist for sufficiently high loop orders.

At one loop, we have found that the prewrapping effect can be incorporated into the dilatation operator of the deformed theory without explicitly calculating Feynman diagrams, simply by removing the deformation whenever the external states match certain criteria. 
This procedure strongly relies on the small number and simple structure of one-loop diagrams, and we doubt that it can be extended to higher 
loops.

We have employed our result to determine the one-loop spectrum of the theory with classical scaling dimension $\Delta_0\leq4.5$.
At this loop order, only the superconformal multiplet with highest-weight state $\tr[\phi^2\phi^3]$ (and its five images under the $\mathds{Z}_3$ symmetry and charge conjugation) are affected by prewrapping. 
The absence of prewrapping for other one-loop candidate multiplets can be traced back to cancellations among Feynman diagrams.
Yet, it would be desirable to understand if there is a deeper principle behind it.

While we have found and analysed prewrapping in the weakly coupled gauge theory, it remains an open problem to understand this effect in the strongly coupled dual string theory. 
As we have argued, the prewrapping-affected state $\tr[\phi^2\phi^3]$ is protected at any order in perturbation theory.
Hence, it should be a supergravity mode and the calculation suggested in \cite{Frolov:2005iq} should yield a vanishing correction 
to its mass at strong coupling. It would be desirable to check this explicitly and to understand how the deformation increases the energy of e.g.\ the state dual to $\tr[\phi^2\phi^3\phi^3]$, which is not protected by prewrapping. 
We hope that this might help to understand the subtleties related to the choice of \UN or \SUN as gauge group on the string theory side. This concerns in particular the role of the \U1{} mode.

\enlargethispage{\baselineskip}
Last but not least, the prewrapping effect has important consequences for the integrability-based descriptions of the $\beta$-deformation. 
The existence of prewrapping means that finite-size effects start one loop order earlier than in the undeformed theory. 
The asymptotic Bethe equations are reliable only up to this lower loop order and need to be supplemented already before the finite-size wrapping effect is incorporated.
A very interesting possibility is that a correct inclusion of prewrapping, i.e.\ the removal of \U1 modes, could cure the divergences at  $L=2$ encountered in the TBA and Y-system equations mentioned in the introduction.  
In fact, our result allows for a patch-work solution to the spectral problem in all compact closed subsectors: the anomalous dimension of the states $\tr[\phi^i\phi^j]$, $\tr[\bar\phi^i\bar\phi^j]$ are zero at all loop orders and the anomalous dimensions of all remaining ones can be computed with the current approach of integrability. 
It remains an important challenge to reproduce these results from a homogeneous, purely integrability-based approach.
Conclusive tests of possible modifications to incorporate prewrapping into the integrability-based description do, however, require to work in non-compact subsectors or the complete theory, where $L\geq3$ prewrapping candidates exist.
The outcome of such tests will show whether the  $\beta$-deformation with gauge group \SUN is indeed as integrable as its undeformed parent theory.%
\footnote{A correct incorporation of prewrapping would also have important consequences for the possible application of integrability beyond conformality in the non-conformal $\gamma_i$-deformation. See \cite{Fokken:2013aea} for a discussion.}
Clearly, it is important to collect more, higher-loop field-theory results to guide and test these modifications of integrability.

\section*{Acknowledgements}

It is a pleasure to thank Sergey Frolov and Matthias Staudacher for various discussions and remarks on the manuscript.
We are grateful to Zoltan Bajnok, Massimo Bianchi, Burkhard Eden, Vladimir Mitev, Elli Pomoni, Radu Roiban, Henning Samtleben and Stijn van Tongeren for discussions as well as communications on different aspects of this paper.
We thank the Kavli Institute for the Physics and Mathematics of the Universe in Tokyo, where we pursued a part of our research, for warm hospitality.
Our work was supported by DFG, SFB 647 \emph{Raum -- Zeit -- Materie. Analytische und Geometrische Strukturen} and by the Marie Curie International Research Staff Exchange Network UNIFY (FP7-People-2010-IRSES under grant agreement number 269217).
J.F.\ und M.W.\ danken der Studienstiftung des deutschen Volkes f\"ur Promotionsf\"orderungsstipendien.

\appendix

\section{Table of anomalous dimensions}
\label{app: spectrum tables}

In this appendix, we provide the anomalous dimensions of all primary states with classical scaling dimensions $\Delta_0=4,\frac{9}{2}$. 
The following table should be understood as continuation of Table~\ref{tab: anomalous dimensions 922} in Section \ref{sec: spectrum}.

\begin{center}
\footnotesize
\begin{longtable}{c|@{\rule[-1.25ex]{0pt}{3.25ex} }l|l| p{10.2cm} }
$\Delta_0$ &  $[j,\bar\jmath]_{(q^1,q^2,q^3)}$ &$L$ & $E$ \\ \hline
\endfirsthead
\multicolumn{4}{c}%
{{Continued from previous page}} \\
$\Delta_0$ &  $[j,\bar\jmath]_{(q^1,q^2,q^3)}$ &$L$ & $E$ \\ \hline
\endhead
\multicolumn{4}{l}{}\\[-0.1cm]
  \multicolumn{4}{l}{\normalsize {\bfseries \tablename\ \thetable{}} -- continued on next page} 
\endfoot
\multicolumn{4}{l}{}\\[-0.5cm]
\caption{Anomalous dimensions of all primary states with classical scaling dimension $\Delta_0 =4, \frac92$ for gauge group $\SUN$ in the notation introduced after \eqref{eq: notation}. The dependence on the deformation parameter is encoded in $\specpar{n}=\sin^2\frac{n\beta}{2}$.
Highest-weight states of the free theory that join lower-lying multiplets in the interacting theory are marked with a $\joinsymbol$. Each state has to be supplemented by its images under the $\Sthree$ symmetry and charge conjugation.
}
\endlastfoot
$4$ & $[0,0]_{(-2,0,2)}  $ & $4$ & $\{12,-32\specpar{2}\}$  \\
$4$ & $[0,0]_{(-1,-1,2)}  $ & $4$ & $\{20,-96,128\specpar{3}\}$  \\
$4$ & $[0,0]_{(-1,0,1)}  $ & $4$ & $\{8\}$, $\{8\}$, $\{8\}$, $\{8\}$, $\{12,-32\specpar{1}\}$,\\*
& & & $\{56,-1232,64(207+4\specpar{1}+\specpar{2}),
-128(540+59\specpar{1}+18\specpar{2}+\specpar{3})$, \\*
& & &
$\phantom{{}\{{}}1024(135+68\specpar{1}+25\specpar{2}+4\specpar{3}),
-1024(190\specpar{1}+88\specpar{2}+26\specpar{3}+\specpar{4})\}$  \\
$4$ & $[0,0]_{(-1,0,3)}  $ & $4$ & $\{8\specpar{1}\}$  \\
$4$ & $[0,0]_{(-1,1,2)}  $ & $4$ & $\{4(3-2\specpar{1})\}^\joinsymbol$, $\{8(1+\specpar{1}),-16(2\specpar{1}+\specpar{2})\}$  \\
$4$ & $[0,0]_{(0,0,0)}  $ & $4$ & $\{0\}$, $\{0\}$, $\{0\}$, $\{8\}$, $\{8\}$, $\{8\}$, $\{8\}$, $\{8\}$, $\{8\}$, $\{12\}$, $\{12\}$, $\{12\}$, $\{12\}$, $\{12\}$, $\{12\}$, $\{12\}$, $\{12\}$, $\{12\}$, $\{20,-80\}$, $\{20,-80\}$, $\{26,-128\}$  \\
$4$ & $[0,0]_{(0,0,2)}  $ & $3$ & $\{12\}$  \\
$4$ & $[0,0]_{(0,0,2)}  $ & $4$ & $\{20,-96,128\specpar{2}\}^\joinsymbol$,
$\{28,-240,640,-512\specpar{2}\}$  \\
$4$ & $[0,0]_{(0,0,4)}  $ & $4$ & $\{0\}$  \\
$4$ & $[0,0]_{(0,1,1)}  $ & $3$ & $\{20,-96,128\specpar{1}\}$  \\
$4$ & $[0,0]_{(0,1,1)}  $ & $4$ & $\{8\}$, $\{8\}$, $\{8\}^\joinsymbol$, $\{8\}^\joinsymbol$, $\{12\}^\joinsymbol$, $\{12\}^\joinsymbol$, $\{12,-32\specpar{1}\}^\joinsymbol$, \\*
& & & $\{32,-320,64(15+4\specpar{1}),-1536\specpar{1}\}$  \\
$4$ & $[0,0]_{(0,1,3)}  $ & $4$ & $\{8\specpar{1}\}^\joinsymbol$  \\
$4$ & $[0,0]_{(0,2,2)}  $ & $4$ & $\{12,-32\specpar{2}\}^\joinsymbol$  \\
$4$ & $[0,0]_{(1,1,2)}  $ & $3$ & $\{0\}$  \\
$4$ & $[0,0]_{(1,1,2)}  $ & $4$ & $\{20,-96,128\specpar{1}\}^\joinsymbol$ \\
$4$ & $[\tablefrac{1}{2},\tablefrac{1}{2}]_{(-1,0,2)}  $ & $3$ & $\{4(5-\specpar{1}),-8(12-4\specpar{1}-\specpar{2})\}$  \\
$4$ & $[\tablefrac{1}{2},\tablefrac{1}{2}]_{(-1,1,1)}  $ & $3$ & $\{8\}$, $\{8\}$, $\{12\}$, $\{12\}$, $\{12\}^\joinsymbol$  \\
$4$ & $[\tablefrac{1}{2},\tablefrac{1}{2}]_{(0,0,1)}  $ & $3$ & $\{8\}$, $\{8\}$, $\{12\}$, $\{12\}^\joinsymbol$, $\{12\}^\joinsymbol$, $\{20,-96,128\specpar{1}\}$, \\*
& & & $\{35,-396,12(120-\specpar{1})\}$, $\{35,-396,12(120-\specpar{1})\}$ \\
$4$ & $[1,0]_{(-1,-1,2)}  $ & $3$ & $\{12\}$  \\
$4$ & $[1,0]_{(-1,0,1)}  $ & $3$ & $\{32,-16(21-\specpar{1}),16(72-8\specpar{1}-\specpar{2})\}$  \\
$4$ & $[1,0]_{(0,0,0)}  $ & $3$ & $\{8\}$, $\{8\}$, $\{8\}$, $\{12\}$, $\{12\}$, $\{12\}$, $\{12\}^\joinsymbol$, $\{18\}$  \\
$4$ & $[1,0]_{(0,0,2)}  $ & $3$ & $\{8\}$, $\{12\}^\joinsymbol$  \\
$4$ & $[1,0]_{(0,1,1)}  $ & $3$ & $\{8\}$, $\{8\}$, $\{12\}$, $\{12\}^\joinsymbol$, $\{12\}^\joinsymbol$  \\
$4$ & $[1,0]_{(1,1,2)}  $ & $3$ & $\{12\}^\joinsymbol$  \\
$4$ & $[1,1]_{(-1,0,1)}  $ & $2$ & $\{\frac{4}{3} (9+2\specpar{1})\}$  \\
$4$ & $[1,1]_{(0,0,0)}  $ & $2$ & $\{12\}$, $\{12\}$, $\{12),\{\frac{50}{3}\}$  \\
$4$ & $[\tablefrac{3}{2},\tablefrac{1}{2}]_{(0,0,1)}  $ & $2$ & $\{12\}$ \\
$\frac{9}{2}$ & $[\frac{1}{2},0]_{(-\frac{5}{2},-\frac{1}{2},\frac{3}{2})} $ & $4$ & $\{8\}$  \\
$\frac{9}{2}$ & $[\frac{1}{2},0]_{(-\frac{5}{2},\frac{1}{2},\frac{1}{2})} $ & $3$ & $\{12\}$  \\
$\frac{9}{2}$ & $[\frac{1}{2},0]_{(-\frac{5}{2},\frac{1}{2},\frac{1}{2})} $ & $4$ & $\{28$, $-240$, $640$, $-512 \specpar{3}\}$  \\
$\frac{9}{2}$ & $[\frac{1}{2},0]_{(-\frac{3}{2},-\frac{3}{2},\frac{3}{2})} $ & $4$ & $\{8\}$, $\{8\}$, $\{12\}^\joinsymbol$  \\
$\frac{9}{2}$ & $[\frac{1}{2},0]_{(-\frac{3}{2},-\frac{1}{2},\frac{1}{2})} $ & $3$ & $\{32$, $-28 (12-\specpar{1})$, $8 (144-46 \specpar{1}+\specpar{2})\}$  \\
$\frac{9}{2}$ & $[\frac{1}{2},0]_{(-\frac{3}{2},-\frac{1}{2},\frac{1}{2})} $ & $4$ &
$\{8\}$, $\{32$, $-16 (21-\specpar{1})$, $16 (72-8 \specpar{1}-\specpar{2})\}^\joinsymbol$, \\*
& & & $\{106$, $-4985-32 \specpar{1}$, $4 (34219+744 \specpar{1}+16 \specpar{2})$, \\*
& & & $\phantom{{}\{{}}-16 (151779+7540 \specpar{1}+320 \specpar{2})$, \\*
& & & $\phantom{{}\{{}}64 (454285+43714 \specpar{1}+2759 \specpar{2}+2 \specpar{3})$, \\*
& & & $\phantom{{}\{{}}-64 (3715060+639132 \specpar{1}+53612 \specpar{2}+176 \specpar{3}+\specpar{4})$, \\*
& & & $\phantom{{}\{{}}1024 (1280225+381886 \specpar{1}+40154 \specpar{2}+357 \specpar{3}+8 \specpar{4})$, \\*
& & & $\phantom{{}\{{}}-1024 (4553400+2387478 \specpar{1}+304021 \specpar{2}+5682 \specpar{3}+242 \specpar{4})$, \\*
& & & $\phantom{{}\{{}}4096 (2358000+2353306 \specpar{1}+355064 \specpar{2}+11892 \specpar{3}+755 \specpar{4}+2 \specpar{5})$,\\*
& & & $\phantom{{}\{{}}-4096 (2160000+5310120 \specpar{1}+935079 \specpar{2}+50086 \specpar{3}+4234 \specpar{4}$, \\*
& & & $\hphantom{{}\{{}-4096(}+50\specpar{5}+\specpar{6})$, \\*
& & & $\phantom{{}\{{}}32768 (653112 \specpar{1}+132666 \specpar{2}+10408 \specpar{3}+1105 \specpar{4}+32 \specpar{5}+2 \specpar{6})\}$  \\
$\frac{9}{2}$ & $[\frac{1}{2},0]_{(-\frac{3}{2},-\frac{1}{2},\frac{5}{2})} $ & $4$ & $\{20-8 \specpar{1}$, $-16 (6-4 \specpar{1}-\specpar{2})\}$  \\
$\frac{9}{2}$ & $[\frac{1}{2},0]_{(-\frac{3}{2},\frac{1}{2},\frac{3}{2})} $ & $4$ & $\{80$, $-4 (692+5 \specpar{1}-4 \specpar{2})$, $8 (6760+177 \specpar{1}-109 \specpar{2}+\specpar{3})$, \\*
& & &
$\phantom{{}\{{}}-16 (40768+2630 \specpar{1}-1187 \specpar{2}+42 \specpar{3})$, \\*
& & & $\phantom{{}\{{}}32 (155360+21264 \specpar{1}-6464 \specpar{2}+648 \specpar{3}+33 \specpar{4})$, \\*
& & & $\phantom{{}\{{}}-64 (365184+100981 \specpar{1}-17319 \specpar{2}+4983 \specpar{3}+581 \specpar{4}+8 \specpar{5})$, \\*
& & & $\phantom{{}\{{}}128 (483840+281360 \specpar{1}-14562 \specpar{2}+20629 \specpar{3}+3800 \specpar{4}+131 \specpar{5}-4 \specpar{6})$, \\*
& & & $\phantom{{}\{{}}-512 (138240+212579 \specpar{1}+12194 \specpar{2}+22021 \specpar{3}+5472 \specpar{4}+344 \specpar{5}$ \\
& & & $\phantom{{}\{{}-512 (}-12 \specpar{6})$, \\*
& & & $\phantom{{}\{{}}512 (268176 \specpar{1}+42376 \specpar{2}+38000 \specpar{3}+11729 \specpar{4}+1156 \specpar{5}-20 \specpar{6}-4 \specpar{7})\}$  \\
$\frac{9}{2}$ & $[\frac{1}{2},0]_{(-\frac{1}{2},-\frac{1}{2},-\frac{1}{2})}  $ & $3$ & $\{0\}$, $\{8\}$, $\{8\}$, $\{8\}$, $\{12\}$, $\{12\}$, $\{12\}$, $\{15\}$, $\{15\}$  \\
$\frac{9}{2}$ & $[\frac{1}{2},0]_{(-\frac{1}{2},-\frac{1}{2},-\frac{1}{2})}  $ & $4$ & $\{0\}$, $\{8\}$, $\{8\}$, $\{8\}$, $\{8\}$, $\{8\}$, $\{8\}$, $\{8\}^\joinsymbol$, $\{8\}^\joinsymbol$, $\{8\}^\joinsymbol$, $\{12\}$, $\{12\}$, $\{12\}^\joinsymbol$, $\{12\}^\joinsymbol$,  $\{12\}^\joinsymbol$, $\{12\}^\joinsymbol$, $\{15\}$, $\{15\}$, $\{15\}$, $\{15\}$, $\{18\}^\joinsymbol$, \\*
& & & $\{20$, $-80\}$, $\{20$, $-80\}$, $\{20$, $-80\}$  \\
$\frac{9}{2}$ & $[\frac{1}{2},0]_{(-\frac{1}{2},-\frac{1}{2},\frac{3}{2})} $ & $4$ & $\{8\}$, $\{8\}$, $\{12\}$, $\{20$, $-96$, $128 \specpar{2}\}$, \\*
& & & $\{35$, $-396$, $12 (120-\specpar{2})\}$, $\{35$, $-396$, $12 (120-\specpar{2})\}$, \\*
& & & $\{40$, $-576$, $3520$, $-64 (120-\specpar{2})\}$, $\{40$, $-576$, $3520$, $-64 (120-\specpar{2})\}$  \\
$\frac{9}{2}$ & $[\frac{1}{2},0]_{(-\frac{1}{2},\frac{1}{2},\frac{1}{2})} $ & $4$ & $\{8\}$, $\{8\}$, $\{8\}$, $\{8\}$, $\{8\}$, $\{8\}$, $\{12\}$, $\{12\}$, $\{12\}$, $\{12\}$, $\{12\}$, $\{12\}$, \\*
& & & $\{38$, $-456$, $64 (27-\specpar{1})\}$, $\{32$, $-320$, $64 (15+4 \specpar{1})$, $-1536 \specpar{1}\}$, \\*
& & & $\{58$, $-16 (80-\specpar{1})$, $13376-640 \specpar{1}$, $-128 (515-48 \specpar{1})$, $4096 (30+\specpar{1})$, \\*
& & & $\phantom{{}\{{}}-8192 (12 \specpar{1}+\specpar{2})\}$, \\*
& & & $\{78$, $-2561$, $45808$, $-256 (1880-\specpar{1})$, $2965440-9216 \specpar{1}$, \\*
& & & $\phantom{{}\{{}}-15360 (645-7 \specpar{1})$, $1024 (13500-396 \specpar{1}+\specpar{2})\}$, \\*
& & & $\{78$, $-2561$, $45808$, $-256 (1880-\specpar{1})$, $2965440-9216 \specpar{1}$, \\*
& & & $\phantom{{}\{{}}-15360 (645-7 \specpar{1})$, $1024 (13500-396 \specpar{1}+\specpar{2})\}$  \\
$\frac{9}{2}$ & $[\frac{1}{2},0]_{(-\frac{1}{2},\frac{1}{2},\frac{5}{2})} $ & $4$ & $\{4 (5-\specpar{1})$, $-8 (12-4 \specpar{1}-\specpar{2})\}^\joinsymbol$, \\*
& & & $\{28$, $-48 (5+\specpar{1})$, $640 (1+\specpar{1})$, $-64 (31 \specpar{1}+2 \specpar{2}+\specpar{3})\}$  \\
$\frac{9}{2}$ & $[\frac{1}{2},0]_{(-\frac{1}{2},\frac{3}{2},\frac{3}{2})} $ & $4$ & $\{8\}$, $\{8\}$, $\{8\}^\joinsymbol$, $\{8\}^\joinsymbol$, $\{12\}^\joinsymbol$, $\{12\}^\joinsymbol$, \\*
& & & $\{32$, $-320$, $64 (15+4 \specpar{2})$, $-1536 \specpar{2}\}$  \\
$\frac{9}{2}$ & $[\frac{1}{2},0]_{(\frac{1}{2},\frac{1}{2},\frac{3}{2})} $ & $3$ & $\{8\}$, $\{12\}$  \\
$\frac{9}{2}$ & $[\frac{1}{2},0]_{(\frac{1}{2},\frac{1}{2},\frac{3}{2})} $ & $4$ & $\{8\}^\joinsymbol$, $\{8\}^\joinsymbol$, $\{12\}^\joinsymbol$, $\{20$, $-96$, $128 \specpar{1}\}^\joinsymbol$, $\{35$, $-396$, $12 (120-\specpar{1})\}^\joinsymbol$, \\*
& & & $\{35$, $-396$, $12 (120-\specpar{1})\}^\joinsymbol$, $\{28$, $-240$, $640$, $-512 \specpar{1}\}$, \\*
& & & $\{40$, $-576$, $3520$, $-64 (120-\specpar{1})\}$, $\{40$, $-576$, $3520$, $-64 (120-\specpar{1})\}$  \\
$\frac{9}{2}$ & $[\frac{1}{2},0]_{(\frac{1}{2},\frac{1}{2},\frac{7}{2})} $ & $4$ & $\{0\}$  \\
$\frac{9}{2}$ & $[\frac{1}{2},0]_{(\frac{1}{2},\frac{3}{2},\frac{5}{2})} $ & $4$ & $\{20$, $-32 (3+\specpar{1})$, $32 (8 \specpar{1}+\specpar{2})\}^\joinsymbol$  \\
$\frac{9}{2}$ & $[\frac{1}{2},0]_{(\frac{3}{2},\frac{3}{2},\frac{3}{2})} $ & $4$ & $\{0\}$, $\{8\}^\joinsymbol$, $\{8\}^\joinsymbol$, $\{8\}^\joinsymbol$, $\{12\}^\joinsymbol$, $\{12\}^\joinsymbol$  \\
$\frac{9}{2}$ & $[1,\frac{1}{2}]_{(-\frac{3}{2},-\frac{1}{2},\frac{3}{2})} $ & $3$ & $\{4(5+2\specpar{1})$, $-4(24+25\specpar{1}-2\specpar{2})\}$  \\
$\frac{9}{2}$ & $[1,\frac{1}{2}]_{(-\frac{3}{2},\frac{1}{2},\frac{1}{2})} $ & $3$ & $\{35$, $-396$, $12 (120-\specpar{2})\}$, $\{35$, $-396$, $12 (120-\specpar{2})\}$  \\
$\frac{9}{2}$ & $[1,\frac{1}{2}]_{(-\frac{1}{2},-\frac{1}{2},\frac{1}{2})} $ & $3$ & $\{8\}$, $\{8\}$, $\{12\}$, $\{12\}$, $\{12\}^\joinsymbol$, $\{\frac{86}{3}$, $-\frac{8}{3} (75+2 \specpar{1})\}$, \\*
& & & $\{38$, $-465$, $24 (75+2 \specpar{1})\}$, $\{38$, $-465$, $24 (75+2 \specpar{1})\}$, \\*
& & & $\{38$, $-456$, $64 (27-\specpar{1})\}$  \\
$\frac{9}{2}$ & $[1,\frac{1}{2}]_{(-\frac{1}{2},-\frac{1}{2},\frac{5}{2})} $ & $3$ & $\{8\}$  \\
$\frac{9}{2}$ & $[1,\frac{1}{2}]_{(-\frac{1}{2},\frac{1}{2},\frac{3}{2})} $ & $3$ & $\{\frac{8 \specpar{1}}{3}+12\}^\joinsymbol$, \\*
& & &
$\{66$, $-3 (595+4 \specpar{1})$, $4 (6329+128 \specpar{1}+4 \specpar{2})$, \\*
& & & $\phantom{{}\{{}}-12 (16552+667 \specpar{1}+36 \specpar{2})$, $\phantom{{}\{{}}48 (17040+1131 \specpar{1}+78 \specpar{2}+\specpar{3})$, \\*
& & & $\phantom{{}\{{}}-32 (43200+4216 \specpar{1}+324 \specpar{2}+16 \specpar{3}-\specpar{4})\}$  \\
$\frac{9}{2}$ & $[1,\frac{1}{2}]_{(\frac{1}{2},\frac{1}{2},\frac{1}{2})} $ & $3$ & $\{8\}$, $\{8\}$, $\{8\}$, $\{8\}$, $\{8\}$, $\{8\}$, $\{12\}$, $\{12\}$, $\{12\}$, 
$\{12\}^\joinsymbol$, $\{12\}^\joinsymbol$, $\{12\}^\joinsymbol$, $\{15\}$, $\{15\}$, $\{15\}$, $\{15\}$, $\{15\}$, $\{15\}$, $\{\frac{50}{3}\}^\joinsymbol$  \\
$\frac{9}{2}$ & $[\frac{3}{2},0]_{(-\frac{1}{2},-\frac{1}{2},\frac{3}{2})} $ & $3$ & $\{12\}^\joinsymbol$  \\
$\frac{9}{2}$ & $[\frac{3}{2},0]_{(-\frac{1}{2},\frac{1}{2},\frac{1}{2})} $ & $3$ & $\{38$, $-456$, $64 (27-\specpar{1})\}$  \\
$\frac{9}{2}$ & $[\frac{3}{2},0]_{(\frac{1}{2},\frac{1}{2},\frac{3}{2})} $ & $3$ & $\{8\}$, $\{12\}$  \\
$\frac{9}{2}$ & $[\frac{3}{2},0]_{(\frac{3}{2},\frac{3}{2},\frac{3}{2})} $ & $3$ & $\{12\}^\joinsymbol$  \\
$\frac{9}{2}$ & $[\frac{3}{2},1]_{(-\frac{1}{2},\frac{1}{2},\frac{1}{2})} $ & $2$ & $\{\frac{86}{3}$, $-\frac{8}{3}(75+2 \specpar{1})\}$
 \label{tab: anomalous dimensions 922}
\end{longtable}
\end{center}

\section{\texorpdfstring{Representation content of the $\beta$-deformation}{Representation content of the beta-deformation}}
\label{app: representation content}

In this appendix, we describe how to determine the representation content of the $\beta$-deformation.
In Subsection~\ref{subsec: review of N=1}, we review some facts about the unitary representations of the $\cN=1$ superconformal algebra and give the explicit formulae for their characters.
In Subsection~\ref{subsec: adaption to the beta deformation}, we show how to adapt them to the $\beta$-deformation.
In Subsection~\ref{eq: super-sieve representation content}, we describe how to use the refined partition function and the characters to determine the representation content via the so-called Eratosthenes super-sieve algorithm.

\subsection{Representations and characters of \texorpdfstring{$\su{2,2|1}$}{su(2,2|1)}}
\label{subsec: review of N=1} 
In this subsection, we summarise some facts about the unitary representations of the $\cN=1$ superconformal algebra, which were first classified in \cite{Flato:1983te}. We stick to the notation of \cite{Dolan:2008qi}, which uses the same as the work \cite{Bianchi:2006ti} for the undeformed \NfSYMt.
We only give the final results necessary to understand our calculations and refer the reader to the literature for their derivations as well as the underlying theory.

The $\cN=1$ superconformal algebra $\su{2,2|1}$ is generated by the Lorentz transformations $M_{\mu\nu}$, the translations $P_\mu$, the dilatation $D$, the special conformal transformations $K_\mu$, the supersymmetry transformations $Q_\alpha,\bar Q_{\dot\alpha}$, the special superconformal transformations $S_\alpha, \bar S_{\dot\alpha}$ and the $\U{1}_R$-symmetry generator $R$.
In the spinor basis, the Lorentz generators $M_{\mu\nu}$ can be written in terms of the $\su{2}\times\overline{\mathfrak{su}}(2)$ generators $J_+,J_-,J_3$ and $\bar{J}_+,\bar{J}_-,\bar{J}_3$. Using the Pauli matrices $\sigma^\mu$, the translations and special conformal transformations can be rephrased as $P_{\alpha\dot\alpha}=\sigma^\mu_{\alpha\dot\alpha} P_\mu$, $K_{\alpha\dot\alpha}=\sigma^\mu_{\alpha\dot\alpha} K_\mu$.

A highest-weight (superconformal primary) state $\ket{\Delta,r,j,\bar{\jmath}}_{\text{hw}}$ of $\su{2,2|1}$  is specified by the requirements
\begin{equation}
 \begin{aligned}
(K_{\alpha\dot\alpha},S_{\alpha},\bar{S}_{\dot\alpha},J_+,\bar{J}_+)\ket{\Delta,r,j,\bar\jmath}_{\text{hw}}&=0 \eqncom\\
(D,R,J_3,\bar{J}_3)\ket{\Delta,r,j,\bar\jmath}_{\text{hw}}&=(\Delta,r,j,\bar\jmath)\ket{\Delta,r,j,\bar\jmath}_{\text{hw}} \eqndot
\end{aligned}
\end{equation}
All other states in the multiplet can be obtained by acting on this state with the lowering operators $P_{\alpha\dot\alpha}$, $Q_{\alpha}$, $\bar{Q}_{\dot\alpha}$, $J_-$ and $\bar{J}_-$.

The character of a representation can be thought of as a (refined) partition function, counting the number of states in the multiplet with a specified set of quantum numbers. The highest-weight state $\ket{\Delta,r,j,\bar\jmath}_{\text{hw}}$ is represented by the monomial $\sDO^{2\Delta}\,u^{r}\,x^{2j}\,\bar{x}^{2\bar\jmath}$. 
The action of the lowering operators increases or decreases the quantum numbers according to $P_{\alpha\dot\alpha} \sim \sDO^2 \,x^{\pm 1} \,\bar{x}^{\mp 1} $,
$Q_\alpha \sim \sDO \, u^{-1}\, x^{\pm1}$, $\bar{Q}_{\dot\alpha} \sim \sDO \, u \, \bar{x}^{\mp 1}$,
where $\alpha=1,2$ correspond to $x,x^{-1}$ and $\dot\alpha=\dot1,\dot2$ to $\bar{x}^{-1},\bar{x}$.
We denote the corresponding character by $\chi^{t,\bar t}_{(\Delta,r,j,\bar\jmath)}(\sDO,u,x,\bar{x})$.
Here, $t,\bar t$ are the fractions of the respective supercharges $Q_\alpha,\bar{Q}_{\dot\alpha}$ that cannot be used to generate new states within the multiplet. They are connected to possible constraints that will be specified later.

For a generic long representation, we have $t=\bar t=0$ and the character is 
\begin{equation}
 \chi^{0,0}_{(\Delta,r,j,\bar\jmath)}(\sDO,u,x,\bar{x})=\sDO^{2\Delta}\,u^{r}\,\chi_{2j+1}(x)\,
\chi_{2\bar\jmath+1}(\bar x)\, \cP(\sDO,x,\bar x)\,\cQ(\sDO\,u^{-1},x)\,{\cQ}(\sDO\,u,\bar x)\eqncom
\end{equation}
where
\begin{align}
 \cP(\sDO,x,\bar x)=\prod_{\epsilon,\eta=\pm 1}\frac{1}{(1-\sDO^2\, x^\epsilon \, \bar{x}^\eta)}\eqncom
\qquad \cQ(\sDO,x)=\prod_{\epsilon=\pm 1}(1+ \sDO\, x^\epsilon)\eqncom
\end{align}
and 
\begin{equation}
 \chi_n(x)= \frac{x^n-x^{-n}}{x-x^{-1}} 
\end{equation}
is the character of the usual $n$-dimensional representation of $\su2$.
Note that $\cP(\sDO,x,\bar x)$, $\cQ(\sDO\,u^{-1},x)$ and $\cQ(\sDO\,u,\bar x)$ account for the bosonic and fermionic -- but otherwise unconstrained -- action of $P_{\alpha\dot\alpha}$, $Q_{\alpha}$ and $\bar{Q}_{\dot\alpha}$, respectively, on the highest-weight state.

For unitary representations, two different kinds of constraints may occur for $Q_{\alpha}$ and $\bar{Q}_{\dot\alpha}$, respectively. They also lead to constraints on the representation labels $\Delta,r,j,\bar\jmath$.
The first kind is called shortening conditions and reads 
\begin{equation}\label{eq: shortening conditions}
\begin{aligned}
  & \bar{t}= 1: \qquad \Delta = +\frac{3}{2} r \eqncom \qquad 
\bar{Q}_{\dot\alpha} \ket{\Delta,r,j,0}_{\text{hw}}=0 \eqncom \\
& t = 1: \qquad \Delta = -\frac{3}{2} r \eqncom \qquad 
Q_\alpha \ket{\Delta,r,0,\bar\jmath}_{\text{hw}}=0 \eqncom
\end{aligned}
\end{equation}
i.e.\ $\bar{Q}_{\dot\alpha}$ and $Q_\alpha$, respectively, act as zero.
In the corresponding characters, the respective factors of $\cQ(\sDO\,u,\bar x)$ and $\cQ(\sDO\,u^{-1},x)$ are absent:
\begin{equation}
 \begin{aligned}
  \chi^{0,1}_{(+\frac{3}{2}r,r,j,0)}(\sDO,u,x,\bar{x})&{}= \sDO^{+3r}\,u^r\,
 \chi_{2j+1}(x)\,\cP(\sDO,x,\bar x)\, \cQ(\sDO\,u^{-1} ,x)\eqncom  & r \geq +\frac{2}{3}(j+1)\eqncom  \\
 \chi^{1,0}_{(-\frac{3}{2}r,r,0,\bar\jmath)}(\sDO,u,x,\bar{x})&{}=
 \sDO^{-3r}\,u^{r} \,\chi_{2\bar\jmath+1}(\bar x)\, \cP(\sDO,x,\bar x)\,{\cQ}(\sDO\,u,\bar x) \eqncom & r \leq -\frac{2}{3}(\bar\jmath+1) \eqndot 
 \end{aligned}
\end{equation}
The second kind is called semi-shortening conditions and reads 
\begin{equation}
\begin{aligned}
 \bar t =\frac{1}{2}: \quad \Delta = +\frac{3}{2}r+2\bar\jmath+2 \eqncom & \quad &\Big( \bar{Q}_{\dot 1} + \frac{1}{2\bar\jmath}   \bar{Q}_{\dot 2} \bar{J}_- \Big)
\ket{\Delta,r,j,\bar\jmath}_{\text{hw}} &= 0 \,\text{ for }\bar\jmath > 0 \eqncom \\
 &&\bar{Q}_{\dot 1} \ket{\Delta,r,j,0}_{\text{hw}} &= 0 \eqncom  \\
 t=\frac{1}{2}: \quad \Delta = -\frac{3}{2} r+2j+2 \eqncom &\quad &\Big ( Q_2 - \frac{1}{2j}  Q_1 J_- \Big)
\ket{\Delta,r,j,\bar\jmath}_{\text{hw}}  &= 0 \,\text{ for }  j >0 \eqncom \\
 & & Q_2 \ket{\Delta,r,0,\bar\jmath}_{\text{hw}} &= 0 \eqncom
\end{aligned}
\end{equation}
i.e.\ the action of $\bar{Q}_{\dot 1}$, or respectively $Q_{2}$, yields a state that can also be obtained via the other lowering operators. Accordingly, the monomials corresponding to these states must only appear once in the character, and the additionally occurring monomials capturing the action of $\bar{Q}_{\dot 1}$, or respectively $Q_2$, have to be removed:
\begin{equation}
 \begin{aligned}
 \chi^{0,\frac{1}{2}}_{(+\frac{3}{2}r+2\bar\jmath+2,r,j,\bar\jmath)}(\sDO,u,x,\bar{x})
 &=\sDO^{+3r+4\bar\jmath+4}\,u^{r}\,\chi_{2j+1}(x)\big(\chi_{2\bar\jmath+1}(\bar{x})+\sDO \,u\,
 \chi_{2\bar\jmath+2}(\bar x) \big) \\
&\phaneq \cP(\sDO,x,\bar x)\,\cQ(\sDO\,u^{-1} ,x)\eqncom \qquad \qquad r \geq {\frac{2}{3}}(j-\bar\jmath) \eqncom \\
 \chi^{\frac{1}{2},0}_{(-\frac{3}{2}r+2j+2,r,j,\bar\jmath)}(\sDO,u,x,\bar{x})
 &=\sDO^{-3r+4j+4}\,u^{r}\,\big(\chi_{2j+1}(x)+\sDO\, u^{-1}\,\chi_{2j+2}(x)\big) 
 \chi_{2\bar\jmath+1}(\bar x) \\ & \phaneq \cP(\sDO,x,\bar x)\,{\cQ}(\sDO\, u,\bar x) \eqncom 
 \qquad \qquad r \leq {\frac{2}{3}}(j-\bar\jmath) \eqndot
\end{aligned}
\end{equation}

So far, (semi-)shortening conditions have only been applied for either $Q_\alpha$ or $\bar{Q}_{\dot\alpha}$. If both $t$ and $\bar t$ are nonzero, the algebra relation $\acomm{Q_\alpha}{\bar{Q}_{\dot\alpha}}=2P_{\alpha\dot\alpha}$ requires that the contributions from the respective $P_{\alpha\dot\alpha}$ are also removed. In particular, the contribution from $P_{2\dot1}$ has to be eliminated for $t=\bar t=\frac{1}{2}$. In this case, we have 
\begin{equation}
 \begin{aligned}
  \chi^{\frac{1}{2},\frac{1}{2}}_{{(j+\bar\jmath+2,\frac{2}{3}(j-\bar\jmath),j,\bar\jmath)}}(\sDO,u,x,\bar{x})=
u^{\frac{2}{3}(j-\bar\jmath)}\big(&{}
\cD_{j,\bar\jmath}(\sDO,x,\bar x)+u^{-1}\,\cD_{j+\frac{1}{2},\bar\jmath}(\sDO,x,\bar x)\\
&{}+ u \, \cD_{j,\bar\jmath+\frac{1}{2}}(\sDO,x,\bar x)+\cD_{j+\frac{1}{2},\bar\jmath+\frac{1}{2}}(\sDO,x,\bar x)\big)\eqncom
 \end{aligned}
\end{equation}
where
\begin{equation}
 \begin{aligned}
  \cD_{j,\bar\jmath}(\sDO,x,\bar x)=\sDO^{2(j+\bar\jmath+2)}\big(\chi_{2j+1}(x)\,\chi_{2\bar\jmath+1}(\bar x)-\sDO^2
\chi_{2j}(x)\, \chi_{2\bar\jmath}(\bar x)\big) \cP(\sDO,x,\bar x)\eqndot
 \end{aligned}
\end{equation}
The remaining cases are 
\begin{equation}
 \begin{aligned}
  \chi^{\frac{1}{2},1}_{{(j+1,+\frac{2}{3}(j+1),j,0)}}(\sDO,u,x,\bar{x})&{}=u^{+\frac{2}{3}(j+1)}
\big(\cE_{j}(\sDO,x,\bar x)+ u^{-1}\,\cE_{j+\frac{1}{2}}(\sDO,x,\bar x)\big)\eqncom \\
\chi^{1,\frac{1}{2}}_{{(\bar\jmath+1,-\frac{2}{3}(\bar\jmath+1),0,\bar\jmath)}}(\sDO,u,x,\bar{x})&{}
=u^{-\frac{2}{3}(\bar\jmath+1)}\big({\bar \cE}_{\bar\jmath}(\sDO,x,\bar x)+u
\,{\bar \cE}_{\bar\jmath+\frac{1}{2}}(\sDO,x,\bar x)\big)\eqncom
 \end{aligned}
\end{equation}
where
\begin{equation}\label{eq: definition EEbar}
 \begin{aligned}
  \cE_{j}(\sDO,x,\bar x)&{}=\sDO^{2j+2}\big(\chi_{2j+1}(x)-\sDO^2
\chi_{2j}(x)\,\chi_{2}(\bar x)+\sDO^4 \chi_{2j-1}(x)\big)
\cP(\sDO,x,\bar x)\eqncom \\
{\bar \cE}_{\bar\jmath}(\sDO,x,\bar x)&{}=\sDO^{2\bar\jmath+2}\big(\chi_{2\bar\jmath+1}(\bar x)-
\sDO^2 \chi_{2}(x)\,\chi_{2\bar\jmath}(\bar x)+
\sDO^4 \chi_{2\bar\jmath-1}(\bar x)\big)\cP(\sDO,x,\bar x)\eqndot
 \end{aligned}
\end{equation}

Unitarity requires that $\Delta \geq \max( 2+2\bar\jmath+\frac{3}{2}r,2+2j-\frac{3}{2}r)$, unless
one of the shortening conditions in \eqref{eq: shortening conditions} is fulfilled, in which case $\Delta = \pm\frac{3}{2}r $. At this so-called unitary threshold, otherwise irreducible representations become reducible, which is also reflected in the characters. The resulting reductions  of the characters that alter the value of $\bar t$ are given by 
\begin{equation}\label{eq: splitting}
 \begin{aligned}
  \chi^{0,0}_{(+\frac{3}{2}r+2\bar\jmath+2,r,j,\bar\jmath)}(\sDO,u,x,\bar x) &= 
 \chi^{0,\frac{1}{2}}_{(+\frac{3}{2}r+2\bar\jmath+2,r,j,\bar\jmath)}(\sDO,u,x,\bar x)\\
 &\phaneq+ \chi^{0,\frac{1}{2}}_{(+\frac{3}{2}r+2\bar\jmath+\frac{5}{2},r+1,j,\bar\jmath-\frac{1}{2})}(\sDO,u,x,\bar x)\qquad\qquad\text{ for }\bar\jmath>0 \eqncom \\
\chi^{0,0}_{(+\frac{3}{2}r+2,r,j,0)}(\sDO,u,x,\bar x) &=  
 \chi^{0,\frac{1}{2}}_{(+\frac{3}{2}r+2,r,j,0)}(\sDO,u,x,\bar x)+
\chi^{0,1}_{(+\frac{3}{2}r+3,r+2,j,0)}(\sDO,u,x,\bar x) \eqncom \\
 \chi^{\frac{1}{2},0}_{{(j+\bar\jmath+2,\frac{2}{3}(j-\bar\jmath),j,\bar\jmath)}}(\sDO,u,x,\bar x)&= 
 \chi^{\frac{1}{2},\frac{1}{2}}_{{(j+\bar\jmath+2,\frac{2}{3}(j-\bar\jmath),j,\bar\jmath)}}(\sDO,u,x,\bar x) \\
 &\phaneq+  \chi^{0,\frac{1}{2}}_{{(j+\bar\jmath+\frac{5}{2},\frac{2}{3}(j-\bar\jmath)+1,j,\bar\jmath-\frac{1}{2})}}(\sDO,u,x,\bar x) \qquad\qquad\text{ for }\bar\jmath>0  \eqncom \\
 \chi^{\frac{1}{2},0}_{{(j+2,\frac{2}{3}j,j,0)}}(\sDO,u,x,\bar x)&= 
 \chi^{\frac{1}{2},\frac{1}{2}}_{{(j+2,\frac{2}{3}j,j,0)}}(\sDO,u,x,\bar x) +  \chi^{0,1}_{{(j+3,\frac{2}{3}j+2,j,0)}}(\sDO,u,x,\bar x)   \eqncom
 \end{aligned}
\end{equation}
and they can be translated directly to the respective reductions of the representations.
The analogous relations that alter $t$ have been omitted and can be obtained by replacing 
\begin{equation}
\chi^{\tilde{t},\tilde{\bar t}}_{(\tilde\Delta(r,j,\bar\jmath),\tilde r(r,j,\bar\jmath),\tilde j(j,\bar\jmath),\tilde{\bar\jmath}(j,\bar\jmath))}
\to
\chi^{\tilde{\bar t},\tilde{t}}_{(\tilde\Delta(-r,\bar\jmath,j),-\tilde r(-r,\bar\jmath,j),\tilde{\bar\jmath}(\bar\jmath,j),\tilde j(\bar\jmath,j))}
\eqncom
\end{equation}
where the quantities with tildes stand for the abstract functions of $r,j,\bar \jmath$ that are specified in $\chi$.
If a multiplet acquires an anomalous dimension, it moves away from the unitary threshold\footnote{Recall that $\Delta= \Delta_0+\gamma$, where $\Delta_0$ denotes the classical scaling dimension and $\gamma=g^2 E+\cO(g^3)$ the anomalous one.} and the pairs of representations on the right hand sides of \eqref{eq: splitting} join again.

\subsection{Adaption to the \texorpdfstring{$\beta$}{beta}-deformation}
\label{subsec: adaption to the beta deformation}

The results reviewed above can be easily applied to the $\beta$-de\-for\-ma\-tion. 
We identify $Q_\alpha$ with the $\cN=4$ supercharge $Q^{A=4}_\alpha$ and the $\U{1}_R$-symmetry charge $r$ with the combination of $\su{4}$ Cartan charges specified in Table~\ref{tab: su(4) charges}.
Furthermore, the generators of the superconformal algebra $\su{2,2|1}$ are supplemented by the generators $Q^1$ and $Q^2$ of the global $\U{1}_{Q^1}\times\U{1}_{Q^2}$ symmetry. In the characters, they are represented by the fugacities $v$ and $w$. As all irreducible representations of abelian groups are one-dimensional, the above character formulae simply have to be supplemented by $v^{Q^1}\,w^{Q^2}$.

\subsection{The super-sieve algorithm}
\label{eq: super-sieve representation content}

Using the above characters, the $\cN=1$ representation content of the free $\beta$-de\-for\-ma\-tion can be determined directly from the refined partition function in analogy to the \NfSYMt case \cite{Bianchi:2003wx}, see also \cite{Bianchi:2006ti}.\footnote{Note that the free $\beta$-deformation is identical to the free \NfSYMt.}

The refined partition function of single-trace operators in the  $\beta$-deformation can be obtained via Polya theory, see e.g.\ \cite{Bianchi:2006ti}. It is given by\footnote{The sign factor in front of $\sDO^k$  takes care of the fact that a state with integer or half-integer classical scaling dimension obeys the Bose-Einstein or Fermi-Dirac statistic, respectively. Note that we organise signs slightly different than the authors of \cite{Bianchi:2006ti}.}
\begin{equation}
\label{eq: polya}
\begin{aligned}
Z(\sDO,x,\bar x, u,v,w)&=-\sum_{k=1}^\infty \frac{\varphi(k)}{k}\ln \left[1-z((-1)^{k+1}\sDO^k,x^k,\bar x^k, u^k,v^k,w^k)\right]\\
&\phaneq-\colors\, z(\sDO,x,\bar x, u,v,w)
 \eqncom
\end{aligned}
\end{equation}
where $\varphi(k)$ is the Euler totient function giving the number of positive integers less than or equal to $k$ that are relative prime to $k$. The single-site partition function can be found in analogy to \cite{Bianchi:2006ti}, and it reads
\begin{equation}\label{eq: single-site partition function}
\begin{aligned}
z(\sDO,x,\bar x, u,v,w)&= (u^{\frac{2}{3}}v^{1}+ u^{\frac{2}{3}}v^{-1}w^{1}+u^{\frac{2}{3}}w^{-1})\cE_0(\sDO,x,\bar x) \\
&\phaneq +(u^{-\frac{2}{3}}v^{-1}+ u^{-\frac{2}{3}}v^{1}w^{-1}+u^{-\frac{2}{3}}w^{1})\bar{\cE}_0(\sDO,x,\bar x)  \\
&\phaneq +(u^{-\frac{1}{3}}v^{1}+ u^{-\frac{1}{3}}v^{-1}w^{1}+u^{-\frac{1}{3}}w^{-1}+u^1)\cE_{\frac{1}{2}}(\sDO,x,\bar x)\\
&\phaneq +(u^{\frac{1}{3}}v^{-1}+ u^{\frac{1}{3}}v^{1}w^{-1}+u^{\frac{1}{3}}w^{1}+u^{-1})\bar{\cE}_{\frac{1}{2}}(\sDO,x,\bar x)\\
&\phaneq + \cE_1(\sDO,x,\bar x) + \bar{\cE}_1(\sDO,x,\bar x) \eqncom
\end{aligned}
\end{equation}
where $\cE$ and $\bar\cE$ were defined in \eqref{eq: definition EEbar}. The first line in \eqref{eq: single-site partition function} accounts for the scalars, the second one for the anti-scalars, the third one for the fermions, the fourth one for the anti-fermions and the last one for the self-dual and anti-self-dual component of the field strength, cf.\ Table~\ref{tab: su(4) charges}. 

Using the refined partition function and the characters, the representation content can be determined via the Eratosthenes super-sieve algorithm proposed in \cite{Bianchi:2003wx}. Starting with the refined partition function \eqref{eq: polya}, one identifies as a highest-weight state the state corresponding to the monomial with the smallest exponent of $\sDO$ and subordinately largest exponent of $x$, $\bar x$. One then determines the character of the representation containing this state as highest-weight state and subtracts this character from the refined partition function. The result of this subtraction serves as input for the next iteration. In this way, the refined partition function can be uniquely expressed as the sum of the characters of the representations in the (free) theory.

For the \NfSYMt, the $\cN=4$ representation content was determined for $\Delta_0\leq4$ in \cite{Bianchi:2003wx}. Hence, as an alternative to the method described above, one can use the super-sieve algorithm to decompose the $\cN=4$ characters of these representations into $\cN=1$ characters.\footnote{Note that the \su4 Cartan charges used in \cite{Bianchi:2006ti} have to be translated to the basis $\U1_{Q^1}\times\U1_{Q^2}\times\U1_{R}$, which leads to the following replacements in the character formulae of \cite{Bianchi:2003wx}: 
\begin{equation}
 u_1\to u^{-\frac{1}{3}}\,v^1 \eqncom u_2\to u^{-\frac{1}{3}}\,v^{-1}\,w^{1} \eqncom u_3 \to u^{-\frac{1}{3}}\,w^{-1}\eqncom u_4 \to u^{1} \eqndot                                                                                                                                                                                                                                                                                                                         \end{equation}
}
The results of both methods agree.

\section{The harmonic action}
\label{subsec: Mathematica implementation of the harmonic action}

In this appendix, we present a completely explicit expression of the harmonic action in a form suitable for an implementation e.g.\ in {\tt Mathematica}. Although apparently different, it is nevertheless equivalent to the one given in \cite{Beisert03}.

The single-site states of the spin chain of \NfSYMt are taken from the alphabet given in \eqref{eq: alphabet}.
The occurring covariant derivatives have been translated from Minkowski indices to spinor indices using the Pauli matrices $(\sigma^\mu)_{\alpha\dot\alpha}$: $\D_{\alpha\dot\alpha}=\D_\mu (\sigma^\mu)_{\alpha\dot\alpha}$. Using the antisymmetric products of Pauli matrices $\sigma^{\mu\nu}$ and $\bar\sigma^{\mu\nu}$, the field strength is translated to the spinor basis and split into its self-dual and anti-self-dual part as $\cF_{\alpha\beta}=(\sigma^{\mu\nu})_{\alpha\beta} F_{\mu\nu}$ and $\bar\cF_{\dot\alpha\dot\beta}=(\bar\sigma^{\mu\nu})_{\dot\alpha\dot\beta} F_{\mu\nu}$, respectively.

In terms of the bosonic $\su2$ and $\overline{\mathfrak{su}}(2)$ oscillators $\aoscdag_\alpha$, $\alpha=1,2$, and $\boscdag_{\dot\alpha}$, $ \dot\alpha=\dot1,\dot2$, as well as the fermionic $\su4$ oscillators $\coscdag_A$, $A=1,2,3,4$, the fields of the alphabet can be written as 
\begin{equation}\label{eq: fields}
\begin{aligned}
\D^k \cF^{\phantom{ABC}} &\mathrel{\widehat{=}} 
  (\aoscdag)^{k+2} 
  (\boscdag)^{k\phantom{+0}}
  \vac \eqncom \\
\D^k \ferm^{A\phantom{BC}} &\mathrel{\widehat{=}}     
  (\aoscdag)^{k+1} 
  (\boscdag)^{k\phantom{+0}}
  \cosc^{\dagger}_A 
  \vac \eqncom \\
\D^k \varphi^{AB\phantom{C}} &\mathrel{\widehat{=}}     
  (\aoscdag)^{k\phantom{+0}} 
  (\boscdag)^{k\phantom{+0}} 
  \cosc^{\dagger}_A \cosc^{\dagger} _B
  \vac \eqncom \\
\D^k \bar\ferm^{ABC} &\mathrel{\widehat{=}} 
  (\aoscdag)^{k\phantom{+0}} 
  (\boscdag)^{k+1} 
  \cosc^{\dagger}_A \cosc^{\dagger}_B \cosc^{\dagger}_C
  \vac \eqncom \\
\D^k \bar\cF^{\phantom{ABC}} &\mathrel{\widehat{=}}   
  (\aoscdag)^{k\phantom{+0}}
  (\boscdag)^{k+2} 
  \cosc^{\dagger}_1 \cosc^{\dagger}_2 \cosc^{\dagger}_3 \cosc^{\dagger}_4 
  \vac \eqncom
\end{aligned}
\end{equation}
where $\phi^i\propto \varphi^{i4}$, $\bar\phi_i\propto\epsilon_{ABi4}\varphi^{AB}$, with antisymmetric $\varphi$, and $\bar\ferm^{ABC}=\frac{1}{3!}\epsilon_{ABCD}\bar\ferm^{D}$.\footnote{The precise constants of proportionality are of no importance here, as they only lead to a change of basis. The corresponding similarity transformation leaves the spectrum of the dilatation operator invariant.}
We denote the numbers of $\aoscdag_1$, $\aoscdag_2$, $\boscdag_{\dot1}$, $\boscdag_{\dot2}$, $\coscdag_1$, $\coscdag_2$, $\coscdag_3$, $\coscdag_4$ oscillators at spin-chain site $i$ by $\akindsite[1]{i},\akindsite[2]{i},\bkindsite[\dot1]{i},\bkindsite[\dot2]{i},\ckindsite[1]{i},\ckindsite[2]{i},\ckindsite[3]{i},\ckindsite[4]{i}$.
These are connected to the spins $j$ and $\bar\jmath$ as $j=\frac12(\akind[1]$-$\akind[2])$ and $\bar\jmath=\frac12(\bkind[\dot2]$-$\bkind[\dot1])$.

For two initial and final single-site states defined by canonically ordered oscillators with the occupation numbers 
\begin{equation}
 \begin{aligned}
 A_{(1)}&=(\akindsite[1]{1},\akindsite[2]{1},\bkindsite[\dot1]{1},\bkindsite[\dot2]{1},\ckindsite[1]{1},\ckindsite[2]{1},\ckindsite[3]{1},\ckindsite[4]{1}) \eqncom\\
 A_{(2)}&=(\akindsite[1]{2},\akindsite[2]{2},\bkindsite[\dot1]{2},\bkindsite[\dot2]{2},\ckindsite[1]{2},\ckindsite[2]{2},\ckindsite[3]{2},\ckindsite[4]{2}) \eqncom
\end{aligned}
\end{equation}
and 
\begin{equation}
 \begin{aligned}
 A_{(3)}&=(\akindsite[1]{3},\akindsite[2]{3},\bkindsite[\dot1]{3},\bkindsite[\dot2]{3},\ckindsite[1]{3},\ckindsite[2]{3},\ckindsite[3]{3},\ckindsite[4]{3}) \eqncom\\
 A_{(4)}&=(\akindsite[1]{4},\akindsite[2]{4},\bkindsite[\dot1]{4},\bkindsite[\dot2]{4},\ckindsite[1]{4},\ckindsite[2]{4},\ckindsite[3]{4},\ckindsite[4]{4})
\eqncom
\end{aligned}
\end{equation}
respectively, such that $A_{(1)}+A_{(2)}=A_{(3)}+A_{(4)}$, the concrete expression of the harmonic action reads
\begin{equation}\label{eq: harminic action}
 \begin{aligned}
 (\diladensity^{\cN=4}_2)_{A_{(1)}A_{(2)}}^{A_{(3)}A_{(4)}}&=
\prod_{\alpha=1}^2\left(\sum_{\akind[\alpha]=\maxset{\akindsite[\alpha]{3} - \akindsite[\alpha]{2},
0}}^{\minset{\akindsite[\alpha]{1}, \akindsite[\alpha]{3}}}
\binom{\akindsite[\alpha]{1}}{\akind[\alpha]}
\binom{\akindsite[\alpha]{2}}{\akindsite[\alpha]{3} - \akind[\alpha]} \right)\\
&\phaneq \prod_{\alphadot=\dot1}^{\dot2}\left(\sum_{\bkind[\alphadot]=\maxset{\bkindsite[\alphadot]{3} - \bkindsite[\alphadot]{2},
0}}^{\minset{\bkindsite[\alphadot]{1}, \bkindsite[\alphadot]{3}}}
 \binom{\bkindsite[\alphadot]{1}}{\bkind[\alphadot]}
\binom{\bkindsite[\alphadot]{2}}{\bkindsite[\alphadot]{3} - \bkind[\alphadot]}\right) \\
&\phaneq \prod_{a=1}^4\left(\sum_{\ckind[a]=\maxset{\ckindsite[a]{3} - \ckindsite[a]{2},
0}}^{\minset{\ckindsite[a]{1}, \ckindsite[a]{3}}}
\binom{\ckindsite[a]{1}}{\ckind[a]}
\binom{\ckindsite[a]{2}}{\ckindsite[a]{3} - \ckind[a]}\right)\\
 & \phaneq c\Big[{\textstyle \sum_{i=1}^2 \big(\sum_{\beta=1}^2 \akindsite[\beta]{i}+\sum_{\betadot=\dot1}^{\dot2} \bkindsite[\betadot]{i}+\sum_{B=1}^4 \ckindsite[B]{i}\big)}
,\\
&\phaneq \hphantom{c\Big[} {\textstyle \sum_{\beta=1}^2 (\akindsite[\beta]{1}-\akind[\beta])+\sum_{\betadot=\dot1}^{\dot2} (\bkindsite[\betadot]{1}-\bkind[\betadot])+\sum_{B=1}^4 (\ckindsite[B]{1}-\ckind[B])}
,\\
&\phaneq\hphantom{c\Big[}{\textstyle \sum_{\beta=1}^2 (\akindsite[\beta]{3}-\akind[\beta])+\sum_{\betadot=\dot1}^{\dot2} (\bkindsite[\betadot]{3}-\bkind[\betadot])+\sum_{B=1}^4 (\ckindsite[B]{3}-\ckind[B]) }
\Big]\\
 &\phaneq(-1)^{(\ckindsite[1]{1} - \ckind[1]
      + \ckindsite[2]{1} - \ckind[2]
      + \ckindsite[3]{1} - \ckind[3]
      + \ckindsite[4]{1} - \ckind[4])
    (\ckindsite[1]{3} - \ckind[1]
      + \ckindsite[2]{3} - \ckind[2]
      + \ckindsite[3]{3} - \ckind[3]
      + \ckindsite[4]{3} - \ckind[4])}\\
&\phaneq(-1)^{( \ckind[2]
         + \ckind[3]
         + \ckind[4])
        (\ckindsite[1]{1}+\ckindsite[1]{3})}\\
 &\phaneq(-1)^{(\ckindsite[1]{2} - \ckindsite[1]{3} + \ckind[1]
         + \ckind[3]
         + \ckind[4])
        (\ckindsite[2]{1}+\ckindsite[2]{3})}\\
 &\phaneq(-1)^{(\ckindsite[1]{2} - \ckindsite[1]{3} + \ckind[1]
         +\ckindsite[2]{2} - \ckindsite[2]{3} + \ckind[2]
         + \ckind[4])
        (\ckindsite[3]{1}+\ckindsite[3]{3})}\\
 &\phaneq(-1)^{(\ckindsite[1]{2} - \ckindsite[1]{3} + \ckind[1]
         +\ckindsite[2]{2} - \ckindsite[2]{3} + \ckind[2]
         +\ckindsite[3]{2} - \ckindsite[3]{3} + \ckind[3])
        (\ckindsite[4]{1} + \ckindsite[4]{3})} 
\eqndot
\end{aligned}
\end{equation}
The coefficients $c[n,n_{12},n_{21}]$ are given in terms of the harmonic numbers $h(k)=\sum_{i=1}^k\frac{1}{i}$ and the Euler gamma function $\Gamma$ as 
\begin{equation}\label{eq: harmonic action coefficient}
 c[n,n_{12},n_{21}]=\begin{cases}2 h(\frac12 n) & \text{if }n_{12}=n_{21}=0 \eqncom\\
                      2(-1)^{1+n_{12}n_{21}} 
\frac{\Gamma(\frac12(n_{12}+n_{21}))\Gamma(1+\frac12(n-n_{12}-n_{21}))}{\Gamma(1+\frac12
n)} & \text{else.}
                     \end{cases}
\end{equation}

\newpage
\section{A cancellation mechanism}
\label{app: argument}
 
In this appendix, we argue that in the tower of conformal primary states built from $n$ covariant derivatives $\D_{\alpha\dot\alpha}$ distributed on $\tr[\phi^2\phi^3]$ only the lowest ($n=0$) state is affected by prewrapping as the contributions from s-channel diagrams to all elements with $n\geq1$ vanish due to cancellations in the spacetime part.

A conformal primary state is, by definition, annihilated by all raising operators. In the oscillator picture of \cite{Beisert03}, these are 
\begin{equation}
 J_+=\aoscdag_1\aosc^2\eqncom\quad \bar{J}_+=\boscdag_{\dot{2}}\bosc^{\dot{1}}\quad \text{ and } \quad K^{\alpha\dot\alpha}=\aosc^\alpha\bosc^{\dot\alpha}\eqndot
\end{equation}
It is easy to see that the operators
\begin{equation}\label{eq: conformal primary}
 \sum_{k=0}^n\frac{(-1)^k}{k!^2(n-k)!^2}\tr[(\D_{1\dot2})^{k}\phi^2(\D_{1\dot2})^{n-k}\phi^3] 
\end{equation}
form indeed a tower of conformal\footnote{Note that the corresponding \emph{super}conformal primaries have classical scaling dimensions $n+\frac{3}{2}$ if $n\geq1$.} primaries if we identify $\D_{\alpha\dot\alpha}=\aoscdag_\alpha\boscdag_{\dot\alpha}$ and recall that $\phi^2,\phi^3$ are built from only $\coscdag$-oscillators, cf.\ Appendix~\ref{subsec: Mathematica implementation of the harmonic action}.

In dimensional regularisation with $D=4-2\varepsilon$, the relevant one-loop tensor integrals are
\begin{equation}\label{eq: general integral}
\frac{p_{(\alpha_1{\alphadot}_1}\dots p_{\alpha_n{\alphadot}_n)}}{p^{2(2-\frac{D}{2})}} G_{(n)}(1,1)=\int \frac{\de^D l}{(2\pi)^D} \frac{l_{(\alpha_1{\alphadot}_1}\dots l_{\alpha_n{\alphadot}_n)}}{l^2(p-l)^2}
 \eqncom
\end{equation}
where
\begin{equation}
 G_{(n)}(1,1)=\frac{1}{(4\pi)^2\varepsilon}\frac{1}{n+1} + \cO(\varepsilon^0) \eqncom
\end{equation}
and the parenthesis denote total symmetrisation in both kinds of spinor indices \cite{Chetyrkin:1980pr}. Hence, the divergent part of the integrals found from the s-channel Feynman diagrams involving one individual operator from the sum \eqref{eq: conformal primary} are given by
\begin{equation}\label{eq: spacetime integral}
 \int \frac{\de^D l}{(2\pi)^D} \frac{(p_{1\dot2}-l_{1\dot2})^{k}(l_{1\dot2})^{n-k}}{l^2(p-l)^2} \sim \frac{1}{(4\pi)^2\varepsilon} \sum_{m=0}^k\frac{(-1)^m}{n-k+m+1}\binom{k}{m}=\frac{1}{(4\pi)^2\varepsilon}\frac{k!(n-k)!}{(n+1)!}\eqndot
\end{equation}
The divergence of the one-loop s-channel diagrams involving the operators \eqref{eq: conformal primary} is obtained by replacing the trace factor in \eqref{eq: conformal primary} with the result of \eqref{eq: spacetime integral}. This yields zero unless $n=0$.

\bibliographystyle{utcaps}
\bibliography{ThesisINSPIRE}

\end{fmffile}

\end{document}